\documentclass[12pt]{iopart}
\usepackage{color}
\usepackage{epic,eepic} 
\usepackage[mathscr]{eucal} 
\usepackage{amsmath}
\usepackage{amssymb}
\usepackage{amscd}

\usepackage[stdtext,vcentermath]{myyoungtab}

\numberwithin{equation}{section}
\newtheorem{Theorem}{Theorem}[section]
\newtheorem{Prop}[Theorem]{Proposition}
\newtheorem{Lemma}[Theorem]{Lemma}

\newtheorem{Definition}[Theorem]{Definition}
\newtheorem{Example}[Theorem]{Example}
\newtheorem{Remark}[Theorem]{Remark}

\newcommand{\R}{\mathbb{R}}
\newcommand{\Z}{\mathbb{Z}}

\newcommand{\ve}{\varepsilon}

\newcommand{\proof}{\noindent \textit{Proof. }}
\newcommand{\qed}{\hfill $\Box$}

\newcommand\et[1]{\tilde{e}_{#1}}
\newcommand\ft[1]{\tilde{f}_{#1}}

\newcommand\ot{\otimes}

\newcommand{\onebox}[1]{%
  \setlength{\unitlength}{5mm}
  \begin{picture}(1,1)(0,0.3)
  \multiput(0,0)(1,0){2}{\line(0,1){1}}
  \multiput(0,0)(0,1){2}{\line(1,0){1}}
  \put(0,0){\makebox(1,1){${#1}$}}
  \end{picture}
}
\newcommand{\twoboxes}[2]{%
  \setlength{\unitlength}{5mm}
  \begin{picture}(2,1)(0,0.3)
  \multiput(0,0)(1,0){3}{\line(0,1){1}}
  \multiput(0,0)(0,1){2}{\line(1,0){2}}
  \put(0,0){\makebox(1,1){${#1}$}}
  \put(1,0){\makebox(1,1){${#2}$}}
  \end{picture}
}
\newcommand{\threeboxes}[3]{%
  \setlength{\unitlength}{5mm}
  \begin{picture}(3,1)(0,0.3)
  \multiput(0,0)(1,0){4}{\line(0,1){1}}
  \multiput(0,0)(0,1){2}{\line(1,0){3}}
  \put(0,0){\makebox(1,1){${#1}$}}
  \put(1,0){\makebox(1,1){${#2}$}}
  \put(2,0){\makebox(1,1){${#3}$}}
  \end{picture}
}

\newcommand{\fiveboxes}[5]{%
  \setlength{\unitlength}{5mm}
  \begin{picture}(5,1)(0,0.3)
  \multiput(0,0)(1,0){6}{\line(0,1){1}}
  \multiput(0,0)(0,1){2}{\line(1,0){5}}
  \put(0,0){\makebox(1,1){${#1}$}}
  \put(1,0){\makebox(1,1){${#2}$}}
  \put(2,0){\makebox(1,1){${#3}$}}
  \put(3,0){\makebox(1,1){${#4}$}}
  \put(4,0){\makebox(1,1){${#5}$}}
   \end{picture}
}
\newcommand{\sixboxes}[6]{%
 \setlength{\unitlength}{5mm}
 \begin{picture}(6,1)(0,0.3)
 \multiput(0,0)(1,0){7}{\line(0,1){1}}
 \multiput(0,0)(0,1){2}{\line(1,0){6}}
 \put(0,0){\makebox(1,1){${#1}$}}
 \put(1,0){\makebox(1,1){${#2}$}}
 \put(2,0){\makebox(1,1){${#3}$}}
 \put(3,0){\makebox(1,1){${#4}$}}
 \put(4,0){\makebox(1,1){${#5}$}}
 \put(5,0){\makebox(1,1){${#6}$}}
  \end{picture}
}

\newcommand{\batten}[4]{%
\begin{picture}(40,40)(-20,-20)
	\put(-10,0){\line(1,0){20}}
	\thicklines
	\put(0,10){\line(0,-1){20}}
	\put(-11,0){\makebox(0,0)[r]{$#1$}}
	\put(0,11){\makebox(0,0)[b]{$#2$}}
	\put(0,-11){\makebox(0,0)[t]{$#3$}}
	\put(11,0){\makebox(0,0)[l]{$#4$}}
\end{picture}
}

\newcommand{\battennew}[4]{%
\begin{picture}(40,40)(-20,-20)
	\put(-10,0){\line(1,0){20}}
	\thicklines
	\put(0,9){\line(0,-1){18}}
	\put(-11,0){\makebox(0,0)[r]{$#1$}}
	\put(0,11){\makebox(0,0)[b]{$#2$}}
	\put(0,-11){\makebox(0,0)[t]{$#3$}}
	\put(11,0){\makebox(0,0)[l]{$#4$}}
\end{picture}
}
\newcommand{\threeboxesnew}[3]{%
  \setlength{\unitlength}{5mm}
  \begin{picture}(3,1)(0,0.3)
  \multiput(0,-0.1)(1,0){4}{\line(0,1){1.2}}
  \multiput(0,-0.1)(0,1.2){2}{\line(1,0){3}}
  \put(0,0){\makebox(1,1){${#1}$}}
  \put(1,0){\makebox(1,1){${#2}$}}
  \put(2,0){\makebox(1,1){${#3}$}}
  \end{picture}
}

\newcommand{\battendot}[4]{%
\begin{picture}(40,40)(-20,-20)
	\put(-10,0){\line(1,0){20}}
	\put(-2.3,-2.3){$\bullet$}
	\thicklines
	\put(0,9){\line(0,-1){18}}
	\put(-11,0){\makebox(0,0)[r]{$#1$}}
	\put(0,11){\makebox(0,0)[b]{$#2$}}
	\put(0,-11){\makebox(0,0)[t]{$#3$}}
	\put(11,0){\makebox(0,0)[l]{$#4$}}
\end{picture}
}

\begin{document}

\bibliographystyle{amsalpha}

\title[Integrable structure of box-ball systems]{Integrable structure of 
box-ball systems:
crystal, Bethe ansatz, ultradiscretization and tropical geometry}

\vspace{0.7cm}

\begin{center}
{\it Dedicated to the memory of Professor Miki Wadati$\qquad\quad\;$}
\end{center}

\author[R.\ Inoue]{Rei Inoue}
\address{
Department of Mathematics and Informatics, Faculty of Science, 
Chiba University
Inage, Chiba 263-8522, Japan}

\author[A.\ Kuniba]{Atsuo Kuniba}
\address{Institute of Physics, University of Tokyo, 
Komaba, Tokyo 153-8902, Japan}

\author[T.\ Takagi]{Taichiro Takagi}
\address{Department of Applied Physics, National Defense Academy, 
Kanagawa 239-8686, Japan}

\begin{abstract}
The box-ball system is an integrable cellular automaton
on one dimensional lattice.
It arises from either quantum or classical integrable systems by 
the procedures called crystallization and ultradiscretization, respectively.
The double origin of the integrability has endowed the box-ball system with 
a variety of aspects related to
Yang-Baxter integrable models in statistical mechanics,
crystal base theory in quantum groups,
combinatorial Bethe ansatz, geometric crystals, 
classical theory of solitons, tau functions, 
inverse scattering method, 
action-angle variables and invariant tori in completely integrable systems,
spectral curves, tropical geometry and so forth.
In this review article, we demonstrate 
these integrable structures of the box-ball system and 
its generalizations
based on the developments in the last two decades.
\end{abstract}

\maketitle


\section{Introduction}

\subsection{The box-ball system}
\label{i:sec-intro}

The {\it box-ball system} (BBS for short) is a cellular automaton
introduced by Takahashi and Satsuma in 1990 \cite{TS90}.
It is a dynamical system of finitely many balls in 
an infinite number of boxes aligned on a line,
whose time evolution is given by the following rule.
We assume that each box can accommodate one ball at most.
\begin{enumerate}
\item Move the leftmost ball to its nearest right empty box.
\item Move the leftmost ball among the rest 
to its nearest right empty box.
\item Repeat (ii) until all the balls are moved exactly once. 
\end{enumerate} 
This defines an update corresponding to the one time step $t\rightarrow t+1$.
We remark that the above evolution rule is invertible.
Let us show an example. By starting with the following configuration 
at time zero,
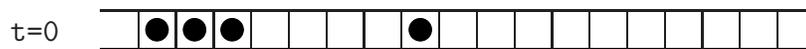
\begin{center}
\unitlength=1.0mm
\begin{picture}(100,8)(5,4)

\put(-2,6){\texttt{t=0}}
\put(10,10){\line(1,0){95}}
\put(10,5){\line(1,0){95}}
\multiput(15,5)(5,0){18}{\line(0,1){5}}

\put(17.5,7.5){\circle*{3}} 
\put(22.5,7.5){\circle*{3}} 
\put(27.5,7.5){\circle*{3}} 
\put(52.5,7.5){\circle*{3}} 

\end{picture}
\end{center}
\noindent
we obtain the configuration at $t=1$ as

\begin{center}
\unitlength=1.0mm
\begin{picture}(100,5)(5,7)

\put(-2,6){\texttt{t=1}}
\put(10,10){\line(1,0){95}}
\put(10,5){\line(1,0){95}}
\multiput(15,5)(5,0){18}{\line(0,1){5}}

\put(32.5,7.5){\circle*{3}} 
\put(37.5,7.5){\circle*{3}} 
\put(42.5,7.5){\circle*{3}} 
\put(57.5,7.5){\circle*{3}} 

\end{picture}
\end{center}
and so on, ...

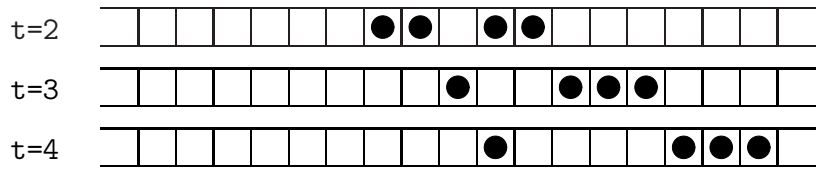
\begin{center}
\unitlength=1.0mm
\begin{picture}(100,22)(5,-6)

\put(-2,11){\texttt{t=2}}
\put(10,15){\line(1,0){95}}
\put(10,10){\line(1,0){95}}
\multiput(15,10)(5,0){18}{\line(0,1){5}}

\put(47.5,12.5){\circle*{3}} 
\put(52.5,12.5){\circle*{3}} 
\put(62.5,12.5){\circle*{3}} 
\put(67.5,12.5){\circle*{3}} 


\put(-2,3){\texttt{t=3}}
\put(10,7){\line(1,0){95}}
\put(10,2){\line(1,0){95}}
\multiput(15,2)(5,0){18}{\line(0,1){5}}

\put(57.5,4.5){\circle*{3}} 
\put(72.5,4.5){\circle*{3}} 
\put(77.5,4.5){\circle*{3}} 
\put(82.5,4.5){\circle*{3}} 

\put(0,-8){
\put(-2,3){\texttt{t=4}}
\put(10,7){\line(1,0){95}}
\put(10,2){\line(1,0){95}}
\multiput(15,2)(5,0){18}{\line(0,1){5}}

\put(62.5,4.5){\circle*{3}} 
\put(87.5,4.5){\circle*{3}} 
\put(92.5,4.5){\circle*{3}} 
\put(97.5,4.5){\circle*{3}} 
}

\end{picture}
\end{center}

\noindent
One observes that ``a series of three balls" and ``a series of one ball"
proceed to the right stably unless they are ``too close" to each other.
The larger one is faster than the smaller one, so they eventually collide.
After the collision, it is non-trivial that they come back 
in the very original shape as $3+1\rightarrow 1+3$, 
instead of being smashed into pieces 
like $3+1 \rightarrow 1+1+1+1$ or getting  
glued together like $3+1\rightarrow 4$.
Moreover, the collision has caused a {\it phase shift};  
observe that the trajectory of the larger (smaller) series has been shifted 
by $2$ to the right (left).

Let us watch another example:

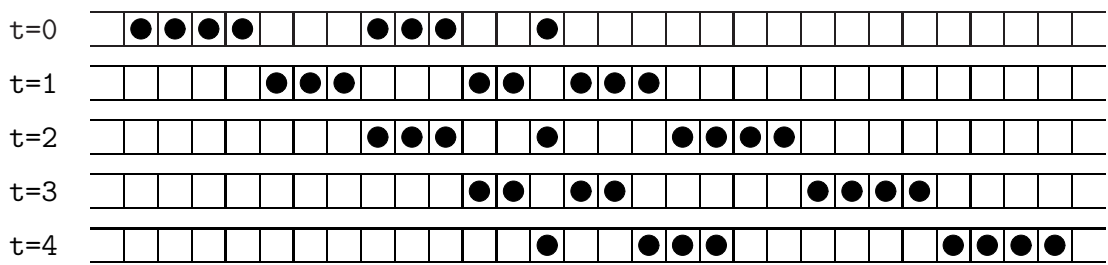
\begin{center}
\unitlength=0.9mm
\begin{picture}(140,42)(10,3)

\put(-2,38){\texttt{t=0}}
\put(10,42){\line(1,0){150}}
\put(10,37){\line(1,0){150}}
\multiput(15,37)(5,0){29}{\line(0,1){5}}

\put(17.5,39.5){\circle*{3}} 
\put(22.5,39.5){\circle*{3}} 
\put(27.5,39.5){\circle*{3}} 
\put(32.5,39.5){\circle*{3}} 

\put(52.5,39.5){\circle*{3}} 
\put(57.5,39.5){\circle*{3}} 
\put(62.5,39.5){\circle*{3}} 

\put(77.5,39.5){\circle*{3}} 


\put(-2,30){\texttt{t=1}}
\put(10,34){\line(1,0){150}}
\put(10,29){\line(1,0){150}}
\multiput(15,29)(5,0){29}{\line(0,1){5}}

\put(37.5,31.5){\circle*{3}} 
\put(42.5,31.5){\circle*{3}} 
\put(47.5,31.5){\circle*{3}} 

\put(67.5,31.5){\circle*{3}} 
\put(72.5,31.5){\circle*{3}} 

\put(82.5,31.5){\circle*{3}} 
\put(87.5,31.5){\circle*{3}} 
\put(92.5,31.5){\circle*{3}} 


\put(-2,22){\texttt{t=2}}
\put(10,26){\line(1,0){150}}
\put(10,21){\line(1,0){150}}
\multiput(15,21)(5,0){29}{\line(0,1){5}}

\put(52.5,23.5){\circle*{3}} 
\put(57.5,23.5){\circle*{3}} 
\put(62.5,23.5){\circle*{3}} 

\put(77.5,23.5){\circle*{3}}
 
\put(97.5,23.5){\circle*{3}} 
\put(102.5,23.5){\circle*{3}} 
\put(107.5,23.5){\circle*{3}} 
\put(112.5,23.5){\circle*{3}} 


\put(-2,14){\texttt{t=3}}
\put(10,18){\line(1,0){150}}
\put(10,13){\line(1,0){150}}
\multiput(15,13)(5,0){29}{\line(0,1){5}}

\put(67.5,15.5){\circle*{3}} 
\put(72.5,15.5){\circle*{3}} 

\put(82.5,15.5){\circle*{3}} 
\put(87.5,15.5){\circle*{3}}
 
\put(117.5,15.5){\circle*{3}} 
\put(122.5,15.5){\circle*{3}} 
\put(127.5,15.5){\circle*{3}} 
\put(132.5,15.5){\circle*{3}} 


\put(-2,6){\texttt{t=4}}
\put(10,10){\line(1,0){150}}
\put(10,5){\line(1,0){150}}
\multiput(15,5)(5,0){29}{\line(0,1){5}}

\put(77.5,7.5){\circle*{3}} 

\put(92.5,7.5){\circle*{3}} 
\put(97.5,7.5){\circle*{3}} 
\put(102.5,7.5){\circle*{3}}
 
\put(137.5,7.5){\circle*{3}} 
\put(142.5,7.5){\circle*{3}} 
\put(147.5,7.5){\circle*{3}} 
\put(152.5,7.5){\circle*{3}} 

\end{picture}
\end{center}
\noindent
Here we have three series of $4$, $3$ and $1$ balls from the left,
and they are interchanged
into the reverse order $1$, $3$ and $4$ during $1 \leq t \leq 3$.
These behaviors of series of balls remind us of {\it solitons}
in the theory of nonlinear waves.
We call the number of balls in a series of balls
before or after collisions
an {\it amplitude} of the soliton.
(A precise definition of solitons and their amplitude will be given later. 
See for example 
(\ref{t:jun10a}) or (\ref{k:ss}).)

\smallskip
One can also set up the BBS with the periodic boundary condition \cite{YT02}.
Let $L$ be the number of boxes aligned on an oriented circle.  
We put $M < L/2$ balls into them.
The balls are moved by the same rule as the previous (i)--(iii) for the 
original (infinite) BBS except a minor adaptation to the fact that 
nothing can be ``leftmost" on a circle.
In (i), the procedure can be started from {\it any} chosen ball.
In (ii), the terms ``leftmost" and ``nearest right" are to be understood 
along the direction of the orientation of the circle. 
Then the  modified evolution rule is well-defined in the sense that 
the result is actually independent of the choice of the first ball to move.
Moreover it is again invertible.
Let us look at an example of $L=12$ and $M=4$ in the following,
where we identify the left and right boundaries (thick lines):

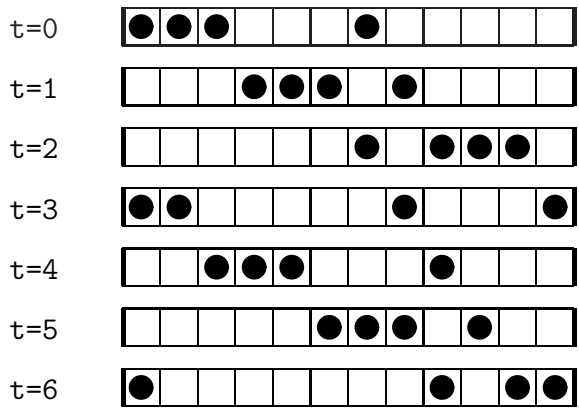
\begin{center}
\unitlength=1.0mm
\begin{picture}(90,55)(0,1)

\put(0,51){\texttt{t=0}}
\put(15,55){\line(1,0){60}}
\put(15,50){\line(1,0){60}}
\multiput(20,50)(5,0){11}{\line(0,1){5}}
\thicklines
\multiput(15,50)(0.2,0){2}{\line(0,1){5}}
\multiput(75,50)(0.2,0){2}{\line(0,1){5}}

\put(17.5,52.5){\circle*{3}} 
\put(22.5,52.5){\circle*{3}} 
\put(27.5,52.5){\circle*{3}} 
\put(47.5,52.5){\circle*{3}} 


\thinlines
\put(0,43){\texttt{t=1}}
\put(15,47){\line(1,0){60}}
\put(15,42){\line(1,0){60}}
\multiput(20,42)(5,0){11}{\line(0,1){5}}
\thicklines
\multiput(15,42)(0.2,0){2}{\line(0,1){5}}
\multiput(75,42)(0.2,0){2}{\line(0,1){5}}

\put(32.5,44.5){\circle*{3}} 
\put(37.5,44.5){\circle*{3}} 
\put(42.5,44.5){\circle*{3}} 
\put(52.5,44.5){\circle*{3}} 


\thinlines
\put(0,35){\texttt{t=2}}
\put(15,39){\line(1,0){60}}
\put(15,34){\line(1,0){60}}
\multiput(20,34)(5,0){11}{\line(0,1){5}}
\thicklines
\multiput(15,34)(0.2,0){2}{\line(0,1){5}}
\multiput(75,34)(0.2,0){2}{\line(0,1){5}}

\put(47.5,36.5){\circle*{3}} 
\put(57.5,36.5){\circle*{3}} 
\put(62.5,36.5){\circle*{3}} 
\put(67.5,36.5){\circle*{3}} 


\thinlines
\put(0,27){\texttt{t=3}}
\put(15,31){\line(1,0){60}}
\put(15,26){\line(1,0){60}}
\multiput(20,26)(5,0){11}{\line(0,1){5}}
\thicklines
\multiput(15,26)(0.2,0){2}{\line(0,1){5}}
\multiput(75,26)(0.2,0){2}{\line(0,1){5}}

\put(52.5,28.5){\circle*{3}} 
\put(72.5,28.5){\circle*{3}} 
\put(17.5,28.5){\circle*{3}} 
\put(22.5,28.5){\circle*{3}} 


\thinlines
\put(0,19){\texttt{t=4}}
\put(15,23){\line(1,0){60}}
\put(15,18){\line(1,0){60}}
\multiput(20,18)(5,0){11}{\line(0,1){5}}
\thicklines
\multiput(15,18)(0.2,0){2}{\line(0,1){5}}
\multiput(75,18)(0.2,0){2}{\line(0,1){5}}

\put(27.5,20.5){\circle*{3}} 
\put(32.5,20.5){\circle*{3}} 
\put(37.5,20.5){\circle*{3}} 
\put(57.5,20.5){\circle*{3}} 


\thinlines
\put(0,11){\texttt{t=5}}
\put(15,15){\line(1,0){60}}
\put(15,10){\line(1,0){60}}
\multiput(20,10)(5,0){11}{\line(0,1){5}}
\thicklines
\multiput(15,10)(0.2,0){2}{\line(0,1){5}}
\multiput(75,10)(0.2,0){2}{\line(0,1){5}}

\put(42.5,12.5){\circle*{3}} 
\put(47.5,12.5){\circle*{3}} 
\put(52.5,12.5){\circle*{3}} 
\put(62.5,12.5){\circle*{3}} 


\thinlines
\put(0,3){\texttt{t=6}}
\put(15,7){\line(1,0){60}}
\put(15,2){\line(1,0){60}}
\multiput(20,2)(5,0){11}{\line(0,1){5}}
\thicklines
\multiput(15,2)(0.2,0){2}{\line(0,1){5}}
\multiput(75,2)(0.2,0){2}{\line(0,1){5}}

\put(57.5,4.5){\circle*{3}} 
\put(67.5,4.5){\circle*{3}} 
\put(72.5,4.5){\circle*{3}} 
\put(17.5,4.5){\circle*{3}} 

\end{picture}
\end{center}
One can observe that the larger soliton overtakes 
the smaller one repeatedly.
The biggest difference from the infinite BBS
is that the system has now a finite configuration space; there are just 
$\binom{L}{M}$ states. 
Thus, any state is cyclic,
i.e. by starting with an arbitrary initial state 
one comes back to itself in a finite time.

Let us motivate our study on BBS from the 
viewpoint of solitons and integrability. 
Nowadays, the term soliton is widely used 
to mean, somewhat loosely,  
various special solutions to nonlinear equations that 
exhibit particle like behavior or certain stability.
In its original context of Zabusky and Kruskal \cite{ZK65} however, it meant  
a solitary wave solution in an
infinite dimensional nonlinear dynamical system (KdV eq. mentioned below)
with more stringent properties as follows:
\begin{enumerate}
\item[(a)]
particle-like propagation (constant velocity, stability under {\it multi}-body collision),
\item[(b)]
factorization of scattering
(pairwise scattering with phase shifts).

\end{enumerate}
Existence of solitons is a signal of 
{\it integrability}, which  at least postulates 
an infinite number of conserved quantities 
(integrals of motion).
The historically important and famous integrable systems of such kind are 
Korteweg-de Vries (KdV) equation 
and Kadomtsev-Petviashvili (KP) equation,
which are prototypes of what is called a soliton equation.
For finite dimensional systems, the notion of integrability is 
clearer, i.e. it implies the existence 
of enough number of conserved quantities 
so that the initial value problem can be solved.
The classic examples as Euler, Lagrange and Kovalevskaya tops belong
to this category.
We remark that the Toda equation is also an important dynamical system
which is integrable either on finite or infinite lattices.

And so what about the BBS in which we have just observed ``solitons"?
Are they really solitons that possess the above mentioned properties?
Is BBS really integrable in some sense?
Is it related to integrable systems known hitherto?
Is there any good mathematical framework to analyse it?

The aim of this review article is to give an introductory exposition on 
a variety of aspects of BBS elucidated in the last two decades, 
where all the above questions will be answered affirmatively.

\subsection{Overview of related mathematics}

It turns out that BBS originates in a {\it quantum} integrable system 
as well as in a {\it classical} integrable system.
It is located at the very special point where the two systems meet   
by the procedures called {\it crystallization} and 
{\it ultradiscretization}, respectively.

By quantum integrable systems, we mean those
associated with the Yang-Baxter relation \cite{B, J89}.
Their symmetry is governed by {\it quantum group}
$U_q=U_q(\mathfrak{g})$ meaning the $q$-deformation of the universal 
enveloping algebra $U(\mathfrak{g})$ of some affine Lie algebra 
$\mathfrak{g}$ \cite{D85, J85}.
Typical examples are solvable lattice models in 
statistical mechanics such as the six-vertex model \cite{B}
for $\mathfrak{g}=\widehat{\mathfrak{sl}}_2$ and its generalizations.
They are spin systems whose Boltzmann weights are
continuous functions of $q$.
The crystallization corresponds to taking the limit $q\rightarrow 0$,
where the models are frozen to the ground state and its 
profile turns out to reproduce the BBS dynamics exactly.

By classical integrable system, we mean here integrable difference equations 
such as discrete KP equation and 
(time-discretized versions of) Lotka-Volterra equation, Toda equation
and so forth.
These equations are already defined on lattices
(discrete space-time), but their 
dynamical variables are yet continuous.
The ultradiscretization is a procedure to transform these nonlinear evolution equations
and their solutions into piecewise-linear forms. 
Leaving technical cautions aside, it is achieved by switching from
the original variable $a$ to $A$ by 
$a=\e^{-A/\varepsilon}$ and taking the limit $\varepsilon \rightarrow +0$.
Being piecewise-linear, the resulting equations 
allow one to restrict the dynamical variables to a certain discrete set. 
In this way one reproduces evolution equations of BBS and 
obtains their solutions.

Having the double, classical as well as quantum,  origins 
of the integrability makes the study of BBS especially rich. 
One can import a variety of notions and techniques
to understand and analyse BBS from the two theories.
For instance from the theory of quantum integrable systems, we have ingredients like
Yang-Baxter relation, quantum $R$ matrices, commuting transfer matrices,
Bethe ansatz, corner transfer matrices and so forth
\cite{B,GRS01,J89,KBI93,Su04,Ta99}.
Similarly, the classical theory provides us with 
solitons, the inverse scattering method, tau functions, 
action-angle variables, isolevel set, spectral curves, 
linearization of flows etc \cite{AS,Au96,BBT,DN93,FT87,H92,MJD00,To89}.
It turns out that they all survive the crystallization or the ultradiscretization
rather miraculously.
Moreover, they allow a systematic (Lie algebraic) generalization beyond the original BBS 
so that there are many kinds of balls or particles/anti-particles, 
boxes with capacity greater than one, and a family of 
commuting time evolutions etc.
(Nonetheless, we will mainly focus on 
the basic type $\widehat{\mathfrak{sl}}_{n+1}$ 
case in this paper to be introductory.)

Compared with traditional integrable systems, 
certainly a  novel feature of BBS is that 
its {\it dependent} (or {\it dynamical}) variables 
have also been discretized.
This fact indicates and actually has led to a fruitful connection to 
the realm of {\it combinatorics}.
From a mathematical point of view,
the crystallization and the ultradiscretization are both 
connected and actually have partly motivated the 
fascinating subjects known as 
{\it crystal base} \cite{Ka1,Ka2} in the 
representation theory of quantum group,
{\it geometric crystals} \cite{BK00} as its geometric counterpart, 
and {\it tropical geometry} \cite{MZ06, SS04} in algebraic geometry.
As the title of the article suggests, 
this review also contains elementary expositions and 
practical applications of these theories to BBS.

Leaving the details to later sections, 
we present a rough schematic view of the relevant 
subjects as a summary. 
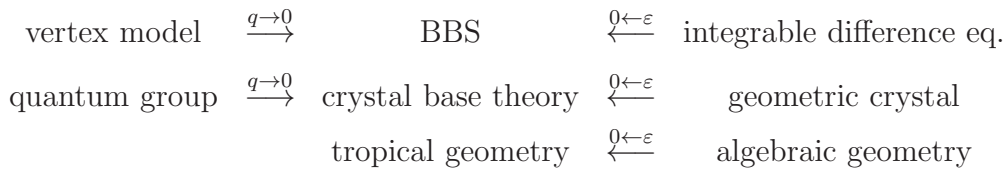
\begin{align*}
\begin{matrix}
\text{vertex model} & \stackrel{q \to 0}{\longrightarrow} & 
\text{BBS} & \stackrel{0 \leftarrow \ve}{\longleftarrow} & 
\text{integrable difference eq.}
\\[2mm]
\text{quantum group} & \stackrel{q \to 0}{\longrightarrow} 
& \text{crystal base theory} & \stackrel{0 \leftarrow \ve}{\longleftarrow} 
& \text{geometric crystal}
\\[1mm]
& & \text{tropical geometry} & \stackrel{0 \leftarrow \ve}{\longleftarrow}
& \text{algebraic geometry} 
\end{matrix}
\end{align*}

\subsection{Contents}

The main contents of each section are as follows:
in \S 2.1 the basic notion of crystallization is illustrated
with the simplest vertex model of $U_q(\widehat{\mathfrak{sl}}_{2})$.
\S 2.2 is an exposition on $\widehat{\mathfrak{sl}}_{n+1}$ 
crystal base theory,
which is applied to describe the infinite BBS in \S 2.3. 
In \S 2.4, we briefly sketch various generalizations of BBS
associated with affine Lie algebras.   
We remark that \S 2.2 and \S 2.3 are essential to study BBS.

In \S 3.1 the physical background of Kerov-Kirillov-Reshetikhin (KKR) 
bijection in the Bethe ansatz is explained.
The definition of KKR bijection for $\widehat{\mathfrak{sl}}_{n+1}$ 
crystal is given in \S 3.2
with the concrete algorithm.
In \S 3.3 we state that KKR bijection linearizes the time evolution 
of BBS, which enables us to solve the initial value problem of BBS.  

In \S 4.1 the notions of tropicalization, ultradiscretization and 
min-plus algebra are introduced.
In \S 4.2, two kinds of evolution equations for BBS are provided,
corresponding to the ``spatial'' and the ``soliton'' descriptions of BBS. 
The equations are the ultradiscretization of known integrable discrete systems.
The first description is studied in \S 4.4 and \S 5.2,
and the second one is studied in \S 6.3.
In \S 4.3, we briefly explain geometric crystal
whose ultradiscretization gives the crystal structure. 
In \S 4.4 the general solution for BBS is given
by the ultradiscrete tau function in a similar way
to many soliton equations.

In \S 5.1 the basic features of periodic BBS is explained
and its general solution is constructed via (modified) KKR bijection in \S 5.2.
A remarkable feature is that the solution can be written in terms of 
tropical theta functions.
In \S 5.3 we discuss more detail of periodic BBS 
from the viewpoint of torus decomposition of the isolevel set
and fundamental periods of the time evolution.

\S 6.1 is an introduction to tropical curve theory
which is the latest mathematical object in this article.
This theory is applied in \S 6.2 to solve the tropical
periodic Toda lattice (trop-pToda). 
In \S 6.3 we show that the isolevel set of the periodic BBS
is embedded in that of trop-pToda.
This embedding bridges two different approaches to the periodic
BBS by the trop-pToda and by the modified KKR bijection in \S 5.2,
from the viewpoint of 
the tropical geometric description of the isolevel sets.

We show a flow chart of the sections in this article: 
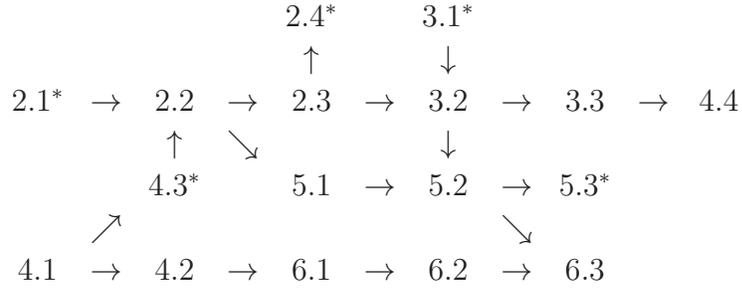
\begin{align*}
\begin{matrix}
& & & & 2.4^\ast & & 3.1^\ast
\\
& & & & \uparrow & & \downarrow
\\
2.1^\ast & \to & 2.2 & \to & 2.3 & \to & 3.2 & \to 
& 3.3 & \to ~~ 4.4
\\
& & \uparrow & \searrow & & & \downarrow
\\
& & 4.3^\ast & & 5.1 & \to & 5.2 & \to & 5.3^\ast
\\
& \nearrow & & & & & & \searrow
\\
4.1 & \to & 4.2 & \to & 6.1 & \to & 6.2 & \to & 6.3
\end{matrix}
\end{align*}
where sections involving somewhat advanced or specialized topics
are indicated by $^\ast$.

We did not intend to make the reference exhaustive.
It is a moderate but sufficient supply for interested readers 
to proceed and find  further references.

\section{BBS and crystals}

\subsection{Crystallization: $q \to 0$ of the vertex model}

For simplicity, we concentrate on the $U_q(\widehat{\mathfrak{sl}}_2)$ case 
in this subsection.
Let us recall the six-vertex model and its fusion.
Consider the two dimensional square lattice, 
where each edge is assigned with a local variable
taking values in $\{1,2\}$.
Around each vertex, we allow the following 6 configurations 
with the respective Boltzmann weights:
\begin{equation}\label{k:6v}
\begin{picture}(300,62)(0,-3)
\multiput(14.5,38)(50,0){6}{
\put(-12,0){\line(1,0){24}}
\put(0,-10){\line(0,1){20}}
}

\put(12,51){1}
\put(-5,34){1}
\put(29,34){1}
\put(12,16){1}

\put(62,51){2}
\put(45,34){2}
\put(79,34){2}
\put(62,16){2}

\put(112,51){1}
\put(95,34){2}
\put(129,34){2}
\put(112,16){1}

\put(162,51){2}
\put(145,34){1}
\put(179,34){1}
\put(162,16){2}

\put(212,51){2}
\put(195,34){1}
\put(229,34){2}
\put(212,16){1}

\put(262,51){1}
\put(245,34){2}
\put(279,34){1}
\put(262,16){2}

\put(0,0){$1-q^2z$}
\put(50,0){$1-q^2z$}
\put(100,0){$q(1-z)$}
\put(150,0){$q(1-z)$}

\put(200,0){$z(1-q^2)$}

\put(255,0){$1-q^2$,}
\end{picture}
\end{equation}
where $z$ is called a {\em spectral parameter}.
The other 10 configurations are assigned with 0 Boltzmann weight.
Let $V={\mathbb C} v_1 \oplus {\mathbb C} v_2$. 
Then (\ref{k:6v}) is arranged in 
the {\em quantum $R$ matrix} $R(z) \in {\rm End}(V\otimes V)$ as
\begin{equation}\label{k:eqa:r}
\begin{split}
&R(z) = a(z)\sum_iE_{ii}\otimes E_{ii} +
b(z)\sum_{i\neq j}E_{ii}\otimes E_{jj} 
+ c(z)\Bigl(z\sum_{i<j}+\sum_{i>j}\Bigr)
E_{ji}\otimes E_{ij},\\
&a(z) = 1-q^2z,\quad b(z) = q(1-z),\quad c(z) = 1-q^2.
\end{split}
\end{equation}
Here the indices run over $\{1,2\}$ and 
$E_{ij}$ is the matrix unit acting as 
$E_{ij}v_k = \delta_{jk}v_i$.
Schematically (\ref{k:eqa:r}) is expressed as
\begin{equation}\label{k:re}
R(z) = \sum_{ijkl}\Bigl(
\begin{picture}(35,22)(-17,-3)
\put(-2,13){$l$}\put(-17,-3){$j$}\put(3,-7){$z$}
\put(12,-3){$i$}\put(-2,-20){$k$}\drawline(0,-3)(3,0)
\put(-10,0){\line(1,0){20}}
\put(0,-10){\line(0,1){20}}
\end{picture}
\Bigr)E_{ij}\otimes E_{kl},
\qquad
{\check R}(z) = \sum_{ijkl}\Bigl(
\begin{picture}(35,22)(-17,-3)
\put(-2,13){$l$}\put(-17,-3){$j$}\put(3,-7){$z$}
\put(12,-3){$k$}\put(-2,-20){$i$}\drawline(0,-3)(3,0)
\put(-10,0){\line(1,0){20}}
\put(0,-10){\line(0,1){20}}
\end{picture}
\Bigr)E_{ij}\otimes E_{kl},
\end{equation}
where the $z$-dependence is exhibited.
The Yang-Baxter equation 
\begin{equation*}
R_{23}(z')R_{13}(z)R_{12}(z/z')
= R_{12}(z/z')R_{13}(z)R_{23}(z')
\end{equation*}
holds \cite{B}, where the indices signify the 
components in the tensor product as
$\overset{1}{V}\otimes \overset{2}{V} \otimes 
\overset{3}{V}$ 
on which the both sides act.
It is depicted as  

\begin{equation}\label{k:ybe}
\begin{picture}(200, 60)(0,-30)

\put(0,10){\line(2,-1){60}}
\put(0,-10){\line(2,1){60}}

\put(46,-30){\line(0,1){60}}

\drawline(46,9)(50,15)

\drawline(24,2)(24,-2)

\drawline(46,-18)(50,-15)

\put(29,-1){$\scriptstyle{z\!/\!z'}$}

\put(50,7){$\scriptstyle{z}$}

\put(48,-24){$\scriptstyle{z'}$}

\put(80,0){\makebox(0,0){$=$}}

\multiput(157,0)(0,0){1}{

\put(-46,-30){\line(0,1){60}}

\put(0,10){\line(-2,-1){60}}

\put(0,-10){\line(-2,1){60}}

\drawline(-46,8)(-42,11)\put(-45,3){$\scriptstyle{z'}$}

\drawline(-16,2)(-16,-2)\put(-11,-1){$\scriptstyle{z\!/\!z'}$}

\drawline(-46,-17)(-42,-11.5)\put(-43,-18){$\scriptstyle{z}$}

\put(10,-28){.}

}

\end{picture}
\end{equation}

The $R$ matrix $R(z)$ is associated with 
the quantum affine algebra $U_q=U_q(\widehat{\mathfrak{sl}}_2)$ \cite{D85,J85}.
There is an algebra homomorphism 
$\Delta: U_q \rightarrow U_q \otimes U_q$ called coproduct,
which enables one to construct the tensor product representation
$V\otimes V'$ from any two representations $V$ and $V'$.
Setting ${\check R}(z):=PR(z)$ with 
$P$ being the transposition of components,
the quantum $R$ matrix is characterized by the 
condition $\Delta(x) R = R \Delta(x)$  
for any  $x \in U_q$. 
The asymmetry between the last two in 
(\ref{k:6v}) is due to the special choice of the coproduct 
$\Delta$ that suits the limit 
$q\rightarrow 0$ that will be considered in what follows.
(The precise form of $\Delta$ is not needed in this paper.)

Starting from the six-vertex model,
one can construct multi-state (``higher spin") solvable vertex models by 
the {\em fusion procedure} \cite{KRS81}.
Let $V_m$ be the irreducible $U_q$ module 
spanned by the $m$ fold $q$-symmetric tensors in 
$V^{\otimes m}$.
It is a $q$-analogue of the spin $\frac{m}{2}$ representation.
Concretely, $V_1=V$ and $V_m$ with 
$m \ge 2$ is realized as the quotient
$V^{\otimes m}/A$,
where 
$A= \sum_j V^{\otimes j} \otimes {\rm Im}\,{\check R}(q^{-2}) 
\otimes V^{\otimes m-2-j}$.
It is easy to see ${\rm Im}\,{\check R}(q^{-2}) 
= {\rm Ker}\,{\check R}(q^{2}) 
= {\mathbb C}(v_1\otimes v_2 - q v_2\otimes v_1)$.
We take the base vector of $V_m$ as  
$v_2^{\otimes x_2}\otimes v_1^{\otimes x_1} \mod A$, where 
$x_i \in \Z_{\ge 0}$ and $x_1 + x_2 = m$.
The base will also be denoted by $x=(x_1, x_2)$ or by 
the sequence $\overbrace{1\ldots 1}^{x_1}\overbrace{2\ldots 2}^{x_2}$.
Obviously $\dim V_m = m+1$.
The outcome of the fusion procedure 
is the fusion $R$ matrix
$R^{(m,1)}(z) \in {\rm End}(V_m \otimes V_1)$ given by
\begin{align}
R^{(m,1)}(z)(x\otimes v_j) 
&= \sum_{k=1,2}\Bigl(
\begin{picture}(35,22)(-17,-3)
\put(-2,14){$j$}\put(-17,-3){$x$}\put(3,-7){$z$}
\put(12,-3){$y$}\put(-2,-20){$k$}\drawline(3,0)(0,-3)
\put(-10,0.2){\line(1,0){20}}
\put(-10,0){\line(1,0){20}}
\put(-10,-0.2){\line(1,0){20}}
\put(0,-10){\line(0,1){20}}
\end{picture}
\Bigr)\;y \otimes v_k,\label{k:r1m}\\
\begin{picture}(35,22)(-17,-3)
\put(-2,14){$j$}\put(-17,-3){$x$}\put(3,-7){$z$}
\put(12,-3){$y$}\put(-2,-20){$k$}\drawline(3,0)(0,-3)
\put(-10,0.2){\line(1,0){20}}\put(-10,0){\line(1,0){20}}
\put(-10,-0.2){\line(1,0){20}}
\put(0,-10){\line(0,1){20}}
\end{picture}
&=
\begin{cases}
q^{m-x_k}-q^{x_k+1}z & j=k,\\
(1-q^{2x_1})z & (j,k)=(2,1),\\
1-q^{2x_2} & (j,k)=(1,2),
\end{cases}\label{k:wjk}
\end{align}
where $y=(y_1,y_2)$ is specified by the weight conservation 
(so called ``ice rule") as 
$y_i = x_i + \delta_{i j}- \delta_{i k}$.
The RHS of (\ref{k:wjk}) is to be understood as $0$ unless
this condition is satisfied.
For $m=1$,
one has $R^{(1,1)}(z)=R(z)$ and 
(\ref{k:wjk}) reduces to (\ref{k:6v}).
For $m=2$, (\ref{k:wjk}) reads explicitly as follows:
\begin{equation}\label{k:wt21}
\begin{picture}(300,162)(0,-103)
\multiput(14.5,38)(65,0){5}{
\put(-12,0){\line(1,0){24}}
\put(-12,0.2){\line(1,0){24}}
\put(-12,-0.2){\line(1,0){24}}
\put(0,-10){\line(0,1){20}}
}

\put(12,51){1}
\put(-11,34){11}
\put(28,34){11}
\put(12,16){1}

\put(77,51){1}
\put(54,34){12}
\put(93,34){11}
\put(77,16){2}

\put(142,51){1}
\put(119,34){22}
\put(158,34){12}
\put(142,16){2}

\put(207,51){1}
\put(184,34){12}
\put(224,34){12}
\put(207,16){1}

\put(272,51){1}
\put(249,34){22}
\put(289,34){22}
\put(272,16){1}

\put(0,0){$1-q^3z$}
\put(65,0){$1-q^2$}
\put(130,0){$1-q^4$}
\put(195,0){$q-q^2z$}
\put(260,0){$q^2-qz$}

\put(0,-80){
\multiput(14.5,38)(65,0){5}{
\put(-12,0){\line(1,0){24}}
\put(-12,0.2){\line(1,0){24}}
\put(-12,-0.2){\line(1,0){24}}
\put(0,-10){\line(0,1){20}}
}

\put(12,51){2}
\put(-11,34){11}
\put(28,34){12}
\put(12,16){1}

\put(77,51){2}
\put(54,34){12}
\put(93,34){22}
\put(77,16){1}

\put(142,51){2}
\put(119,34){22}
\put(158,34){22}
\put(142,16){2}

\put(207,51){2}
\put(184,34){11}
\put(224,34){11}
\put(207,16){2}

\put(272,51){2}
\put(249,34){12}
\put(289,34){12}
\put(272,16){2}

\put(0,0){$(1-q^4)z$}
\put(65,0){$(1-q^2)z$}
\put(130,0){$1-q^3z$}
\put(195,0){$q^2-qz$}
\put(260,0){$q-q^2z$~~.}
}

\end{picture}
\end{equation}
Here we have suppressed $z$ in the diagrams.
(\ref{k:wt21}) is regarded as the (allowed) local configurations
and their Boltzmann weights in a new vertex model where
the horizontal and vertical edges take the 3 states 
$\{11,12,22\}$ and the 2 states $\{1,2\}$, respectively.
The weight conservation of $R^{(2,1)}(z)$ means that 
the total number of letters 1 and 2 are preserved 
from NW to SE.

Let us sketch how (\ref{k:wt21}) is obtained 
from (\ref{k:6v}).
The Yang-Baxter equation (\ref{k:ybe}) with $z'=zq^{2}$ shows that 
${\rm Im}\,{\check R}(q^{-2})
\subset \overset{1}{V}\otimes \overset{2}{V}$ is preserved under the
action of  $R_{13}(zq^2)R_{23}(z)$.
Therefore its action on 
$(\overset{1}{V} \otimes \overset{2}{V})\otimes 
\overset{3}{V}$
can be restricted to 
$V_2\otimes V_1 
= \left((V\otimes V)/{\rm Im}\,{\check R}(q^{-2})\right)
\otimes V$.
This yields the $2\times 1$ fusion leading to $R^{(2,1)}(z)$.
Similarly, $R^{(m,1)}(z)$ can be deduced by restricting 
the composition ($a(z)$ defined in (\ref{k:eqa:r})) 
\begin{equation}\label{k:rcomp}
\frac{R_{1,m+1}(zq^{m-1})
R_{2,m+1}(zq^{m-3})\cdots R_{m,m+1}(zq^{-m+1})}
{a(zq^{m-3})a(zq^{m-5})\cdots a(zq^{-m+1})}
\end{equation}
to $V_m \otimes V_1$. 
One can furthermore fuse $R^{(m,1)}(z)$ 
along the other component of the tensor product 
in a completely parallel fashion.
The result yields the quantum $R$ matrix 
$R^{(m,l)}(z) \in {\rm End}(V_m \otimes V_l)$.
The $R$ matrices so obtained again satisfy the Yang-Baxter equation
in ${\rm End}(V_l \otimes V_m \otimes V_k)$:
\begin{equation}\label{k:ybe2}
R^{(m,k)}_{23}(z')R^{(l,k)}_{13}(z)R^{(l,m)}_{12}(z/z')
= R^{(l,m)}_{12}(z/z')R^{(l,k)}_{13}(z)R^{(m,k)}_{23}(z').
\end{equation}
It is depicted as (\ref{k:ybe}) with the three lines to be interpreted 
as representing $V_l, V_m$ and $V_k$.
The quantum $R$ matrix $R^{(m,l)}(z)$ gives rise to  
a fusion vertex model on a square lattice 
in a similar manner to (\ref{k:re}) and (\ref{k:r1m}).
The local variables on the horizontal and vertical edges are
taken from $V_m$ and $V_l$, respectively.
In terms of the linear operator 
$\check{R}^{(m,l)}(z):=PR^{(m,l)}(z) :  
V_m\otimes V_l \rightarrow V_l \otimes V_m$, which is also called 
an $R$ matrix, the Yang-Baxter equation (\ref{k:ybe2}) takes 
another familiar form:
\begin{equation}\label{k:ybe3}
\begin{split}
&\left(\check{R}^{(m,k)}(z')\otimes 1\right)
\left(1\otimes \check{R}^{(l,k)}(z)\right)
\left(\check{R}^{(l,m)}(z/z')\otimes 1\right)\\
&=\left(1\otimes \check{R}^{(l,m)}(z/z')\right)
\left(\check{R}^{(l,k)}(z)\otimes 1\right)
\left(1\otimes \check{R}^{(m,k)}(z')\right).
\end{split}
\end{equation}
It is $\check{R}^{(m,l)}(z)$ rather than $R^{(m,l)}(z)$ 
that will be directly related 
to the combinatorial or birational $R$ introduced in the later sections.

An important object in the vertex models 
is the (row to row) {\it transfer matrix}.
For simplicity, we consider the basic case 
corresponding to $R^{(m,1)}(z)$.
Then, the transfer matrix $T_m(z)$ is defined by
\begin{equation}\label{k:rtm}
T_m(z)(v_{j_1}\otimes \cdots \otimes v_{j_L})
=\sum_{\{k_i\}} \sum_{\{x^{(i)}\}}\left(
\begin{picture}(155,15)(0,0)

\put(30,0){
\put(-4,15){$j_1$}\put(-4,-20){$k_1$}
\put(-30,-2){$\small{x^{(1)}}$}
\put(-12,0){\line(1,0){24}}
\put(-12,0.2){\line(1,0){24}}
\put(-12,-0.2){\line(1,0){24}}
\put(0,-10){\line(0,1){20}}
\drawline(0,-3)(3,0)\put(3,-7){$z$}}

\put(76,0){
\put(-4,15){$j_2$}\put(-4,-20){$k_2$}
\put(-30,-2){$x^{(2)}$}
\put(-12,0){\line(1,0){24}}
\put(-12,0.2){\line(1,0){24}}
\put(-12,-0.2){\line(1,0){24}}
\put(0,-10){\line(0,1){20}}
\drawline(0,-3)(3,0)\put(3,-7){$z$}}

\put(120,0){
\put(-4,15){$j_L$}\put(-4,-20){$k_L$}
\put(-29,-2.3){$\cdots$}
\put(-12,0){\line(1,0){24}}
\put(-12,0.2){\line(1,0){24}}
\put(-12,-0.2){\line(1,0){24}}
\put(0,-10){\line(0,1){20}}
\put(16,-2){$x^{(1)}$}
\drawline(0,-3)(3,0)\put(3,-7){$z$}}

\end{picture}
 \right)
v_{k_1}\otimes \cdots \otimes v_{k_L},
\end{equation}
where each $k_i$ runs over $\{1,2\}$ and 
$x^{(i)}$ does over the set of base 
of $V_m$ labeled with $\{ 1...11, 1...12,\ldots, 2...22\}$.
The array of the vertex diagrams means the 
product of the corresponding Boltzmann weights (\ref{k:wjk}). 
We have assumed that the (horizontal)  
length of the lattice is $L$ and 
employed the periodic boundary condition.
The transfer matrix allows one to express 
the partition function of the model 
(for $R^{(m,1)}(z)$) on $N\times L$ lattice 
with the periodic boundary condition 
as $Z= \mathrm{Tr}(T_m(z)^N)$.
All the matrices $T_1(z), T_2(z), \ldots$ act on 
the same space $V_1^{\otimes L}$. 
Using (\ref{k:ybe2}), one can show that 
they form a commuting family:
\begin{equation}\label{k:comm}
T_m(z)T_l(w)=T_l(w)T_m(z).
\end{equation}

Now we are ready to discuss the main issue of the present section,
namely, the {\em crystallization} limit $q \rightarrow 0$.
In (\ref{k:wt21}), we see that only the 6 configurations 
in the left 3 columns survive.
In terms of the $R$ matrix, the result 
may be stated that $\check{R}^{(2,1)}(z)$ 
has the following action at $q=0$:
\begin{equation}\label{k:cr21}
\begin{split}
&11 \otimes 1 \mapsto \;\;1 \otimes 11,\quad\;\;\,
12 \otimes 1 \mapsto \;\;2 \otimes 11,\quad\;\;\,
22 \otimes 1 \mapsto 2 \otimes 12,\\
&11 \otimes 2 \mapsto z(1 \otimes 12),\quad
12 \otimes 2 \mapsto z(1 \otimes 22),\quad
22 \otimes 2 \mapsto 2 \otimes 22.
\end{split}
\end{equation}
Here $\{11, 12, 22\}$ and $\{1,2\}$ 
are to be understood as labels of the bases of 
$V_2$ and $V_1$, respectively. 
Apart from the factor $z$, (\ref{k:cr21}) provides a
bijection between the two sets.  
The same feature can be checked easily 
for the general $m$ case (\ref{k:wjk}).
(It is immediately seen for $m=1$ by (\ref{k:6v}).)
The configurations having 
non-vanishing matrix elements (Boltzmann weights)
at $q=0$ are the following:
\begin{equation}\label{k:wtm1}
\begin{picture}(200,93)(0,-70)

\put(-10,0){
\put(3,0){\line(1,0){22}}\put(14,-8){\line(0,1){16}}
\put(11.5,11){1}
\put(-47,-3){$\overbrace{1..........1}^m$}
\put(29,-3){$\overbrace{1..........1}^m$}
\put(11.5,-17){1}}

\put(160,0){
\put(3,0){\line(1,0){22}}\put(14,-8){\line(0,1){16}}
\put(11.5,11){2}
\put(-47,-3){$\overbrace{2..........2}^m$}
\put(29,-3){$\overbrace{2..........2}^m$}
\put(11.5,-17){2}}

\put(-10,-49){
\put(3,0){\line(1,0){22}}\put(14,-8){\line(0,1){16}}
\put(11.5,11){1}
\put(-47,-3){$\overbrace{1...1}^{x_1}\overbrace{2...2}^{x_2}$}
\put(28,-3){$\overbrace{1...1}^{x_1+1}\overbrace{2...2}^{x_2-1}$}
\put(11.5,-17){2}
\put(24,-23){$(0< x_2 \le m)$}}

\put(160,-49){
\put(3,0){\line(1,0){22}}\put(14,-8){\line(0,1){16}}
\put(11.5,11){2}
\put(-47,-3){$\overbrace{1...1}^{x_1}\overbrace{2...2}^{x_2}$}
\put(28,-3){$\overbrace{1...1}^{x_1-1}\overbrace{2...2}^{x_2+1}$}
\put(11.5,-17){1}
\put(24,-23){$(0\le x_2 < m)$~~.}}
\end{picture}
\end{equation}
The limiting Boltzmann weights are all 1 except the bottom right type, 
in which case it is $z$.
The configurations (\ref{k:wtm1}) determine a bijection 
between the data on the NW to SE, generalizing 
(\ref{k:cr21}).
In a physical terminology, the limit $q\rightarrow 0$ 
corresponds to the low temperature limit, where 
{\em crystallization} takes place.
Namely, spins are not allowed to thermally fluctuate and are
frozen to the ground state configuration 
determined from their choice at the boundary of the lattice. 
Here is an example of such a configuration:
\begin{equation}\label{k:hatten}
\begin{picture}(350,64)(-30,-25)
\setlength{\unitlength}{0.33mm}
\multiput(0,0)(0,-29){2}{
\multiput(12,21)(38,0){9}{
\put(-8,0){\line(1,0){16}}\put(0,-7){\line(0,1){14}}
}}

\put(0,2){
\put(9,30){1}\put(47,30){2}\put(85,30){2}
\put(123,30){2}\put(161,30){1}\put(199,30){2}
\put(237,30){1}\put(275,30){1}\put(313,30){1}}

\put(-15,17){111}\put(22,17){111}
\put(59,17){112}\put(97,17){122}
\put(136,17){222}\put(174,17){122}
\put(213,17){222}\put(250,17){122}
\put(288,17){112}\put(325,17){111}

\put(0,-28){
\put(9,30){1}\put(47,30){1}\put(85,30){1}
\put(123,30){1}\put(161,30){2}\put(199,30){1}
\put(237,30){2}\put(275,30){2}\put(313,30){2}}

\put(0,-29){
\put(-15,17){222}\put(22,17){122}
\put(59,17){112}\put(97,17){111}
\put(136,17){111}\put(174,17){112}
\put(213,17){111}\put(250,17){112}
\put(288,17){122}\put(325,17){222}}

\put(0,-57){
\put(9,30){2}\put(47,30){2}\put(85,30){2}
\put(123,30){1}\put(161,30){1}\put(199,30){2}
\put(237,30){1}\put(275,30){1}\put(313,30){1}}

\put(343,-26){.}

\end{picture}
\end{equation}
Regard such a configuration 
on 2 dimensional lattice as successive downward transfer of 
the horizontal array of spins on vertical edges.
Then each step is a deterministic map corresponding to the 
crystallization of the transfer matrix (\ref{k:rtm}).
The example (\ref{k:hatten}) corresponds to $T_3(1)$.
This is an origin of the BBS time evolutions
$T_1, T_2,\ldots$.
The spins on the horizontal edges are ``hidden variables" 
playing the role of {\it carrier} \cite{TM97}.

In the argument so far, one starts with $q$-dependent objects, 
e.g. fusion $R$-matrices and transfer matrices, and 
then consider their crystallization $q \rightarrow 0$.
In the subsequent sections,
we explain how such procedures can be 
simplified and even more systematized by invoking the crystal theory
of the quantum group $U_q$ \cite{Ka1, Ka2, HK02}.
It provides a general framework to set up everything 
at $q=0$ from the outset.
The labeling set of the bases of $V_m$ and 
the quantum $R$ matrix at $q=0$ will be 
formulated as {\em crystal} and {\em combinatorial $R$}, respectively.
The power of $z$ in (\ref{k:cr21}) is called the {\em energy}
(\ref{t:aug1a}), which will also play an important role.
The BBS and its generalizations will be constructed  
as the canonical dynamical systems associated with the 
crystalline vertex models. 
We remark that a similar approach to the BBS by the crystallization of  
the quantum Lotka-Volterra lattice has been 
undertaken in \cite{HIK99}.

\subsection{Elements of crystal base theory}\label{t:2-2}

The theory of crystal bases was founded by Kashiwara \cite{Ka1, Ka2} 
as a representation theory of quantum group $U_q$ at $q=0$.
The notion of {\em crystal} is abstracted from the theory
of crystal base \cite{KKM}. In this subsection
we give a brief description of {\em crystals} and {\em combinatorial $R$}
which are basic ingredients in BBS.
We remark that the notion of crystal lattice in the theory of crystal bases 
(see \cite{HK02} for example) is omitted in this review, and our attention is focused on 
the notion of crystal.

\subsubsection{Definition of crystals.}\label{t:2-2-1}
Let $I$ be an index set.
A crystal $B$ is a set equipped with maps 
$\et{i}, \ft{i} : B \rightarrow B \sqcup \{ 0 \}$ for $i \in I$,
satisfying certain axiom.
In this article we are exclusively concerned with the semiregular case 
of \cite[Def. 4.5.1]{HK02}.
Then the relevant axiom is the following:

\begin{itemize}
\item 
For any $b \in B$ and $i \in I$, there is $n>0$ such that
 $\et{i}^n b = \ft{i}^n b = 0$,
 
\item $\et{i} 0 = \ft{i} 0 = 0$,
\item For $b_1, b_2 \in B$, $\ft{i} b_1 = b_2$ is equivalent to $\et{i} b_2 = b_1$.
\end{itemize}
Here  we have omitted the items in the axiom involving the weight wt not used in this article.
The $\et{i}$ and $\ft{i}$ are called {\em Kashiwara operators}.
They serve as $q=0$ analogues of 
Chevalley generators.
For $b \in B$, we set 
\begin{align}\label{t:ef-def}
\varepsilon_i (b) = \max \{ m | (\et{i})^m (b) \ne 0 \},
\qquad
\varphi_i (b) = \max \{ m | (\ft{i})^m (b) \ne 0 \}.
\end{align}

For our construction of BBS, we use the crystal $B_l$
associated with $\widehat{\mathfrak{sl}}_{n+1}$,   
where we take $I=\{0,1,\ldots,n \}$ with $n \in \Z_{\geq 1}$.
As a set $B_l$ is given by
\begin{equation}\label{t:jun24a}
B_l = \left\{ (x_1, \ldots , x_{n+1} ) \in \Z^{n+1} \bigm| x_i \geq 0, \, \textstyle{\sum_{i=1}^{n+1}} x_i = l \right\}.
\end{equation}
The elements of $B_l$ are also represented by Young tableaux.
For each $x = (x_1, \ldots , x_{n+1} ) \in B_l$ we associate a one-row 
semistandard tableau of length $l$
in which letter $i$ appears $x_i$ times.
For instance let $n=2$ and $l=2$.
Then the crystal 
\begin{equation}\label{t:jun24b}
B_2 = \{ (2,0,0), (0,2,0),(0,0,2),(1,1,0),(1,0,1),(0,1,1) \}
\end{equation}
is also written as
\begin{equation}\label{t:jun24c}
B_2 = \{ \, \twoboxes{1}{1}\,,\twoboxes{2}{2}\,,\twoboxes{3}{3}\,,\twoboxes{1}{2}\,,\twoboxes{1}{3}\,,\twoboxes{2}{3} \, \}.
\end{equation}
In what follows all indices of $x_i, y_i, \ldots$ are interpreted in $\Z_{n+1} = \Z / (n+1) \Z$, namely $x_{i+n+1} = x_i$.
The $B_l$ is the labeling set of the bases of the $l$-fold symmetric 
tensor representation (an example of so-called Kirillov-Reshetikhin modules) of 
$U_q(\widehat{\mathfrak{sl}}_{n+1})$.

For $x = (x_1, \ldots , x_{n+1} ) \in B_l$, 
let $\et{i}, \ft{i} : B_l \rightarrow B_l \sqcup \{ 0 \} \quad (0 \leq i \leq n)$ be maps defined by
\begin{equation}\label{t:jun24e}
\et{i} (x) = (\ldots, x_i+1, x_{i+1}-1, \ldots ), \quad  \ft{i} (x) =(\ldots, x_i-1, x_{i+1}+1, \ldots ),
\end{equation}
if their images fall into $B_l$, or
they are interpreted as $0$ otherwise.
According to \eqref{t:ef-def},
the maps 
$\varepsilon_i, \varphi_i  : B_l \rightarrow \Z \quad (0 \leq i \leq n)$ 
are given by
\begin{equation}\label{t:jun24d}
\varepsilon_i (x) = x_{i+1},  \quad \varphi_{i} (x) = x_{i}.
\end{equation}

For any crystals $B, B'$ one can define their tensor product $B \otimes B'$. 
As a set it is a direct product $B \times B'$, but it also has a crystal structure.
Any $(x,y) \in B \times B'$ determines an element $x \otimes y \in B \otimes B'$, and
we understand $x \otimes 0 = 0 \otimes y = 0$.
For $x \otimes y \in B \otimes B'$ the maps $\varepsilon_i, \varphi_i, \et{i}, \ft{i}$ are given by
\begin{align}
\varepsilon_i (x \otimes y) &= \varepsilon_i (x) + (\varepsilon_i (y) - \varphi_i (x))_+, \label{t:jun24f}\\
\varphi_i (x \otimes y) &= \varphi_i (y) + (\varphi_i (x) - \varepsilon_i (y))_+, \label{t:jun24g}\\
\et{i} (x \otimes y) &= 
\begin{cases}
\et{i} x \otimes y \qquad \mbox{if $\varphi_i (x) \geq \varepsilon_i (y)$}, \\
x \otimes \et{i} y \qquad \mbox{if $\varphi_i (x) < \varepsilon_i (y)$}, 
\end{cases}
\label{t:jun24h}\\
\label{t:jun24i}
\ft{i} (x \otimes y) &= 
\begin{cases}
\ft{i} x \otimes y \qquad \mbox{if $\varphi_i (x) > \varepsilon_i (y)$}, \\
x \otimes \ft{i} y \qquad \mbox{if $\varphi_i (x) \leq \varepsilon_i (y)$}, 
\end{cases}
\end{align}
where $(x)_+ = \max (x,0)$.
The tensor product defined in this way satisfies the axioms of the crystals.
By repeated use of this construction one can define tensor products of more than two crystals, where
the (co)associativity $(B \otimes B') \otimes B'' = B \otimes (B' \otimes B'')$ holds.
In particular, this allows one to
define the $\widehat{\mathfrak{sl}}_{n+1}$ crystal $B_{l_1} \otimes \cdots \otimes B_{l_m}$
for any set of positive integers $l_1, \ldots, l_m$. 

The crystals are represented by colored oriented graphs, known as {\em crystal graphs}.
Let us show an example.
\begin{equation}\label{t:cry:b1b2}
\setlength{\unitlength}{0.43mm}
\begin{picture}(240,30)(-7,0)
\put(-15,10){$B_1:$}
\put(20,10){$\young(1) \qquad \quad\young(2)$} \put(35,9){\vector(1,0){20}}\put(55,15){\vector(-1,0){20}}
\put(43,19){${\scriptstyle 0}$}\put(43,-1){${\scriptstyle 1}$}

\put(86,10){$B_2:$}
\put(120,10){$\young(11) \qquad \quad\young(12) \qquad \quad\young(22)$} \put(148,9){\vector(1,0){20}}\put(168,15){\vector(-1,0){20}}
\put(156,19){${\scriptstyle 0}$}\put(156,-1){${\scriptstyle 1}$}
\put(200,9){\vector(1,0){20}}\put(220,15){\vector(-1,0){20}}
\put(208,19){${\scriptstyle 0}$}\put(208,-1){${\scriptstyle 1}$}

\end{picture}
\end{equation}
Here the arrows with index $i$ represent the actions of $\ft{i}$.
A tensor product of crystals is represented as follows.
\begin{equation}\label{t:cry:b11}
\setlength{\unitlength}{0.4mm}
\begin{picture}(150,70)(0,-35)
\put(-43,0){$B_1 \otimes B_1:$}
\put(20,0){$\young(1)\otimes \young(1)$} \put(40,10){\vector(1,1){15}}\put(42,20){${\scriptstyle 1}$}
\put(55,-23){\vector(-1,1){15}}\put(43,-25){${\scriptstyle 0}$}
\put(60,30){$\young(2)\otimes \young(1)$} \put(100,26){\vector(1,-1){15}}\put(110,20){${\scriptstyle 1}$}
\put(60,-30){$\young(1)\otimes \young(2)$} \put(117,-8){\vector(-1,-1){15}}\put(112,-22){${\scriptstyle 0}$}
\put(100,0){$\young(2)\otimes \young(2)$} 
\end{picture} 
\end{equation} 
Let us show two more examples.
\begin{equation}\label{t:cry:b12}
\setlength{\unitlength}{0.4mm}
\begin{picture}(250,80)(-25,-35)
\put(-40,0){$B_1 \otimes B_2:$}
\put(20,0){$\young(1)\otimes \young(11)$} \put(40,10){\vector(1,1){15}}\put(42,20){${\scriptstyle 1}$}
\put(55,-23){\vector(-1,1){15}}\put(43,-25){${\scriptstyle 0}$}
\put(60,30){$\young(2)\otimes \young(11)$} \put(60,-30){$\young(1)\otimes \young(12)$}

\put(140,30){$\young(2)\otimes \young(12)$} \put(140,-30){$\young(1)\otimes \young(22)$}

\put(112,35){\vector(1,0){20}}\put(120,38){${\scriptstyle 1}$}
\put(132,29){\vector(-1,0){20}}\put(120,18){${\scriptstyle 0}$}

\put(112,-25){\vector(1,0){20}}\put(120,-22){${\scriptstyle 1}$}
\put(132,-31){\vector(-1,0){20}}\put(120,-42){${\scriptstyle 0}$}

\put(190,26){\vector(1,-1){15}}\put(200,20){${\scriptstyle 1}$}
\put(207,-8){\vector(-1,-1){15}}\put(202,-22){${\scriptstyle 0}$}
\put(190,0){$\young(2)\otimes \young(22)$}

\end{picture}
\end{equation}
\begin{equation}\label{t:cry:b21}
\setlength{\unitlength}{0.4mm}
\begin{picture}(250,80)(-25,-35)

\put(-40,0){$B_2 \otimes B_1:$}
\put(20,0){$\young(11)\otimes \young(1)$} \put(40,10){\vector(1,1){15}}\put(42,20){${\scriptstyle 1}$}
\put(55,-23){\vector(-1,1){15}}\put(43,-25){${\scriptstyle 0}$}
\put(60,30){$\young(12)\otimes \young(1)$} \put(60,-30){$\young(11)\otimes \young(2)$}

\put(140,30){$\young(22)\otimes \young(1)$} \put(140,-30){$\young(12)\otimes \young(2)$}

\put(112,35){\vector(1,0){20}}\put(120,38){${\scriptstyle 1}$}
\put(132,29){\vector(-1,0){20}}\put(120,18){${\scriptstyle 0}$}

\put(112,-25){\vector(1,0){20}}\put(120,-22){${\scriptstyle 1}$}
\put(132,-31){\vector(-1,0){20}}\put(120,-42){${\scriptstyle 0}$}

\put(190,26){\vector(1,-1){15}}\put(200,20){${\scriptstyle 1}$}
\put(207,-8){\vector(-1,-1){15}}\put(202,-22){${\scriptstyle 0}$}
\put(190,0){$\young(22)\otimes \young(2)$}

\end{picture}
\end{equation}
\subsubsection{Combinatorial $R$ and its explicit formula.}\label{t:2-2-2}
In general, two crystals $B \otimes B'$ and $B' \otimes B$ share a common crystal structure.
The {\em combinatorial $R$} is the bijection 
between $B \otimes B'$ and $B' \otimes B$
that commutes with the actions of Kashiwara operators.
It is a $q \to 0$ limit of the quantum $R$ matrix ${\check R}(z)$ in \S 2.1.
In other words it is a map 
$R=R_{B B'}: B \otimes B' \rightarrow B' \otimes B$ which satisfies the 
following relations:
\begin{equation}
R (\et{i} (x \otimes y)) = \et{i} (R (x \otimes y)),\qquad
R (\ft{i} (x \otimes y)) = \ft{i}  (R (x \otimes y)).\label{t:jun22_1}
\end{equation}
In all the cases we consider in this article,
the combinatorial $R$ is uniquely determined by demanding the above conditions. 

By the definition, the inversion relation
$R_{B B'}\circ R_{B' B}={\rm Id}_{B'\otimes B}$ holds.
The simplest case is $B=B'$, where 
the combinatorial $R$ reduces to the identity map.
As a non-trivial example, we find that the combinatorial 
$R: B_2 \otimes B_1 \rightarrow B_1 \otimes B_2$ 
for $\widehat{\mathfrak{sl}}_{2}$ is given by (\ref{k:cr21}) by
comparing the crystal graphs (\ref{t:cry:b12}) and (\ref{t:cry:b21}),
modulo the power of $z$ (which will be related to the energy function in (\ref{t:aug1a}) ).
For $m$ general, $R: B_m \otimes B_1 \rightarrow B_1 \otimes B_m$ for 
$\widehat{\mathfrak{sl}}_2$ 
is given by \eqref{k:wtm1} in the notation of \eqref{t:july15a}. 
More generally,
we shall present
a simple algorithm for
$R: B_l \otimes B_{l'} \rightarrow B_{l'} \otimes B_l$ in \S \ref{t:2-2-3}.

For the $\widehat{\mathfrak{sl}}_{n+1}$ crystals there is a piecewise-linear formula for the combinatorial $R$.
Given $x = (x_1, \ldots , x_{n+1} ), y = (y_1, \ldots , y_{n+1} ) \in \Z^{n+1}$ let 
$\tilde{x} = (\tilde{x}_1, \ldots , \tilde{x}_{n+1} )$,
$\tilde{y} = (\tilde{y}_1, \ldots , \tilde{y}_{n+1} ) \in \Z^{n+1}$
be defined by 
\begin{equation}\label{t:udRPP}
\begin{split}
&\tilde{x}_i = x_i - P_i(x,y) + P_{i-1}(x,y), \quad
\tilde{y}_i = y_i + P_i(x,y) - P_{i-1}(x,y), \\
&P_i(x,y) = \max_{1 \leq k \leq n+1} \left( \sum_{j=k}^{n+1} x_{i+j} + \sum_{j=1}^{k} y_{i+j} \right).
\end{split}
\end{equation}
%
\begin{Prop}\label{t:aug2a}
Given $x \in B_l, y \in B_{l'}$ define $\tilde{x}, ~\tilde{y} \in \Z^{n+1}$ by (\ref{t:udRPP}). Then:
\begin{enumerate}
\item All their elements are non-negative, hence $\tilde{x} \in B_l, \tilde{y} \in B_{l'}$.
\item Define $R : B_l \otimes B_{l'} \rightarrow B_{l'} \otimes B_l$ by
$R(x \otimes y) = \tilde{y} \otimes \tilde{x}$.
Then it is the combinatorial $R$ for the $\widehat{\mathfrak{sl}}_{n+1}$ crystals, i.e. it satisfies the
relations (\ref{t:jun22_1}).
\end{enumerate}
\end{Prop}
One can prove it by showing the equivalence of the piecewise-linear formulas (\ref{t:udRPP}) 
with an algorithm for the $R$ in \S \ref{t:2-2-3}
(\cite[Prop. 4.1]{HHIKTT01}).
Another proof will be given,
following an idea in \cite[Th. 4.28]{KOTY03}, as a consequence of
the corresponding assertion in geometric crystals (Proposition \ref{t:aug16b}).

The formula of the
combinatorial $R$ (\ref{t:udRPP}) is characterized by the following relations
\begin{equation}\label{t:udtoda}
x_i+ y_i = \tilde{y}_i + \tilde{x}_i,\qquad 
\max (-x_i, -y_{i+1}) = \max (-\tilde{y}_i, -\tilde{x}_{i+1}),
\end{equation}
with an extra constraint $\sum_{i=1}^{n+1} (x_i - \tilde{x}_i) = \sum_{i=1}^{n+1} (y_i - \tilde{y}_i) = 0$.
The relations (\ref{t:udtoda}) are consequence of 
$\varepsilon_i ({\tilde y} \otimes {\tilde x}) = \varepsilon_i (x \otimes y)$
and
$\varphi_i ({\tilde y} \otimes {\tilde x}) = \varphi_i (x \otimes y)$
which follow from (\ref{t:jun22_1}).

%

A notion related to the combinatorial $R$ is the {\em energy function} $H: B \ot B' \rightarrow \Z$,
For the $\widehat{\mathfrak{sl}}_{n+1}$ crystals,
$H: B_l \otimes B_{l'} \rightarrow \Z$ is explicitly given by 
\begin{equation}\label{t:aug1a}
H(x \otimes y) = P_0(x,y) - \max (l,l').
\end{equation}
Even if $B=B'$ where we have $R = \mbox{Id}$, the energy function is not trivial and plays an important role
in the theory of crystals and its applications \cite{KKM, KMN1}.
For the BBS, $H$ will be used at Proposition \ref{t:prop:1}.

\subsubsection{Algorithm for combinatorial $R$.}\label{t:2-2-3}
There is a simple way to calculate the image of 
the combinatorial $R$ and the energy function 
without drawing the whole crystal graph,
due to Nakayashiki and Yamada \cite{NY97}.
We explain the algorithm along an example:
\begin{align}\label{i:R-ex}
&R \left( \; \fiveboxes{1}{3}{3}{4}{7}
\ot
\threeboxes{1}{3}{5} \;\right)
=
\threeboxes{1}{4}{7}
\ot
\fiveboxes{1}{3}{3}{3}{5}.
\end{align}
Given the left hand side we can obtain the
right hand side by using the following diagram:
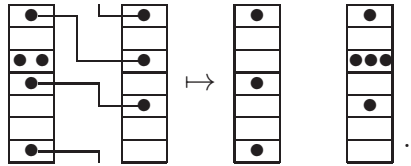
\begin{displaymath}
\setlength{\unitlength}{3mm}
\begin{picture}(17,6)(0,0.7)
\multiput(0,0)(0,1){8}{\line(1,0){2}}
\multiput(5,0)(0,1){8}{\line(1,0){2}}
\multiput(0,0)(2,0){2}{\line(0,1){7}}
\multiput(5,0)(2,0){2}{\line(0,1){7}}
\multiput(10,0)(0,1){8}{\line(1,0){2}}
\multiput(15,0)(0,1){8}{\line(1,0){2}}
\multiput(10,0)(2,0){2}{\line(0,1){7}}
\multiput(15,0)(2,0){2}{\line(0,1){7}}
\put(0.5,0){\makebox(1,1){$\bullet$}}
\put(0.5,3){\makebox(1,1){$\bullet$}}
\multiput(0,4)(1,0){2}{\makebox(1,1){$\bullet$}}
\put(0.5,6){\makebox(1,1){$\bullet$}}
\put(5.5,2){\makebox(1,1){$\bullet$}}
\put(6,2.5){\line(-1,0){2}}
\put(4,2.5){\line(0,1){1}}
\put(4,3.5){\line(-1,0){2.7}}
\put(5.5,4){\makebox(1,1){$\bullet$}}
\put(6,4.5){\line(-1,0){3}}
\put(3,4.5){\line(0,1){2}}
\put(3,6.5){\line(-1,0){1.7}}
\put(5.5,6){\makebox(1,1){$\bullet$}}
\put(6,6.5){\line(-1,0){2}}
\put(4,6.5){\line(0,1){0.5}}
\put(4,0){\line(0,1){0.5}}
\put(4,0.5){\line(-1,0){2.7}}
\put(8,3){\makebox(1,1){$\mapsto$}}
\put(10.5,0){\makebox(1,1){$\bullet$}}
\put(10.5,3){\makebox(1,1){$\bullet$}}
\put(10.5,6){\makebox(1,1){$\bullet$}}
\put(15.5,2){\makebox(1,1){$\bullet$}}
\multiput(14.9,4)(0.66,0){3}{\makebox(1,1){$\bullet$}}
\put(15.5,6){\makebox(1,1){$\bullet$}}
\end{picture} ~.
\end{displaymath}
We suppose $l \geq l'$ but to guess the algorithm in the case $l < l'$ is easy.
Represent $x = (x_1, \ldots, x_{n+1}) \in B_l$ by a pile of $n+1$ boxes in which
there are $x_i$ dots in the $i$th highest box.
Do the similar for $y = (y_1, \ldots, y_{n+1}) \in B_{l'}$ and then juxtapose these piles of boxes.
Repeat the following procedure (1)--(3) to obtain $l'$ pairs of connected dots.
(All the dots are unconnected initially.)
(1) Choose any unconnected dot A in the right pile. 
(2) Look for its partner B in the left pile which is an unconnected dot
in the lowest position but higher than that of A.
If there is no such dots, B is chosen among unconnected dots in the lowest position.
(We call the former case {\em unwinding} and the latter {\em winding}.)
(3) Connect A and B.
At the end we transfer all the unconnected dots from the left pile to
the right one horizontally, yielding the piles for $R (x \otimes y)$.
The energy function (\ref{t:aug1a}) is given by 
\begin{equation}\label{t:aug2b}
H (x \otimes y) = \# \mbox{(winding pairs)}.
\end{equation}
For the above example \eqref{i:R-ex} we have
$H(x \otimes y) = P_0(x,y) - \max(l,l') = 6-5=1.$

The algorithm for $R$ and $H$ will serve as the most substantial tool
to check the examples in \S \ref{t:2-2-4} and \S \ref{t:2-3}.

\subsubsection{Yang-Baxter equation.}\label{t:2-2-4}
The most important property of the combinatorial $R$ is:
\begin{Prop}\label{t:YBeq}
The following relation holds on 
$B \ot B'  \ot B''$:
\begin{equation}\label{t:aug16a}
(R \ot 1)(1 \ot R)(R \ot 1)=(1 \ot R)(R \ot 1)(1 \ot R).
\end{equation}
\end{Prop}
The relation (\ref{t:aug16a}) is known as the {\em Yang-Baxter equation}. 
We depict the relation $R(x \otimes y) = \tilde{y} \otimes \tilde{x}$
by
\begin{equation}\label{t:july15a}
\batten{x}{y}{\tilde{y}}{\tilde{x}}.
\end{equation}

\begin{Example}\label{t:aug1g}
By using the algorithm in \S \ref{t:2-2-3}, 
one can observe that 
the maps in the both sides of (\ref{t:aug16a}) 
send an element
in $B_6 \otimes B_3 \otimes B_1$ to
the same element in $B_1 \otimes B_3 \otimes B_6$.
\begin{align*}
&(R \ot 1)(1 \ot R)(R \ot 1) (\; \sixboxes{2}{2}{3}{4}{5}{5} \otimes \threeboxes{3}{3}{4} \otimes \onebox{6}\;) \\
&= (R \ot 1)(1 \ot R)(\;\threeboxes{2}{2}{3} \otimes \sixboxes{3}{3}{4}{4}{5}{5} \otimes \onebox{6}\;) \\
&= (R \ot 1)(\;\threeboxes{2}{2}{3} \otimes \onebox{5} \otimes \sixboxes{3}{3}{4}{4}{5}{6}\;) \\
&= \onebox{3} \otimes \threeboxes{2}{2}{5} \otimes \sixboxes{3}{3}{4}{4}{5}{6}\;, \\
& \\
&(1 \ot R)(R \ot 1)(1 \ot R)(\;\sixboxes{2}{2}{3}{4}{5}{5} \otimes \threeboxes{3}{3}{4} \otimes \onebox{6}\;) \\
& = (1 \ot R)(R \ot 1)(\;\sixboxes{2}{2}{3}{4}{5}{5} \otimes \onebox{4} \otimes \threeboxes{3}{3}{6}\;) \\
& = (1 \ot R)(\;\onebox{3} \otimes \sixboxes{2}{2}{4}{4}{5}{5} \otimes \threeboxes{3}{3}{6}\;) \\
&= \onebox{3} \otimes \threeboxes{2}{2}{5} \otimes \sixboxes{3}{3}{4}{4}{5}{6}\;.
\end{align*}
It is also depicted as the following diagrams, where the lines represent crystals and
their crossings stand for the combinatorial $R$s.
(See (\ref{t:july15a}).)
\begin{equation}\label{t:july20a}
\setlength{\unitlength}{4mm}
\begin{picture}(30,12)
\multiput(4,1)(0,6){2}{\line(0,1){4}}
\multiput(2,3)(5,2.5){2}{\line(2,1){3}}
\multiput(2,9)(5,-2.5){2}{\line(2,-1){3}}
\put(1,2.5){\small $3$}
\put(6,5){\small $5$}
\put(11,7.5){\small $6$}
\put(0,9.3){\small $223455$}
\put(5,6.8){\small $334455$}
\put(10,4.3){\small $334456$}
\put(3.5,0){\small $225$}
\put(3.5,5.8){\small $223$}
\put(3.5,11.6){\small $334$}
\multiput(24,1)(0,6){2}{\line(0,1){4}}
\multiput(26,3)(-5,2.5){2}{\line(-2,1){3}}
\multiput(26,9)(-5,-2.5){2}{\line(-2,-1){3}}
\put(26,2.3){\small $334456$}
\put(20.4,4.7){\small $224455$}
\put(16,7.3){\small $223455$}
\put(27,9.3){\small $6$}
\put(22,6.8){\small $4$}
\put(17,4.3){\small $3$}
\put(23.5,0){\small $225$}
\put(23.5,5.8){\small $336$}
\put(23.5,11.6){\small $334$}
\end{picture}
\end{equation}
\end{Example}

\subsection{Basic features of BBS}\label{t:2-3}

In this section we introduce a one-dimensional cellular automaton 
associated with $\widehat{\mathfrak{sl}}_{n+1}$ crystals.
For a more extensive presentation,  see \cite{T07}.

\subsubsection{States and time evolutions.}\label{t:2-3-1}
For any positive integer $L$,
we define a dynamical system on $(B_1)^{\otimes L}$ which generalizes the BBS in \S 1.
In the system one may regard  $\onebox{a} \in B_1 \; (a >1)$ as a box of capacity one containing a ball with color $a$ inside it, and
$\onebox{1} \in B_1$ as an empty box of capacity one.
We call our dynamical system an $\widehat{\mathfrak{sl}}_{n+1}$ BBS.
It is a cellular automaton equipped with a family of commuting time evolutions $T_1, T_2, \ldots$ defined in the sequel.

Let $R: B_l \otimes B_1 \rightarrow B_1\otimes B_l$ 
be the combinatorial $R$ 
and define
$R_i = \overbrace{{\rm Id} \otimes \cdots \otimes {\rm Id}}^{i-1} \otimes R \otimes \overbrace{{\rm Id} \otimes \cdots \otimes {\rm Id}}^{L-i}
\quad (1 \leq i \leq L)$, 
which is a map from 
$(B_1)^{\otimes i-1} \otimes B_l \otimes (B_1)^{\otimes L-i+1}$ to
$(B_1)^{\otimes i} \otimes B_l \otimes (B_1)^{\otimes L-i}$.
Then $\mathcal{R} = R_L \circ \cdots \circ R_1$
is a map from $B_{l} \otimes (B_{1})^{\otimes L}$ to $(B_{1})^{\otimes L} \otimes B_l$.
Given an arbitrary $v \otimes b_1 \otimes \cdots \otimes b_L \in B_{l} \otimes (B_{1})^{\otimes L}$
let $\mathcal{R} (v \otimes b_1 \otimes \cdots \otimes b_L) = b'_1 \otimes \cdots \otimes b'_L \otimes v'$.
It is depicted by
\begin{equation}\label{t:apr20}
\batten{v}{b_1}{b_1'}{v_1}\!\!\!
\batten{}{b_2}{b_2'}{v_2}\!\!\!
\batten{}{}{}{\cdots\cdots}
\quad
\batten{}{}{}{v_{L-2}}\,\,
\batten{}{b_{L-1}}{b_{L-1}'}{v_{L-1}}\,\,
\batten{}{b_L}{b_L'}{v',}
\end{equation}
or simply by
\begin{equation}\label{t:hone}
\begin{picture}(150,40)(-20,-13)
\put(-10,7){$v$}
\put(11,30){$b_1$}\put(15,-5){\line(0,1){30}}\put(12,-16){$b'_1$}

\put(36,30){$b_2$}\put(40,-5){\line(0,1){30}}\put(37,-16){$b'_2$}

\put(-2,10){\line(1,0){57}}

\put(67,6){$\cdots$}

\put(90,10){\line(1,0){30}}
\put(107,-5){\line(0,1){32}} 
\put(103,30){$b_L$}\put(104,-16){$b'_L$}
\put(123,7){$v'$.}
\end{picture}
\end{equation}
We assume that the conditions
\begin{align}
L  \gg 1, \qquad
b_i = \onebox{1} & \quad \mbox{for all $i \gg 1$},\label{t:aug1c}
\end{align}
are satisfied in (\ref{t:apr20}), and take
\begin{equation}
v = u_l  := \threeboxes{1}{\scriptstyle \cdots}{1} \; . \label{t:aug1d}
\end{equation}
Then we have $v' = u_l$ and the set $\{b'_1, \ldots ,b'_L \}$ coincides with $\{b_1, \ldots ,b_L \}$ as a set but the order of
its elements gets shuffled.
Under this setting let
$T_l: (B_1)^{\otimes L} \rightarrow (B_1)^{\otimes L}$ and
$E_l: (B_1)^{\otimes L} \rightarrow \Z$ be the maps given by
\begin{align}
&T_l (b_1 \otimes \cdots \otimes b_L) = b'_1 \otimes \cdots \otimes b'_L,\label{t:t}\\
&E_l(b_1 \otimes \cdots \otimes b_L) = \sum_{i=1}^L (1-H (v_{i-1} \otimes b_i)),
\label{t:e}
\end{align}
where $v_0 = v$. 
Call $T_l$ the $l$-th {\em time evolution} and $E_l$ the $l$-th {\em energy}.
We note that every summand of (\ref{t:e}) vanishes for $i \gg 1$ because of
$H ( u_l \otimes \onebox{1})=1$, which ensures the convergence of the energy in the limit $L \rightarrow \infty$.

In what follows we often
use the symbol $\simeq$ to indicate that 
its two sides are transformed to each other by the isomorphism
(composition of combinatorial $R$s) of crystals.
For instance, the relation (\ref{t:july15a}) is expressed as $x \otimes y \simeq \tilde{y} \otimes \tilde{x}$.
This notation allows one to write the relation (\ref{t:t}) as a crystal version of ``Lax equation''
\begin{equation}\label{t:aug31a}
u_l \otimes p \simeq T_l (p) \otimes u_l,
\end{equation}
where $p = b_1 \otimes \cdots \otimes b_L$.
An element in $B_l$ is regarded as a carrier which can carry at most $l$ balls.
(The notion of carrier was introduced in \cite{TM97} 
in the case of $\widehat{\mathfrak{sl}}_{2}$.)
%
In the carrier $x = (x_1,\ldots,x_{n+1}) \in B_l$, the count of balls with label $i(>1)$ is $x_i$.
The $u_l$ (\ref{t:aug1d}) corresponds to a vacant carrier.
In (\ref{t:apr20}), a carrier runs from left to right, changing itself as $v \rightarrow v_1 \rightarrow v_2 \rightarrow \cdots$.
Although it is nothing but a repeated use of the algorithm in \S \ref{t:2-2-3} with $l'=1$,
one can regard it as a successive loading/unloading processes of balls into/out of the carrier.

To illustrate how the carrier works, as well as how the energy (\ref{t:e}) is evaluated,
we show a few examples for (\ref{t:apr20}).
\begin{Example}\label{t:aug31b}
Carriers with capacity $4,3$ and $2$.
Consider the state at $t=3$ in Example \ref{t:exp2} below.
By the time evolution $T_4$ it evolves into the state at $t=4$
as follows.
\footnotesize
\begin{equation*}
\scriptstyle
\battendot{1111}{2}{1}{1112}\,
\battendot{}{2}{1}{1122}\,
\battendot{}{2}{1}{1222}\,
\battendot{}{2}{1}{2222}\,
\battennew{}{1}{2}{1222}\,
\battennew{}{1}{2}{1122}\,
\battendot{}{3}{2}{1123}\,
\battendot{}{2}{1}{1223}\,
\battendot{}{4}{3}{1224}\,
\battendot{}{3}{2}{1234}\,
\battendot{}{3}{2}{1334}
\end{equation*}
\normalsize
We added $\bullet$ to each vertex which scores $+1$ to the energy (\ref{t:e}).
Assuming that there are only $1$'s to the right of the first row, there are no more scoring vertices.
Hence we have $E_4 = 9$.
At the vertex with $\bullet$, no winding pairs occur in the algorithm in \S \ref{t:2-2-3} which makes
the value of the energy function (\ref{t:aug2b}) zero.

If the state were evolved by $T_3$, then the diagram would be as follows. 
\footnotesize
\begin{equation*}
\scriptstyle
\battendot{111}{2}{1}{112}\!
\battendot{}{2}{1}{122}\!
\battendot{}{2}{1}{222}\!
\battennew{}{2}{2}{222}\!
\battennew{}{1}{2}{122}\!
\battennew{}{1}{2}{112}\!
\battendot{}{3}{2}{113}\!
\battendot{}{2}{1}{123}\!
\battendot{}{4}{3}{124}\!
\battendot{}{3}{2}{134}\!
\battendot{}{3}{1}{334}
\end{equation*}
\normalsize
Hence $E_3 = 8$.
In the same way, $T_2$ would change the state as follows. 
\footnotesize
\begin{equation*}
\scriptstyle
\battendot{11}{2}{1}{12}\!\!\!\!
\battendot{}{2}{1}{22}\!\!\!\!
\battennew{}{2}{2}{22}\!\!\!\!
\battennew{}{2}{2}{22}\!\!\!\!
\battennew{}{1}{2}{12}\!\!\!\!
\battennew{}{1}{2}{11}\!\!\!\!
\battendot{}{3}{1}{13}\!\!\!\!
\battendot{}{2}{1}{23}\!\!\!\!
\battendot{}{4}{3}{24}\!\!\!\!
\battendot{}{3}{2}{34}\!\!\!\!
\battennew{}{3}{4}{33}
\end{equation*}
\normalsize
Hence $E_2 = 6$.
It is easy to see $E_1 = 3$ and $E_l = 9$ for $l \geq 4$.
\end{Example}

Based on the Yang-Baxter equation (\ref{t:aug16a}),
one can prove the commutativity of the time evolutions and the conservation of the energy.
\begin{Prop}{\rm \cite[Th.3.2]{FOY00}}\label{t:aug1f}
The following relations are satisfied
\begin{equation}
T_l T_{l'}(p) = T_{l'}T_l(p),\qquad
E_l (T_{l'} (p)) = E_l (p),
\end{equation}
for any $l,l'\ge 1$ and state $p = b_1 \otimes \cdots \otimes b_L$.
\end{Prop}

The $l$-th energy $E_l$ is the conserved quantity associated with 
the time evolution $T_l$.
Under the conditions (\ref{t:aug1c}) we define $T_\infty$
just by taking a formal limit $l \to \infty$ in (\ref{t:aug31a}).
In fact, $T_l(p)=T_\infty(p)$ holds if and only if
$l $ is greater than or equal to the maximum amplitude of the
solitons contained in $p$.
(This is due to Theorem \ref{k:th:lin},  especially \eqref{k:rr}. 
See \eqref{t:jun10a} for how to determine the amplitudes of solitons.) 
Remark that $T_1$ serves as a shift operator
which moves every ball to its right adjacent box.

Time evolutions of a state will be illustrated by drawing 
$T_l^t(p), T_l^{t+1}(p), T_l^{t+2}(p),\ldots$ downwardly. 
For a state 
$T_l^t(p) = \onebox{{a_1^t}} \otimes \cdots \otimes \onebox{a_L^t}$,
we omit the symbol $\otimes$ and write it as $a_1^t \ldots a_L^t$.
Hence $a_i^t$ denotes the value at site $i$ and time $t$.
In what follows we write $a_i^t = .$ (a dot) instead of $a_i^t = 1$ 
for simplicity.
Here is an example of  the time evolution
under $T_\infty$: 
\par\noindent
\begin{verbatim}
t=3   ..................322554433.6.............................
t=4   .....................322...5564433........................
t=5   ........................322..5....654433..................
\end{verbatim}
\par\noindent

The special time evolution $T_\infty$  admits an
elementary algorithm due to Takahashi \cite{T93} generalizing 
the $\widehat{\mathfrak{sl}}_{2}$ case  in \S 1, which is a (non-local) 
description without a carrier.
\begin{Prop}\label{t:jun15c}
$T_\infty = K_2 K_3 \cdots K_{n+1}$ where $K_a$ is an operator
that works as follows.
\begin{enumerate}
\item Exchange the leftmost $\onebox{a}$ with its nearest right $\onebox{1}$.
\item Exchange the leftmost $\onebox{a}$ 
among the rest of the $\onebox{a}$'s
with its nearest right $\onebox{1}$.
\item Repeat (ii) until all of the $\onebox{a}$'s are moved exactly once.
\end{enumerate}
\end{Prop}

\begin{Example}\label{t:apr21}
We apply $T_\infty = K_2 K_3 K_4 K_5 K_6$ to the $t=4$
state in Example \ref{t:exp1} to obtain the $t=5$ state.
\par\noindent
\begin{verbatim}
t=4   .....................322...5564433........................
      .....................322...55.44336.......................
      .....................322.....5443365......................
      .....................322.....5..336544....................
      ......................223....5....654433..................
t=5   ........................322..5....654433..................
\end{verbatim}
\end{Example}
\noindent


A comparison of
our formalism of BBS with
that of the vertex models
is summarized in the following table.
\begin{table}[h]
\begin{center}
\begin{tabular}{c|c|c}
\hfil  & vertex models &  BBS \\
\hline 
&&\vspace{-0.5cm}\\
local states
& $U_q$-module & 
crystal \\
&& \vspace{-0.5cm} \\
\hline
&& \vspace{-0.5cm}\\
local interaction
& quantum $R$
& combinatorial $R$ \\
&& \vspace{-0.5cm} \\
\hline
&& \vspace{-0.5cm}\\
$T_l$ & transfer matrix & time evolution
\end{tabular}
\end{center}
\end{table}

\subsubsection{Solitons.}\label{t:2-3-2}
Now we define {\em solitons} in BBS.
Intuitively,
a pattern like $i_l \ldots i_1$ satisfying the condition $i_l \geq \cdots \geq i_1 > 1$
can be regarded as a soliton of amplitude $l$.
It is denoted by $[i_l \ldots i_1]$.
In Examples \ref{t:jun22_3} and \ref{t:jun22_4} below,
the sequence $554322$ is a soliton of amplitude 6.

\begin{Example}\label{t:jun22_3}
Time evolution by $T_l$ with $l \geq 6$:
\par\noindent
\begin{verbatim}
t=0   .......................554322.............................
t=1   .............................554322.......................
t=2   ...................................554322.................
\end{verbatim}
\end{Example}

\begin{Example}\label{t:jun22_4}
Time evolution by $T_4$:
\par\noindent
\begin{verbatim}
t=0   .......................554322.............................
t=1   ...........................554322.........................
t=2   ...............................554322.....................
\end{verbatim}
\end{Example}
If well separated from the others,
a soliton of amplitude $l$ travels at a speed of $\min (l,k)$ under the time evolution $T_k$.
In particular we have the following for any 1-soliton state:
\begin{Prop}{{\rm \cite[Lemma 4.1]{FOY00}}}\label{t:sep5a}
If there exists only one soliton in the state,
it travels at a speed of $\min (l,k)$ under the time evolution $T_k$ where $l$ is the amplitude of the soliton.
And the value of the associated energy $E_k$ is also given by $\min (l,k)$.
\end{Prop}
%
Under the intuitive definition of solitons, one observes that
the number of solitons of each amplitude may look changing during their scattering processes.
We rather want to treat solitons as conserved quantities in BBS.
For the purpose, we define $m_l$ by
\begin{equation}\label{t:jun10a}
E_l = \sum_{k\ge 1}\min(l,k)m_k \;\,
\text{or equivalently}\;\,
m_l = -E_{l-1}+2E_l-E_{l+1},
\end{equation}
where $E_0 = 0$ is understood.
Since $E_l$s are conserved quantities, so are $m_l$s.
Then one can interpret $m_l$ as the number of solitons of amplitude $l$.
In view of Proposition \ref{t:sep5a},
it is consistent with
the previous definition of solitons
in the case where solitons are well separated.
By the definition, {\it any} state $p$ is an $N$-soliton
state, where $N$ is determined by $N=E_1(p)$.
We note that according to (2.42), $E_1(p)$ equals the number of 
adjacent pairs $\onebox{1} \otimes \onebox{a}$ with 
$a>1$ appearing in $p$, whereas
$E_\infty(p)$ is the total number of balls.

\begin{Example}\label{t:exp2}
A three body scattering process under $T_\infty$.
From  Example \ref{t:aug31b}, we find at $t=3$ that 
$m_l = -E_{l-1} + 2 E_l - E_{l+1} = 1 $ 
for $l=2,3,4$ and $m_l =0$ in the other cases.
\par\noindent
\begin{verbatim}
t=0   ........2222.....332..43..................................
t=1   ............2222....332.43................................
t=2   ................2222...33243..............................
t=3   ....................2222..32433...........................
t=4   ........................222.322433........................
t=5   ...........................22..3224332....................
t=6   .............................22...322.4332................
t=7   ...............................22....322..4332............
t=8   .................................22.....322...4332........
\end{verbatim}
We will see that the nonlinear time evolutions of BBS are transformed into linear ones
on the rigged configurations.
See Example \ref{k:rcd}.
\end{Example}
We remark that, besides $m_l$s, our BBS has additional conserved quantities \eqref{k:av}.
By using the crystals for anti-symmetric representations, one can show that
the color degrees of freedom for any state of 
the BBS can be transformed into a ``word'' which does not change under any $T_l$ \cite{T05}.

\subsubsection{Scattering rules.}
The scattering of solitons in our BBS
consists of the exchange of their internal degrees of freedom
and the phase shifts.
Although it is possible to treat general many-body scattering processes,
we devote ourselves to the case of two-body scatterings for simplicity.
In what follows we assume that the time evolution is given by $T_\infty$.

\begin{Example}\label{t:sept9a}
A scattering process of two solitons
with amplitudes $l = 6$ and $l' = 3$ in $\widehat{\mathfrak{sl}}_{5}$ BBS.
\par\noindent
\begin{verbatim}
t=0   554322.........422......................................
t=1   ......554322......422...................................
t=2   ............554322...422................................
t=3   ..................5543..42222...........................
t=4   ......................553....442222.....................
t=5   .........................553.......442222...............
t=6   ............................553..........442222.........
t=7   ...............................553.............442222...
\end{verbatim}
\end{Example}

Suppose at time $t=0$, the state bears two solitons 
$[i_l \ldots i_1]$ and $[j_{l'} \ldots j_1]$. 
Denote this two-soliton state by $[i_l \ldots i_1]_x \times[j_{l'} \ldots j_1]_y$, 
where $x$ and $y$ are the positions of their leftmost letters.
For instance, we have $[554322]_1 \times[422]_{16}$ in Example \ref{t:sept9a} at $t=0$.
We assume $l > l'$ and $x \ll y$.
Then the former catches up with the latter and they eventually collide.
Before the collision these solitons travel at speeds of $l$ and $l'$ cells per unit time respectively, so at time $t$ we have
$[i_l \ldots i_1]_{x+l t} \times[j_{l'} \ldots j_1]_{y+l't}$.
After the collision we have two-soliton state 
$[\tilde{j}_{l'} \ldots \tilde{j}_1]_{y+l't - \delta} \times
[\tilde{i}_l \ldots \tilde{i}_1]_{x+l t + \delta }$ where $\delta$ is a {\em phase shift}.
(This fact is also a part of the statements in the forthcoming Proposition 
\ref{t:prop:1}.)
By the collision the larger soliton gets pushed forward and the smaller soliton pulled backward by 
an amount of $\delta$ cells.

The exchange of the internal degrees of freedom occurring here is governed by the combinatorial $R$, and
the phase shift is essentially given by the energy function $H$.

\begin{Prop}{\rm \cite[Th.4.6]{FOY00}}\label{t:prop:1}
Any collisions of two solitons asymptotically break up into two solitons.
Let the two-soliton state be  
$[i_l \ldots i_1]_{x+l t} \times [j_{l'} \ldots j_1]_{y+l't}$ 
at time $t$ well before the collision, and 
$[\tilde{j}_{l'} \ldots \tilde{j}_1]_{y+l't - \delta} \times
[\tilde{i}_l \ldots \tilde{i}_1]_{x+l t + \delta }$ for $t$
after the collision.
Then the phase shift $\delta$ is given by
\begin{equation}\label{t:aug2c}
\delta = 
H \left( \, \threeboxes{i_1}{\scriptstyle \cdots}{i_l} \otimes \threeboxes{j_1}{\scriptstyle \cdots}{j_{l'}} \, \right) +l'.
\end{equation}
And the exchange of the internal degrees of freedom 
is described by the combinatorial $R$ for the $\widehat{\mathfrak{sl}}_{n}$ 
crystals\footnote{The elements of crystals which
appeared in the formulas in Proposition \ref{t:prop:1} have no letter ``1".
Hence the combinatorial $R$ and the energy function used there are regarded as those for $\widehat{\mathfrak{sl}}_{n}$ crystals
by reducing the value of all letters in the Young tableaux by $1$. }
\begin{equation}\label{t:sep12b}
R \left( \, \threeboxes{i_1}{\scriptstyle \cdots}{i_l} \otimes \threeboxes{j_1}{\scriptstyle \cdots}{j_{l'}} \, \right) =
\threeboxesnew{\tilde{j}_1}{\scriptstyle \cdots}{\tilde{j}_{l'}} \otimes
\threeboxesnew{\tilde{i}_1}{\scriptstyle \cdots}{\tilde{i}_{l}}\; .
\end{equation}
\end{Prop}
The phase shift can be computed by (\ref{t:aug1a}) or (\ref{t:aug2b}).
It is always positive and take values between $l'$ and $2 l'$.

\begin{Example}\label{t:sept9b}
The scattering process in Example \ref{t:sept9a} is expressed as
$[554322]_{1+6t} \times [422]_{16 + 3t} \rightarrow [553]_{11 + 3t} \times [442222]_{6 + 6t}$.
This  
is described by the combinatorial $R$ for the $\widehat{\mathfrak{sl}}_{4}$ crystals.
By using the algorithm in \S \ref{t:2-2-3}, it can be computed as
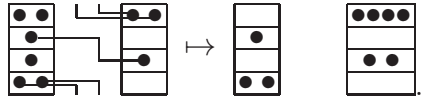
\begin{displaymath}
\setlength{\unitlength}{3mm}
\begin{picture}(18,4)(0,0)
\multiput(0,0)(0,1){5}{\line(1,0){2}}
\multiput(5,0)(0,1){5}{\line(1,0){2}}
\multiput(0,0)(2,0){2}{\line(0,1){4}}
\multiput(5,0)(2,0){2}{\line(0,1){4}}
\multiput(10,0)(0,1){5}{\line(1,0){2}}
\multiput(15,0)(0,1){5}{\line(1,0){3}}
\multiput(10,0)(2,0){2}{\line(0,1){4}}
\multiput(15,0)(3,0){2}{\line(0,1){4}}
\multiput(0,0)(1,0){2}{\makebox(1,1){$\bullet$}}
\put(0.5,1){\makebox(1,1){$\bullet$}}
\put(0.5,2){\makebox(1,1){$\bullet$}}
\multiput(0,3)(1,0){2}{\makebox(1,1){$\bullet$}}
\put(5.5,1){\makebox(1,1){$\bullet$}}
\put(6,1.5){\line(-1,0){2}}
\put(4,1.5){\line(0,1){1}}
\put(4,2.5){\line(-1,0){2.7}}
\multiput(5,3)(1,0){2}{\makebox(1,1){$\bullet$}}
\put(5.5,3.6){\line(-1,0){1.5}}
\put(4,3.6){\line(0,1){0.4}}
\put(4,0){\line(0,1){0.6}}
\put(4,0.6){\line(-1,0){2.7}}
\put(6.5,3.4){\line(-1,0){3.5}}
\put(3,3.4){\line(0,1){0.6}}
\put(3,0){\line(0,1){0.4}}
\put(3,0.4){\line(-1,0){2.7}}
%
\put(8,1.5){\makebox(1,1){$\mapsto$}}
\multiput(10,0)(1,0){2}{\makebox(1,1){$\bullet$}}
\put(10.5,2){\makebox(1,1){$\bullet$}}
\multiput(15,3)(0.66,0){4}{\makebox(1,1){$\bullet$}}
\multiput(15.5,1)(1,0){2}{\makebox(1,1){$\bullet$}}
\end{picture}.
\end{displaymath}
The value of the energy function ($= \#$ (winding pairs)) is $2$
and we observe that the phase shift is given by $\delta = 2 + 3 = 5$.
\end{Example}
\begin{Example}\label{t:exp1}
Scattering processes of three solitons.
There are three solitons [554322], [433], [6] at time $t=0$, and
again three solitons [3], [522], [654433] at time $t=8$.
\par\noindent
\begin{verbatim}
t=0   554322......433..........6................................
t=1   ......554322...433........6...............................
t=2   ............554322433......6..............................
t=3   ..................322554433.6.............................
t=4   .....................322...5564433........................
t=5   ........................322..5....654433..................
t=6   ...........................3225.........654433............
t=7   ..............................3522............654433......
t=8   ...............................3..522...............654433
\end{verbatim}

\par\noindent
\begin{verbatim}
t=0   554322...............433..6...............................
t=1   ......554322............4336..............................
t=2   ............554322.........4633...........................
t=3   ..................554322....4..633........................
t=4   ........................5543242...633.....................
t=5   .............................3.554422633..................
t=6   ..............................3......522654433............
t=7   ...............................3........522...654433......
t=8   ................................3..........522......654433
\end{verbatim}
Both processes have the same kinds of solitons at $t=0$. 
The orders of collisions occurring in subsequent times are different.
The fact that the outcomes at $t=8$ share a common soliton content
can be viewed as a consequence of the Yang-Baxter equation (\ref{t:aug16a}).
See Example \ref{t:aug1g}.
%
\end{Example}
Due to the commutativity of the time evolutions (Proposition \ref{t:aug1f}), the scattering rule remains unchanged
when $T_\infty$ is replace by $T_k$ for any $k > l'$.
In this case we have the two body scattering 
$[i_l \ldots i_1]_{x+\min(l,k) t} \times [j_{l'} \ldots j_1]_{y+l't} \rightarrow [\tilde{j}_{l'} \ldots \tilde{j}_1]_{y+l't - \delta} \times
[\tilde{i}_l \ldots \tilde{i}_1]_{x+\min(l,k) t + \delta }$.
Here the phase shift and the exchange of internal degrees of freedom are still 
given by
(\ref{t:aug2c}) and (\ref{t:sep12b}). 
\subsubsection{$\mathfrak{sl}_n$ symmetry.}
So far we have not mentioned the role of 
the Kashiwara operators acting on the states of BBS.
Their significance is recognized as the $\mathfrak{sl}_n$ symmetry in the system.
Let $p$ be a state of BBS.
Suppose $\et{i} p \ne 0$ for some $i \in \{ 2, \ldots, n \}$.
Then we have
\begin{equation}\label{t:aug2f}
T_l (\et{i} p) = \et{i} T_l (p), \qquad E_l(\et{i} p) = E_l (p),
\end{equation}
for any $l$ \cite{FOY00}.
The first relation is a manifestation of the 
$\mathfrak{sl}_n$ symmetry of the 
time evolution.
Due to (\ref{t:jun10a}) and the second relation in (\ref{t:aug2f}),
this transformation does not change
the amplitudes of solitons but alters their internal labels.

A conserved quantity associated with the $\mathfrak{sl}_n$ symmetry is defined as follows.
Given a state $p = \onebox{{a_1}} \otimes \cdots \otimes \onebox{a_L}$,
let $w_1 \ldots w_k$ be the word which is obtained from $a_1 \ldots a_L$ by ignoring every $1$.
Denote by $P(p)$ the $P$-symbol $P (w_k \ldots w_1)$
and $Q(p)$ the $Q$-symbol $Q (w_k \ldots w_1)$
(semi-standard Young tableaux) obtained from the opposite word $w_k \ldots w_1$ by
the Robinson-Schensted-Knuth correspondence \cite{Ful}.
Explicitly the $P$-symbol is defined as
\begin{eqnarray*}
P(w_k \ldots w_1) &=& w_k \rightarrow ( w_{k-1} \rightarrow ( \ldots (w_3 \rightarrow (w_2 \rightarrow w_1) )\ldots) ) \\
&=& ((\ldots ((w_k \leftarrow w_{k-1}) \leftarrow w_{k-2}) \ldots ) \leftarrow w_{2}) \leftarrow w_{1},
\end{eqnarray*}
where $\rightarrow$ implies the column insertion and 
$\leftarrow$ does the row insertion \cite{Ful}.
The $Q$-symbol is the standard tableau consisting of $\{1,\ldots,k\}$ that records
the growth history of the $P$-symbol.
We have the following:
\begin{Prop}{\rm \cite[Th.3.1]{Fuk}}\label{t:jun10b}
The $P$-symbol $P(p)$ is a conserved quantity of BBS, i.e.
$P(T_l^t(p))$ is independent of $t$ for any $l$.
\end{Prop}

We note that the time evolution of the BBS is attributed to the dynamics of $Q$-symbol 
\cite[Th.5.1]{Fuk}.

For instance consider Example \ref{t:exp1}.
The opposite words $w_{10} \ldots w_1$ are $6334223455$ for $t=0$ and
$3344562253$ for $t=8$.
Both words share a common $P$-symbol
\begin{equation*}
P(T_\infty^t(p)) =
\begin{array}{l}
\vspace{-2mm} 223455 \\
\vspace{-2mm} 334 \\
6
\end{array}.
\end{equation*}
The growth pattern of the pair $(P,Q)$
by the successive row insertions for the word $6334223455$ looks as follows:
\begin{eqnarray*}
&&
\begin{array}{l}
\vspace{-2mm} 6 \\
{}
\end{array}
\begin{array}{l}
\vspace{-2mm} 1 \\
{}
\end{array},\quad
\begin{array}{l}
\vspace{-2mm} 3 \\
6
\end{array}
\begin{array}{l}
\vspace{-2mm} 1 \\
2
\end{array},\quad
\begin{array}{l}
\vspace{-2mm} 33 \\
6
\end{array}
\begin{array}{l}
\vspace{-2mm} 13 \\
2
\end{array},\quad
\begin{array}{l}
\vspace{-2mm} 334 \\
6
\end{array}
\begin{array}{l}
\vspace{-2mm} 134 \\
2
\end{array},\\
&&
\begin{array}{l}
\vspace{-2mm} 234 \\
\vspace{-2mm} 3 \\
6
\end{array}
\begin{array}{l}
\vspace{-2mm} 134 \\
\vspace{-2mm} 2 \\
5
\end{array},\quad
\begin{array}{l}
\vspace{-2mm} 224 \\
\vspace{-2mm} 33 \\
6
\end{array}
\begin{array}{l}
\vspace{-2mm} 134 \\
\vspace{-2mm} 26 \\
5
\end{array},\quad
\begin{array}{l}
\vspace{-2mm} 223 \\
\vspace{-2mm} 334 \\
6
\end{array}
\begin{array}{l}
\vspace{-2mm} 134 \\
\vspace{-2mm} 267 \\
5
\end{array},\\
&&
\begin{array}{l}
\vspace{-2mm} 2234 \\
\vspace{-2mm} 334 \\
6
\end{array}
\begin{array}{l}
\vspace{-2mm} 1348 \\
\vspace{-2mm} 267 \\
5
\end{array},\quad
\begin{array}{l}
\vspace{-2mm} 22345 \\
\vspace{-2mm} 334 \\
6
\end{array}
\begin{array}{l}
\vspace{-2mm} 13489 \\
\vspace{-2mm} 267 \\
5
\end{array},\quad
\begin{array}{l}
\vspace{-2mm} 223455 \\
\vspace{-2mm} 334 \\
6
\end{array}
\begin{array}{l}
\vspace{-2mm} 13489 \text{{\small X}} \\
\vspace{-2mm} 267 \\
5
\end{array}.
\end{eqnarray*}
Here  {\small X} denotes $10$. 

In the case of $\mathfrak{sl}_2$ it is known that one can introduce
another $P$-symbol whose shape, which represents the list of amplitudes of solitons,
is a conserved quantity \cite{A2,TTS}.

\subsection{Various generalizations}

The BBS has been generalized extensively.
Here we present a few prototype examples.

\noindent
$\bullet$ Generalizations in $\widehat{\mathfrak{sl}}_{n+1}$ case. 
The original BBS consists of boxes with capacity $1$ only, which
corresponds to the fact that the states belong to 
$\cdots \otimes B_1 \otimes B_1 \otimes \cdots$.
A natural generalization is to replace it with  
 $\cdots  \otimes B_{k_i} \otimes B_{k_{i+1}}\cdots \otimes \cdots$.
The commuting family of time evolutions $\{T_l\}$ 
are defined in the same way as before, 
where the vertical lines in (\ref{t:hone}) now represent
$B_{k_i}$s.
The resulting dynamical system is BBS involving 
a box with capacity $k_i$ at the site $i$ \cite{HHIKTT01,TTM00}. 
The basic features of the system, e.g. 
solitons, scattering rules, conserved quantities, linearization scheme, etc. 
remain the same as the capacity one case.
See \cite{HHIKTT01,KOSTY06,KSY}.
The BBS with a periodic boundary condition will be treated in \S 5. 

One can further use crystals other than the family $\{B_l\}$.
Examples of such kind are a BBS with reflecting end \cite{KOY} 
and a BBS associated with anti-symmetric tensor representations of 
$U_q(\widehat{\mathfrak{sl}}_{n+1})$ \cite{DY}.

\noindent
$\bullet$ $\mathfrak{g}_n$-automaton.
Similarly to the $\widehat{\mathfrak{sl}}_{n+1} (= A_n^{(1)})$ case, 
integrable cellular automata associated with 
the non-exceptional affine Lie algebra
$\mathfrak{g}_n = B_n^{(1)}$, $C_n^{(1)}$, 
$D_n^{(1)}$, $A_{2n-1}^{(2)}$, 
$A_{2n}^{(2)}$ and $D_{n+1}^{(2)}$ 
have been constructed \cite{HKT1} and 
the soliton scattering rule determined \cite{HKOTY2}.
The dynamics allows a neat description in terms of particles and anti-particles 
that undergo pair creations and annihilations \cite{HKT01}.
The BBS turns out to be the special case in which no anti-particle is present.
Let us demonstrate the $D_4^{(1)}$ case.
Each local state takes values in 
$\{1,2,3,4,\bar{1},\bar{2},\bar{3},\bar{4}\}$,
where $\bar{2},\bar{3},\bar{4}$ are anti-particles of $2,3,4$, respectively.
As in the BBS, $1$ represents an empty box whereas $\bar{1}$
plays the role of 
particle \& anti-particle bound state.
The prototype time evolution $T_\infty$ is given by
\begin{equation}\label{k:tkkk}
T_\infty = K_2 K_3 K_4 K_{\bar 4} K_{\bar 3} K_{\bar 2},
\end{equation}
where each $K_a$ is defined by the following algorithm
(We understand $\bar{\bar{2}}=2$ etc.):
\begin{enumerate}
\item Replace each $\bar{1}$ by a pair $a, \bar{a}$ within a box.
\item Move the leftmost $a$ (if any) to the nearest right 
box which is empty or containing just ${\bar a}$.
(Boxes involving the pair $a, \bar{a}$ are prohibited as the destination.)

\item  Repeat (ii) until all of $a$'s are moved exactly once.
\item Replace the pair $a, \bar{a}$ within a box (if any) by $\bar{1}$.
\end{enumerate}
 
When anti-particles are absent, (i) and (iv) become void and 
the algorithm reduces to the one for BBS in Proposition \ref{t:jun15c}.
 
\begin{Example}\label{i:ex:d}
$D_4^{(1)}$-automaton. 
We write $., a,b,c, d$ for 
$1, \bar{1}, \bar{2}, \bar{3}, \bar{4}$, respectively.
\begin{verbatim}
  t=0  ..b22....3................     | t=2    .....b223...........
  t=1  .....b22..3...............     |        ......a23...........
  t=2  ........b223..............     |        ......32a...........
  t=3  ...........3c32...........     |        ......324d..........
  t=4  ............3..c32........     |        ......32.a..........
  t=5  .............3....c32.....     |        .......23c3.........
  t=6  ..............3......c32..     | t=3    ........3c32........
\end{verbatim}
A soliton is a consecutive array of the form
$\bar{2}^{\nu_{\bar 2}}\bar{3}^{\nu_{\bar 3}}\bar{4}^{\nu_{\bar 4}}
4^{\nu_4}3^{\nu_3}2^{\nu_2}$, where $\nu_i$'s are nonnegative integers 
such that $\nu_{\bar 4}\nu_4 = 0$.
{\rm [Left]:} Successive time evolutions under $T_\infty$, where
pair annihilation/creation b2 $\rightarrow$ c3 takes place 
in the scattering. 
{\rm [Right]:} The intermediate states between $t=2$ and $t=3$  
corresponding to (\ref{k:tkkk}), where 
the procedures (i)--(iv) can be checked.
\end{Example}

In general, it is expected that so-called Kirillov-Reshetikhin module  
has the crystal base \cite{O07} and 
one can use its crystal to construct the corresponding generalization of BBS.

\noindent
$\bullet$ Supersymmetric case.
The supersymmetric automaton given by the crystal 
for the super Lie algebra $A(m,n)$
was introduced in \cite{HI00}.
We have {\it fermionic} balls labeled by $m+2, \ldots, m+n+1$,
besides the empty boxes labeled by $1$ and 
the ({\it bosonic}) balls labeled by $2, \ldots, m+1$.
The time evolution rule is the same as that for the 
$\widehat{\mathfrak{sl}}_{n+m+1}$-automaton
in Proposition \ref{t:jun15c}, 
except for the step to move a ball with a fermionic label $a$. 
For a fermionic label $a$, we replace (iii) of Proposition \ref{t:jun15c} with 
\\[1mm]
(iii') Exchange the leftmost $a$ 
among the rest of the $a$'s
with its nearest right $1$ if this $a$ has not been overtaken by the
previously moved $a$. 
\\[1mm]
This rule denotes that each soliton can contain 
at most one fermionic ball of each label.
\begin{Example}
$A(1,1)$-automaton. 
The constraint in (iii') with $a=3$ is relevant 
in the steps from  $t=1$ to $t=2$ and from $t=2$ to $t=3$.
\begin{verbatim}
     t=0   ...322..3...............
     t=1   ......3223..............
     t=2   .........3322...........
     t=3   ..........3..322........
     t=4   ...........3....322.....
\end{verbatim}
\end{Example}


\section{Bethe ansatz approach}\label{sec:baa}
\subsection{Introduction}

The Kerov-Kirillov-Reshetikhin (KKR) bijection \cite{KKR,KR} is a
one to one correspondence 
\begin{equation}\label{k:rh}
\{\text{rigged configurations}\}
\underset{\phi}{\overset{\phi^{-1}}{\rightleftarrows}}
\{\text{highest paths}\}.
\end{equation}
It originates in Bethe's consideration on 
the completeness of the Bethe ansatz 
under the {\em string hypothesis} \cite{Be}.
We shall explain (\ref{k:rh}) 
after a brief exposition on the background along the 
simplest example from $\widehat{\mathfrak{sl}}_2$.

Consider the spin $\frac{1}{2}$ Heisenberg chain
with the Hamiltonian acting on $({\mathbb C}^2)^{\otimes L}$:
\begin{equation}\label{k:ha}
{\mathcal H} = \sum_{k=1}^L(\sigma^x_k\sigma^x_{k+1}
+\sigma^y_k\sigma^y_{k+1}
+\sigma^z_k\sigma^z_{k+1}-1).
\end{equation}
Here $\sigma^\alpha_k$ is a Pauli matrix acting on 
the $k$-th site and the periodic boundary condition 
$\sigma^\alpha_{L+1}=\sigma^\alpha_1$ is assumed.
The model possesses 
the (global) ${\mathfrak{sl}}_2$ symmetry in the sense that 
$\sigma^\alpha:=\sum_k\sigma^\alpha_k$ satisfies the
defining relations of ${\mathfrak{sl}}_2$, and $[\sigma^\alpha,{\mathcal H}]=0$.
Let ${\mathbb C}^2 = {\mathbb C}v_1 \oplus {\mathbb C} v_2$,
where $v_1$ and $v_2$ are regarded as spin up and down local states,
respectively.
As a consequence of the ${\mathfrak{sl}}_2$ symmetry,
the Hamiltonian ${\mathcal H}$ preserves the number of down 
(hence up as well) spins, so one may concentrate on
a subspace $W_r$ with $r$ down spins and $L-r$ up spins.
The diagonalization of ${\mathcal H}$ is done by the 
Bethe ansatz \cite{Be}.
It reduces the task to finding the solutions of  
the {\em Bethe equation} $(r\le L/2)$:
\begin{equation}\label{k:be1}
\left(\frac{u_j+{\rm i}}{u_j-{\rm i}}\right)^L =  
-\prod_{k=1}^r
\frac{u_j - u_k + 2{\rm i}}{u_j - u_k - 2{\rm i}}\;\;\quad
(j=1,\ldots, r).
\end{equation}
In term of the {\em Bethe roots} $\{u_1,\ldots, u_r\}$, 
one can construct the eigenvector  $|u_1, \ldots, u_r\rangle \in W_r$,
called {\em Bethe vector}, of 
${\mathcal H}$ whose eigenvalue is given by 
$\sum_{j=1}^r\frac{-8}{u_j^2+1}$.
It is known that the Heisenberg Hamiltonian is contained in the
commuting transfer matrices $\{T_m(z)\}_{m\ge 1}$ (\ref{k:rtm}) 
with $q=1$
as $T_1(z) = T_1(1)(1+\text{const}(z-1){\mathcal H}+\cdots)$
(cf. \cite[\S 10.14]{B}).
Thus the Bethe equation (\ref{k:be1})
is actually relevant to their joint spectrum and therefore to the 
``diagonalization" of the commuting time evolutions $T_m$ in the BBS
(although the latter corresponds to $q=0$ rather than $q=1$).
 
Back to (\ref{k:be1}), 
the variety of eigenvalues is provided by the variety of solutions to the 
Bethe equation.
Thus a basic question arises; how many solutions should there be 
for the completeness of the Bethe ansatz ?
The answer is $\binom{L}{r}-\binom{L}{r-1}$\footnote{
In this argument, independence of the 
associated Bethe vectors has not been taken into account, 
and all the Bethe roots are 
supposed to be finite.}.
The decrement from $\dim W_r = \binom{L}{r}$
is due to the fact that the Bethe vectors are by construction 
{\em highest weight} vectors annihilated by the 
${\mathfrak{sl}}_2$ raising operator.
Namely, it has the property 
$\sigma^+|u_1, \ldots, u_r\rangle=0$ with  
$\sigma^+=(\sigma^x+{\rm i}\sigma^y)/2$ by construction \cite{FT, Be}.
By virtue of the ${\mathfrak{sl}}_2$ symmetry,
the other eigenvectors can be produced by applying the 
lowering operator $\sigma^-=(\sigma^x-{\rm i}\sigma^y)/2$ successively.
Thus one should be content with capturing all the highest weight vectors 
as Bethe vectors.

Let us observe an example $L=6, r=3$.
There are certainly $\binom{6}{3}-\binom{6}{2}=5$ solutions
as given below.

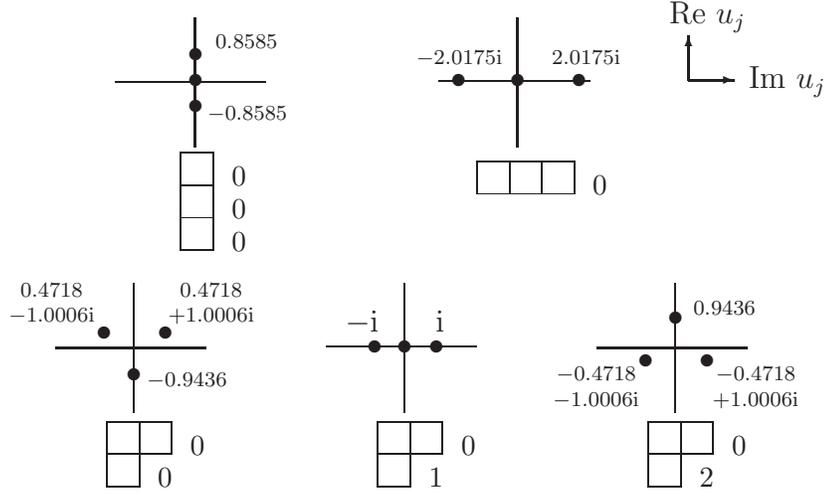
\begin{picture}(300,194)(-75,-108)
\setlength{\unitlength}{0.2mm}

\multiput(10,-8)(0,0){1}{
\multiput(30,95)(0,0){1}{
\put(-48,4){\line(1,0){100}}
\put(4.9,47){\line(0,-1){86}}
\put(0,17){$\bullet\; {0.8585\atop }$}\put(0,0){$\bullet$}
\put(0,-17){$\bullet$}\put(11,-14){${\atop -0.8585}$}
\put(-5,-80){\small $\yngrc(1,{0},1,{0},1,{0})$}}

\multiput(245,95)(0,0){1}{
\put(-48,4.9){\line(1,0){100}}
\put(4,47){\line(0,-1){86}}
\put(40,0){$\bullet$}\put(25,23){${\atop 2.0175{\rm i}}$}
\put(-0.9,0){$\bullet$}
\put(-40,0){$\bullet$}\put(-65,23){${\atop-2.0175{\rm i}}$}
\put(-23,-64){\small $\yngrc(3,{0})$}

\put(18,-25){
\put(100,30){\vector(0,1){30}}\put(87,68){Re $u_j$}
\put(100,30){\vector(1,0){30}}\put(140,25){Im $u_j$}
}

}

}

\multiput(0,-90)(0,0){1}{
\put(-48,4){\line(1,0){100}}
\put(4,47){\line(0,-1){86}}
\put(-20.9,9.4){$\bullet$}\put(-81,29){${0.4718 \atop -1.0006{\rm i}}$}
\put(20,9.4){$\bullet$}\put(25,29){${0.4718\atop +1.0006{\rm i}}$}
\put(-1,-19){$\bullet$}\put(11,-14){${\atop -0.9436}$}
\put(-14,-71){\small $\yngrc(2,{0},1,{0})$}
}

\multiput(180,-90)(0,0){1}{
\put(-48,4.9){\line(1,0){100}}
\put(4,47){\line(0,-1){86}}
\put(-21,0){$\bullet$}\put(-35,15){$-{\rm i}$}
\put(-1,0){$\bullet$}
\put(20,0){$\bullet$}\put(25,15){${\rm i}$}
\put(-14,-71){\small $\yngrc(2,{0},1,{1})$}
}

\multiput(360,-90)(0,0){1}{
\put(-48,4){\line(1,0){100}}
\put(4,47){\line(0,-1){86}}
\put(-20.8,-9.4){$\bullet$}
\put(-79,-27){${-0.4718\atop -1.0006{\rm i}}$}
\put(20,-9.4){$\bullet$}
\put(27,-27){${-0.4718\atop +1.0006{\rm i}}$}
\put(-0.9,19){$\bullet$}\put(14,17){${0.9436\atop }$}
\put(-14,-71){\small $\yngrc(2,{0},1,{2})$}
}

\end{picture}

Here each Bethe root $u_j$ is depicted as $\bullet$.
Within each solution, they are grouped into {\em strings}.
A string is an array of $\bullet$'s which is symmetric with 
respect to the real axis and equidistant of difference $2{\rm i}$
with possibly ``negligible" distortions.
Strings consisting of $k$ $\bullet$'s are called $k$-strings.
In the top left (right) solution, there are three $1$-strings 
(one $3$-string).
The three solutions in the bottom line consist of a $1$-string 
and a $2$-string with different real parts (called centers).
These features are conveniently symbolized in 
a Young diagram (called {\em configuration}) where each row is
attached with a nonnegative integer (called {\em rigging}) 
as shown in the figure.
They are examples of {\em rigged configurations}.
Each row including the rigging signifies the length and 
the center of the string encoded as an integer.
They are to obey a certain selection rule that will be specified later 
in a more general setting.
(See (\ref{k:rp}). The way to find the rigging will also be explained 
in section \ref{subsec:kkr}.)
To summarize so far, rigged configurations are 
combinatorial analogue of the pattern of Bethe roots under the string hypothesis.

Let us turn to the RHS of (\ref{k:rh}).
Bethe vectors have the form  
$|u_1, \ldots, u_r\rangle = 
\sum c_{i_1,\ldots, i_L}v_{i_1}\otimes \cdots \otimes v_{i_L} \in W_r$,
where the sum runs over $i_1,\ldots, i_L\in \{1,2\}$
such that $\#_1\{i_1,\ldots, i_L\}=L-r$ and 
$\#_2\{i_1,\ldots, i_L\}=r$.
Highest paths are combinatorial analogue of them 
represented as the sequence  $i_1,\ldots, i_L\in \{1,2\}^L$
satisfying the same condition as above and 
\begin{equation}\label{k:hp1}
\#_1\{i_1,\ldots, i_k\}\ge  \#_2\{i_1,\ldots, i_k\}
\;\;\text{for } 1 \le k \le L.
\end{equation} 
This is a remnant of the highest condition
$\sigma^+|u_1, \ldots, u_r\rangle = 0$.
There are $\binom{L}{r}-\binom{L}{r-1}$ highest paths 
as expected.
In our example $L=6, r=3$,
the highest paths and the corresponding 
rigged configurations in (\ref{k:rh}) are given as follows:
\begin{equation}\label{k:ist}
\begin{picture}(280,62)(0,28)
\put(30,60){
\multiput(0,0)(0,9){4}{
\put(0,0){\line(1,0){9}}}
\multiput(0,0)(9,0){2}{
\put(0,0){\line(0,1){27}}}
\put(12,18){0}\put(12,9){0}\put(12,0){0}
\put(25,9){$\longleftrightarrow$}
\put(48,9){121212}}

\put(150,69){
\multiput(0,-1)(0,9){2}{
\put(0,0){\line(1,0){27}}}
\multiput(0,-1)(9,0){4}{
\put(0,0){\line(0,1){9}}}
\put(30,0){0}
\put(42.6,0){$\longleftrightarrow$}
\put(66.2,0){111222}}

\put(-8,27){
\put(0,0){\line(0,1){18}}\put(9,0){\line(0,1){18}}
\put(18,9){\line(0,1){9}}
\put(0,18){\line(1,0){18}}
\put(0,9){\line(1,0){18}}
\put(0,0){\line(1,0){9}}
\put(21,9){0}
\put(12,0){0}
\put(32,5.5){$\longleftrightarrow$}
\put(55,5){121122}}

\put(95,27){
\put(0,0){\line(0,1){18}}\put(9,0){\line(0,1){18}}
\put(18,9){\line(0,1){9}}
\put(0,18){\line(1,0){18}}
\put(0,9){\line(1,0){18}}
\put(0,0){\line(1,0){9}}
\put(21,9){0}
\put(12,0){1}
\put(32,5.5){$\longleftrightarrow$}
\put(55,5){112122}}

\put(198,27){
\put(0,0){\line(0,1){18}}\put(9,0){\line(0,1){18}}
\put(18,9){\line(0,1){9}}
\put(0,18){\line(1,0){18}}
\put(0,9){\line(1,0){18}}
\put(0,0){\line(1,0){9}}
\put(21,9){0}
\put(12,0){2}
\put(32,5.5){$\longleftrightarrow$}
\put(55,5){112212.}}

\end{picture}
\end{equation}
This is an example of the KKR bijection.
The arrows $\rightarrow$ here, or equivalently 
the map $\phi^{-1}$ in (\ref{k:rh}),  is a combinatorial analogue 
of the Bethe ansatz which produces Bethe vectors from 
Bethe roots as 
$\{u_1, \ldots, u_r\} \mapsto |u_1,\ldots, u_r \rangle$.

The (vague) claim that any solution of the Bethe equation 
can be described as a collection of 
strings is called string hypothesis. 
It is known that string hypothesis is not always valid literally 
(see for example \cite{EKS}).
Nevertheless, as we shall illustrate below for $\widehat{\mathfrak{sl}}_{n+1}$ case, 
one can define rigged configurations and highest paths 
and establish their bijective correspondence mathematically.

\subsection{KKR bijection}\label{subsec:kkr}

Let $B_1$ be the $\widehat{\mathfrak{sl}}_{n+1}$ crystal (\ref{t:jun24a})
with $l=1$.
For simplicity we shall exclusively consider 
the crystal of the form $B_{1}^{\otimes L}$ 
($L \in \Z_{\ge 1}$) and call its elements as paths.
For a Young diagram $\lambda$ with 
$|\lambda | = L$ and depth at most $n+1$, elements of the set 
\begin{equation}\label{k:hp2}
{\mathcal P}_+(L,\lambda)
=\{p \in B_{1}^{\otimes L} \mid 
\tilde{e}_ip = 0\;\;(1 \le i \le n),\;
\mathrm{wt}\, p=\lambda\}
\end{equation}
are called {\em highest paths} with weight $\lambda$.
Setting  $p = \onebox{i_1}\otimes \cdots \otimes \onebox{i_L}$, 
the highest condition 
$\tilde{e}_ip = 0\;(1 \le i \le n)$ is concretely described as 
\begin{equation}\label{k:hp3}
\#_1\{i_1,\ldots, i_k\}\ge  \#_2\{i_1,\ldots, i_k\}
\ge \cdots \ge \#_{n+1}\{i_1,\ldots, i_k\}
\;\;\text{for } 1 \le k \le L,
\end{equation} 
which is a generalization of (\ref{k:hp1}).
By the condition $\mathrm{wt}\, p=\lambda$ we mean
\begin{equation}\label{k:wt}
\#_a\{i_1,\ldots, i_L\} = \lambda_a\qquad(1 \le a \le n+1).
\end{equation}

Let us proceed to the definition of the rigged configurations.
Let $\mu^{(0)}, \mu^{(1)},\ldots, \mu^{(n)}$ 
be an $(n+1)$-tuple of Young diagrams.
We will always take $\mu^{(0)}=(1^L)$ ($L \in \Z_{\ge 1}$)
which will match the choice of the crystal $B_1^{\otimes L}$ in (\ref{k:hp2}).
Denote by $m^{(a)}_j$ 
the number of length $j$ rows in $\mu^{(a)}$ and introduce the following:
\begin{align}
p^{(a)}_j &= q^{(a-1)}_j - 2q^{(a)}_j + q^{(a+1)}_j
\quad (1 \le a \le n),\label{k:paj}\\
q^{(a)}_j &= \sum_{k\ge 1}
\min(j,k)m^{(a)}_k \quad (q^{(n+1)}_j = 0).\label{k:qaj}
\end{align}
By the definition $m^{(0)}_j=L\delta_{j,1}$ and 
$q^{(0)}_j = L$ for $j \ge 1$.
In general $q^{(a)}_j$ is the number of cells 
in the left $j$ columns of $\mu^{(a)}$.
The integer $p^{(a)}_j$ is called a {\em vacancy} and will play an 
important role in what follows.

An $(n+1)$-tuple of Young diagrams 
$(\mu^{(0)},\ldots, \mu^{(n)})$ is a {\em configuration}
if $p^{(a)}_j \ge 0$ for any $1 \le a \le n$ and $j\in \Z_{\ge 1}$
such that $m^{(a)}_j \ge 1$\footnote{This condition is known actually 
to ensure that 
$p^{(a)}_j  \ge 0$ for {\em all} $j \in \Z_{\ge 1}$.}.
Such a pair $(a,j)$ (i.e. the $m^{(a)}_j \times j$ rectangle 
constituting $\mu^{(a)}$) will be referred to as a {\em block}. 

Given a configuration $(\mu^{(0)},\ldots, \mu^{(n)})$,
we attach a {\em rigging}  $J^{(a)}_{j,\alpha}\in \Z_{\ge 0}$ 
to every row in 
$\mu^{(a)}$ except $\mu^{(0)}=(1^L)$
as follows (shown for a block $(a,j)$).

\begin{equation}\label{k:zu}
\unitlength 0.1in
\begin{picture}( 18.0000, 18)(8,-22.0000)
%
\special{pn 8}%
\special{pa 1400 600}%
\special{pa 2400 600}%
\special{fp}%
\special{pa 2400 800}%
\special{pa 2200 800}%
\special{fp}%
\special{pa 2400 600}%
\special{pa 2400 800}%
\special{fp}%
\special{pa 2200 800}%
\special{pa 2200 1600}%
\special{fp}%
\special{pa 1800 1600}%
\special{pa 1800 2000}%
\special{fp}%
\special{pa 1600 2000}%
\special{pa 1600 2200}%
\special{fp}%
\special{pa 1400 2200}%
\special{pa 1400 600}%
\special{fp}%
\special{pa 1400 2200}%
\special{pa 1600 2200}%
\special{fp}%
\special{pa 1600 2000}%
\special{pa 1800 2000}%
\special{fp}%
\special{pa 1800 1600}%
\special{pa 2200 1600}%
\special{fp}%
\put(23.1000,-13.1000){\makebox(0,0)[lb]{$\vdots$}}%
\put(22.4000,-11.4000){\makebox(0,0)[lb]{$J^{(a)}_{j,m^{(a)}_j}$}}%
\put(22.4000,-15.9000){\makebox(0,0)[lb]{$J^{(a)}_{j,1}$}}%
\put(19.1000,-4.1000){\makebox(0,0){$(\mu^{(a)}, J^{(a)})$}}%
\put(17.8000,-12.0000){\makebox(0,0){$j$}}%
\put(10.9500,-12.1000){\makebox(0,0){$m^{(a)}_j$}}%
%
\special{pn 8}%
\special{pa 1046 1360}%
\special{pa 1046 1602}%
\special{fp}%
\special{sh 1}%
\special{pa 1046 1602}%
\special{pa 1066 1536}%
\special{pa 1046 1550}%
\special{pa 1026 1536}%
\special{pa 1046 1602}%
\special{fp}%
%
\special{pn 8}%
\special{pa 1046 1080}%
\special{pa 1046 800}%
\special{fp}%
\special{sh 1}%
\special{pa 1046 800}%
\special{pa 1026 868}%
\special{pa 1046 854}%
\special{pa 1066 868}%
\special{pa 1046 800}%
\special{fp}%
%
\special{pn 8}%
\special{pa 1870 1200}%
\special{pa 2176 1200}%
\special{fp}%
\special{sh 1}%
\special{pa 2176 1200}%
\special{pa 2110 1180}%
\special{pa 2124 1200}%
\special{pa 2110 1220}%
\special{pa 2176 1200}%
\special{fp}%
%
\special{pn 8}%
\special{pa 1706 1200}%
\special{pa 1400 1200}%
\special{fp}%
\special{sh 1}%
\special{pa 1400 1200}%
\special{pa 1468 1220}%
\special{pa 1454 1200}%
\special{pa 1468 1180}%
\special{pa 1400 1200}%
\special{fp}%
\end{picture}%
\end{equation}

We group the rigging as
$J=(J^{(1)}, \ldots, J^{(n)})$
where $J^{(a)} = (J^{(a)}_{j,\alpha})_{j\ge 1, 1 \le \alpha \le m^{(a)}_j}$
is the one attached to $\mu^{(a)}$.
A configuration 
$((1^L),\mu^{(1)}, \ldots, \mu^{(n)})$ attached with 
a rigging $J=(J^{(1)}, \ldots, J^{(n)})$
will be denoted by 
$(\mu, J)_L$
with  $\mu=(\mu^{(1)}, \ldots, \mu^{(n)})$.
We say 
$(\mu, J)_L$ is a {\em rigged configuration} if the condition 
\begin{equation}\label{k:rp}
0 \le J^{(a)}_{j,1} \le J^{(a)}_{j,2} \le \cdots \le J^{(a)}_{j,m^{(a)}_j}
\le p^{(a)}_j
\end{equation}
is satisfied for all the blocks $(a,j)$\footnote{Equivalently, 
one may only 
impose $0 \le J^{(a)}_{j,1}, \ldots, J^{(a)}_{j,m^{(a)}_j} \le p^{(a)}_j$ 
and identify their permutations.}.

\begin{Example}\label{k:ex1}
We list all the rigged configurations 
having the configuration $((1^8), \mu^{(1)})$ 
with $\mu^{(1)}=(2,1,1)$.
To save the space, only $(\mu^{(1)}, J^{(1)})$ part is given.
\begin{equation}\label{k:rc1}
{\small
\begin{picture}(100,40)(90,0)
\setlength{\unitlength}{0.35mm}
\multiput(0,0)(50,0){6}{
\put(0,0){\line(1,0){10}}
\put(0,10){\line(1,0){10}}
\put(0,20){\line(1,0){20}}
\put(0,30){\line(1,0){20}}
\put(0,0){\line(0,1){30}}
\put(10,0){\line(0,1){30}}
\put(20,20){\line(0,1){10}}
\put(-8,7.4){\rm 2} \put(-8,22.8){\rm 0}
\put(22,23){\rm 0}}

\put(12,0){\rm 0}\put(12,10){\rm 0} 
\put(62,0){\rm 0}\put(62,10){\rm 1} 
\put(112,0){\rm 0}\put(112,10){\rm 2}
\put(162,0){\rm 1}\put(162,10){\rm 1}
\put(212,0){\rm 1}\put(212,10){\rm 2}
\put(262,0){\rm 2}\put(262,10){\rm 2}
\end{picture}}
\end{equation}
For later convenience, we have exhibited the vacancy 
$p^{(1)}_2=0$ and $p^{(1)}_1=2$ to the left of the 
relevant blocks.
\end{Example}

\begin{Example}\label{k:ex2}
An $n=3$ example. 
Again, vacancies, e.g. $p^{(1)}_2=5$, are exhibited.
\begin{equation}
\unitlength 10pt
\begin{picture}(19,5)(1,10.4)

\put(-1.4,14.5){$\mu^{(0)}$}
\put(-1.8,13){$(1^{14})$}

\multiput(4,11)(1,0){3}{\line(0,1){3}}
\put(4,11){\line(1,0){2}}
\put(4,12){\line(1,0){3}}
\put(4,13){\line(1,0){4}}
\put(4,14){\line(1,0){4}}
\put(7,12){\line(0,1){2}}
\put(8,13){\line(0,1){1}}
\put(5.5,14.5){$\mu^{(1)}$}
\put(3.2,13.15){\rm 0}
\put(3.2,12.15){\rm 2}
\put(3.2,11.15){\rm 5}
\put(8.3,13.15){\rm 0}
\put(7.3,12){\rm 2}
\put(6.3,11){\rm 3}
\multiput(11,12)(1,0){2}{\line(0,1){2}}
\put(11,12){\line(1,0){1}}
\multiput(11,13)(0,1){2}{\line(1,0){3}}
\multiput(13,13)(1,0){2}{\line(0,1){1}}
\put(12,14.5){$\mu^{(2)}$}
\put(10.2,13.15){\rm 1}
\put(10.2,12.15){\rm 0}
\put(14.3,13){\rm 1}
\put(12.3,12){\rm 0}
\multiput(17,13)(1,0){2}{\line(0,1){1}}
\multiput(17,13)(0,1){2}{\line(1,0){1}}
\put(17,14.5){$\mu^{(3)}$}
\put(16.2,13.15){\rm 0}
\put(18.3,13.15){\rm 0}
\end{picture}
\end{equation}
\end{Example}
 
A weight of a rigged configuration $(\mu,J)_L$ is the 
Young diagram $\lambda=(\lambda_1,\ldots, \lambda_{n+1})$
specified by
\begin{equation}\label{k:wt3}
|\mu^{(a)}| = \lambda_{a+1}+\lambda_{a+2}+\cdots + \lambda_{n+1}
\quad (0 \le a \le n).
\end{equation} 
We write it as 
$\mathrm{wt} \bigl((\mu,J)_L\bigr) = \lambda$,
which is actually dependent only on the configuration.
The inequality $\lambda_1\ge \cdots \ge \lambda_{n+1}\ge 0$ is 
guaranteed by the condition 
$p^{(a)}_\infty(= |\mu^{(a-1)}| -2|\mu^{(a)}| +
|\mu^{(a+1)}| ) \ge 0$ for $1 \le a \le n$.
Note also that $|\lambda | = L$.
Let $\mathrm{RC}(L,\lambda)$ be the set of 
rigged configurations of weight $\lambda$.
\begin{Theorem}
For any $L \in \Z_{\ge 1}$ and a Young 
diagram $\lambda$ with $|\lambda | = L$, there is a bijection 
\begin{equation}\label{k:rcp}
\mathrm{RC}(L,\lambda)
\underset{\phi}{\overset{\phi^{-1}}{\rightleftarrows}}
{\mathcal P}_+(L,\lambda).
\end{equation}
\end{Theorem}
The original KKR bijection \cite{KKR, KR} is the one
between rigged configurations and Littlewood-Richardson tableaux.
Its ultimate generalization for type $\widehat{\mathfrak{sl}}_{n+1}$
is available in \cite{KSS,S07}.
In the simple setting of this paper,
the Littlewood-Richardson tableaux are
in one-to-one correspondence with highest paths 
via the Robinson-Schensted correspondence \cite{Ful}.

We regard a rigged configuration 
$(\mu, J)_L$ as a multiset of {\em strings}.
A string corresponds to a row in (\ref{k:zu}).
It is a triple $\big(a, j, J^{(a)}_{j,\alpha}\big)$ consisting of
color $a$, length $j$ and rigging $J^{(a)}_{j,\alpha}$.
A string is {\em singular} if $J^{(a)}_{j,\alpha}= p^{(a)}_j$,
namely if the rigging attains the allowed maximum in (\ref{k:rp}).
We regard the highest path 
$p = \onebox{i_1}\otimes \cdots \otimes \onebox{i_L}$
as a word $i_1i_2\ldots i_L \in \{1,\ldots, n+1\}^L$.
(The Littlewood-Richardson tableau mentioned in the above 
is the $Q$-symbol \cite{Ful} of this word.)

For simplicity, we first explain the algorithm for 
$\phi^{\pm 1}$ for $n=1$ case.
Even in this case, it may look formidably complicated 
at first glance.
However, it is a very well-designed algorithm, and  
the readers will be impressed and get familiarized with it pretty well 
by working out a few examples.
The $m^{(1)}_j$ and the vacancy $p^{(1)}_j$ 
will be denoted by $m_j$ and $p_j$.
Thus the definition (\ref{k:paj}) becomes
$p_j = L -2\sum_k\min(j,k)m_k$.
It is useful to remember it as $p_j = L-
2(\text{number of cells in the left $j$ columns in the Young diagram})$.

\smallskip\noindent
{\mathversion{bold} {\bf Algorithm of $\phi$ for $n=1$}}.

Given a highest path $i_1\ldots i_L\in \{1,2\}^L$,
we construct the
rigged configuration $\phi(i_1\ldots i_L) = (\mu, J)_L$
inductively with respect to $L$.
When $L=0$, we understand that $\phi(\cdot)$
is an empty Young diagram.
Suppose that
$\phi(i_1\ldots i_L) = (\mu, J)_L$ has been obtained.
We are to construct
$(\mu', J')_{L+1}=\phi(i_1\ldots i_L i_{L+1})$ from
$(\mu, J)_L$ and $i_{L+1}\in \{1,2\}$.

Case $i_{L+1}=1$. One has
$(\mu', J')_{L+1}=(\mu, J)_{L+1}$, which means that
no change should be made in the length and rigging of the strings.
(By the definition, their vacancies $p_j$ increase uniformly by one.)

Case $i_{L+1}=2$.
(a) If there is no singular string in $(\mu, J)_L$, just
additionally create a length 1 
singular string with respect to the new configuration.
(Its rigging is therefore $L+1-2\sum_k\min(1,k)(m_k+\delta_{k1})$.)
(b) If there exist singular strings,
pick a longest one among them 
and let $\ell$ be its length. (Any choice is OK
when there are more than one such strings.)
Then $(\mu', J')_{L+1}$ is obtained by 
extending the string to length $\ell+1$ and making it  
singular with respect to the new configuration.
(Its rigging is therefore $L+1-2\sum_k\min(\ell+1,k)
(m_k-\delta_{k,\ell}+\delta_{k,\ell+1})$.)
In either case of (a) and (b), 
keep the other strings unchanged.

\smallskip\noindent
{\mathversion{bold} {\bf Algorithm of $\phi^{-1}$ for $n=1$}}.

Given a rigged configuration $(\mu, J)_L$,
we construct a highest path
$i_1\ldots i_L = \phi^{-1}((\mu, J)_L)$
inductively with respect to $L$.
We are to determine $i_L\in \{1,2\}$
and $(\mu', J')_{L-1}$ such that
$\phi^{-1}((\mu, J)_L) =
\phi^{-1}((\mu', J')_{L-1}) \,i_L$.

If $(\mu,J)_L$ contains no singular string, 
then $i_L = 1$ and 
$(\mu', J')_{L-1}=(\mu,J)_{L-1}$.
The latter means no change should be made in any string. 
(By the definition, their vacancies $p_j$ decrease uniformly by one.)
If $(\mu,J)_L$ contains singular strings, then $i_L=2$.
Pick a shortest singular string and let $\ell$ be its length.
(Any choice is OK when there are more than one such strings.)
Then $(\mu', J')_{L-1}$ is obtained by 
shortening the string to length $\ell-1$ and making it singular 
with respect to the new configuration.
(Its rigging is therefore 
$L-1-2\sum_k\min(\ell-1,k)(m_k+\delta_{k,\ell-1}-\delta_{k,\ell})$.)
The other strings are kept unchanged.

\begin{Example}
For the rigged configurations in Example \ref{k:ex1},
the algorithm of $\phi^{-1}$ proceeds along the arrows.
The algorithm of $\phi$ proceeds backward.
To save the space, $L$ is given in the first line. 
\begin{equation*}
\setlength{\unitlength}{0.26mm}
\begin{picture}(450,310)(50,-5)
\put(503,297){$\mathrm{Im}\phi^{-1}$}
\put(490,272){\rm 12121122,}

\put(490,222){\rm 12112122,}

\put(490,172){\rm 12112212,}

\put(490,122){\rm 11212122,}

\put(490,72){\rm 11212212,}

\put(490,22){\rm 11221212.}

\multiput(0,0)(0,50){6}{
\put(0,0){\line(1,0){10}}
\put(0,10){\line(1,0){10}}
\put(0,20){\line(1,0){20}}
\put(0,30){\line(1,0){20}}
\put(0,0){\line(0,1){30}}
\put(10,0){\line(0,1){30}}
\put(20,20){\line(0,1){10}}
\put(-9,7){\small {\rm 2}} \put(-9,21){\small {\rm 0}}
\put(22,22){\small {\rm 0}}

\multiput(35,16)(65,0){5}{$\rightarrow$}}
\multiput(0,0)(0,50){6}{
\multiput(350,16)(55,0){2}{$\rightarrow$}}

\multiput(442,16)(0,50){6}{$\rightarrow$}

\multiput(0,0)(65,0){2}{
\multiput(0,0)(0,50){2}{
\put(65,10){\line(1,0){10}}
\put(65,20){\line(1,0){20}}
\put(65,30){\line(1,0){20}}
\put(65,10){\line(0,1){20}}
\put(75,10){\line(0,1){20}}
\put(85,20){\line(0,1){10}}}
\multiput(0,0)(0,150){2}{
\put(65,10){\line(1,0){10}}
\put(65,20){\line(1,0){20}}
\put(65,30){\line(1,0){20}}
\put(65,10){\line(0,1){20}}
\put(75,10){\line(0,1){20}}
\put(85,20){\line(0,1){10}}}}

\put(65,100){\line(1,0){10}}
\put(65,110){\line(1,0){10}}
\put(65,120){\line(1,0){10}}
\put(65,130){\line(1,0){10}}
\put(65,100){\line(0,1){30}}
\put(75,100){\line(0,1){30}}
\multiput(0,100)(0,50){2}{
\put(65,100){\line(1,0){10}}
\put(65,110){\line(1,0){10}}
\put(65,120){\line(1,0){10}}
\put(65,130){\line(1,0){10}}
\put(65,100){\line(0,1){30}}
\put(75,100){\line(0,1){30}}}

\multiput(0,0)(0,50){4}{
\put(130,110){\line(1,0){10}}
\put(130,120){\line(1,0){10}}
\put(130,130){\line(1,0){10}}
\put(130,110){\line(0,1){20}}
\put(140,110){\line(0,1){20}}}
\multiput(65,-50)(0,50){5}{
\put(130,110){\line(1,0){10}}
\put(130,120){\line(1,0){10}}
\put(130,130){\line(1,0){10}}
\put(130,110){\line(0,1){20}}
\put(140,110){\line(0,1){20}}}
\multiput(130,150)(0,00){1}{
\put(130,110){\line(1,0){10}}
\put(130,120){\line(1,0){10}}
\put(130,130){\line(1,0){10}}
\put(130,110){\line(0,1){20}}
\put(140,110){\line(0,1){20}}}

\multiput(0,0)(65,0){2}{
\put(195,20){\line(1,0){20}}
\put(195,30){\line(1,0){20}}
\put(195,20){\line(0,1){10}}
\put(205,20){\line(0,1){10}}
\put(215,20){\line(0,1){10}}}

\multiput(0,0)(0,50){4}{
\put(260,70){\line(1,0){10}}
\put(260,80){\line(1,0){10}}
\put(260,70){\line(0,1){10}}
\put(270,70){\line(0,1){10}}}
\multiput(65,-50)(0,50){6}{
\put(260,70){\line(1,0){10}}
\put(260,80){\line(1,0){10}}
\put(260,70){\line(0,1){10}}
\put(270,70){\line(0,1){10}}}
\multiput(122,100)(0,50){3}{
\put(260,70){\line(1,0){10}}
\put(260,80){\line(1,0){10}}
\put(260,70){\line(0,1){10}}
\put(270,70){\line(0,1){10}}}

\multiput(0,0)(0,50){3}{\put(385,21){\small $\emptyset$}}
\multiput(-8,0)(40,0){2}{
\multiput(50,0)(0,50){6}{\put(385,21){\small $\emptyset$}}}

\put(7,298){\small {\rm 8}}
\put(67,298){\small {\rm 7}}
\put(132,298){\small {\rm 6}}
\put(197,298){\small {\rm 5}}
\put(260,298){\small {\rm 4}}
\put(325,298){\small {\rm 3}}
\put(386,298){\small {\rm 2}}
\put(436,298){\small {\rm 1}}
\put(467,298){\small {\rm 0}}

\put(12,9){\small {\rm 2}}\put(12,-2){\small {\rm 2}}

\put(12,59){\small {\rm 2}}\put(12,48){\small {\rm 1}}

\put(12,109){\small {\rm 1}}\put(12,98){\small {\rm 1}}

\put(12,159){\small {\rm 2}}\put(12,148){\small {\rm 0}}

\put(12,209){\small {\rm 1}}\put(12,198){\small {\rm 0}}

\put(12,259){\small {\rm 0}}\put(12,248){\small {\rm 0}}

\multiput(0,0)(0,50){6}{\put(38,24){$\scriptstyle{2}$}}

\put(57,22){\small {\rm 1}}\put(57,10){\small {\rm 3}}
\put(87,22){\small {\rm 0}}\put(77,9){\small {\rm 2}}

\put(57,72){\small {\rm 1}}\put(57,60){\small {\rm 3}}
\put(87,72){\small {\rm 0}}\put(77,59){\small {\rm 1}}

\put(57,112){\small {\rm 1}}
\put(77,121){\small {\rm 1}}\put(77,110){\small {\rm 1}}
\put(77,99){\small {\rm 1}}

\put(57,172){\small {\rm 1}}\put(57,160){\small {\rm 3}}
\put(87,172){\small {\rm 0}}\put(77,158){\small {\rm 0}}

\put(57,212){\small {\rm 1}}
\put(77,221){\small {\rm 1}}\put(77,210){\small {\rm 1}}
\put(77,198.5){\small {\rm 0}}

\put(57,262){\small {\rm 1}}
\put(77,271){\small {\rm 1}}\put(77,260){\small {\rm 0}}
\put(77,248.5){\small {\rm 0}}

\put(103,23){$\scriptstyle{1}$}
\put(103,73){$\scriptstyle{1}$}
\put(103,123){$\scriptstyle{2}$}
\put(103,173){$\scriptstyle{1}$}
\put(103,223){$\scriptstyle{2}$}
\put(103,273){$\scriptstyle{2}$}

\put(121,22){\small {\rm 0}}\put(121,10){\small {\rm 2}}
\put(152,20){\small {\rm 0}}\put(142,9){\small {\rm 2}}

\put(121,72){\small {\rm 0}}\put(121,60){\small {\rm 2}}
\put(152,72){\small {\rm 0}}\put(142,59){\small {\rm 1}}

\put(121,117){\small {\rm 2}}
\put(142,120.5){\small {\rm 1}}\put(142,109){\small {\rm 1}}

\put(121,172){\small {\rm 0}}\put(121,160){\small {\rm 2}}
\put(152,172){\small {\rm 0}}\put(142,159){\small {\rm 0}}

\put(121,217){\small {\rm 2}}
\put(142,220){\small {\rm 1}}\put(142,209){\small {\rm 0}}

\put(121,267){\small {\rm 2}}
\put(142,272){\small {\rm 0}}\put(142,260){\small {\rm 0}}

\put(168,23){$\scriptstyle{2}$}
\put(168,73){$\scriptstyle{2}$}
\put(168,123){$\scriptstyle{1}$}
\put(168,173){$\scriptstyle{2}$}
\put(168,223){$\scriptstyle{1}$}
\put(168,273){$\scriptstyle{1}$}

\put(187,20){\small {\rm 1}}
\put(218,20){\small {\rm 0}}

\put(187,67){\small {\rm 1}}
\put(207,70){\small {\rm 1}}\put(207,59){\small {\rm 1}}

\put(187,117){\small {\rm 1}}
\put(207,120){\small {\rm 1}}\put(207,109){\small {\rm 1}}

\put(187,167){\small {\rm 1}}
\put(207,170){\small {\rm 1}}\put(207,159){\small {\rm 0}}

\put(187,217){\small {\rm 1}}
\put(207,220){\small {\rm 1}}\put(207,209){\small {\rm 0}}

\put(187,267){\small {\rm 1}}
\put(207,270){\small {\rm 0}}\put(207,259){\small {\rm 0}}

\put(233,23){$\scriptstyle{1}$}
\put(233,73){$\scriptstyle{2}$}
\put(233,123){$\scriptstyle{2}$}
\put(233,173){$\scriptstyle{2}$}
\put(233,223){$\scriptstyle{2}$}
\put(233,273){$\scriptstyle{1}$}

\put(252,20){\small {\rm 0}}
\put(282,20){\small {\rm 0}}

\put(252,70){\small {\rm 2}}
\put(272,70){\small {\rm 1}}

\put(252,120){\small {\rm 2}}
\put(272,120){\small {\rm 1}}

\put(252,170){\small {\rm 2}}
\put(272,170){\small {\rm 0}}

\put(252,220){\small {\rm 2}}
\put(272,220){\small {\rm 0}}

\put(252,267){\small {\rm 0}}
\put(272,270.5){\small {\rm 0}}
\put(272,259.4){\small {\rm 0}}

\put(298,23){$\scriptstyle{2}$}
\put(298,73){$\scriptstyle{1}$}
\put(298,123){$\scriptstyle{1}$}
\put(298,173){$\scriptstyle{1}$}
\put(298,223){$\scriptstyle{1}$}
\put(298,273){$\scriptstyle{2}$}

\put(317,20){\small {\rm 1}}
\put(337,20){\small {\rm 1}}

\put(317,70){\small {\rm 1}}
\put(337,70){\small {\rm 1}}

\put(317,120){\small {\rm 1}}
\put(337,120){\small {\rm 1}}

\put(317,170){\small {\rm 1}}
\put(337,170){\small {\rm 0}}

\put(317,220){\small {\rm 1}}
\put(337,220){\small {\rm 0}}

\put(317,270){\small {\rm 1}}
\put(337,270){\small {\rm 0}}

\put(353,23){$\scriptstyle{2}$}
\put(353,73){$\scriptstyle{2}$}
\put(353,123){$\scriptstyle{2}$}
\put(353,173){$\scriptstyle{1}$}
\put(353,223){$\scriptstyle{1}$}
\put(353,273){$\scriptstyle{1}$}

\put(-3,0){
\put(376,170){\small {\rm 0}}
\put(397,170){\small {\rm 0}}

\put(376,220){\small {\rm 0}}
\put(397,220){\small {\rm 0}}

\put(376,270){\small {\rm 0}}
\put(397,270){\small {\rm 0}}

\put(1.5,0){
\put(409,23){$\scriptstyle{1}$}
\put(409,73){$\scriptstyle{1}$}
\put(409,123){$\scriptstyle{1}$}
\put(409,173){$\scriptstyle{2}$}
\put(409,223){$\scriptstyle{2}$}
\put(409,273){$\scriptstyle{2}$}}
}

\multiput(446,23)(0,50){6}{\put(0,0){$\scriptstyle{1}$}}
\end{picture}
\end{equation*}
Note that one should keep updating the vacancies with $L$.
\end{Example}

Now we proceed to the $n$ general case.
The basic idea is to apply the removal/addition procedure for the $n=1$ case 
recursively in the direction of color.
\smallskip\noindent
{\mathversion{bold} {\bf Algorithm of $\phi$ for general $n$}}.

Given a highest path $i_1\ldots i_L$,
we construct the
rigged configuration $\phi(i_1\ldots i_L) = (\mu, J)_L$
inductively with respect to $L$.
When $L=0$, we understand that $\phi(\cdot)$
is the array of empty Young diagrams.
Suppose that
$\phi(i_1\ldots i_L) = (\mu, J)_L$ has been obtained.
Denote $i_{L+1} \in \{1,\ldots, n+1\}$ simply by $d$.
We are to construct
$(\mu', J')_{L+1}=\phi(i_1\ldots i_L d)$ from
$(\mu, J)_L$ and $d$.
If $d=1$, then
$(\mu', J')_{L+1}=(\mu, J)_{L+1}$, which means that
no change should be made on any string.
(By the definition (\ref{k:paj}), the vacancies $p^{(a)}_j$ 
increase by $\delta_{a 1}$.)
Suppose $d \ge 2$.
\begin{enumerate}\itemsep=0pt

\item[$(i)$]
Set $\ell^{(d)}=\infty$.
For $c=d-1, d-2, \ldots, 1$ in this order, proceed as follows.
Find the color $c$ singular string whose length $\ell^{(c)}$
is largest within the condition $\ell^{(c)}\le \ell^{(c+1)}$.
If there are more than one such strings, pick any one of them.
If there is no such string with color~$c$, set $\ell^{(c)} = 0$.
Denote these selected strings by
$\big(c, \ell^{(c)}, J^{(c)}_\ast\big)$ with $c=d-1,d-2,\ldots, 1$,
where it is actually void when $\ell^{(c)}=0$.

\item[$(ii)$]
 Replace the selected string
$\big(c, \ell^{(c)}, J^{(c)}_\ast\big)$ by
$\big(c, \ell^{(c)}+1, J^{(c)}_\bullet\big)$
for all $c=d-1,d-2,\ldots, 1$ leaving the other strings unchanged.
Here the new rigging $J^{(c)}_\bullet$ is to be chosen so that
the extended string $\big(c, \ell^{(c)}+1, J^{(c)}_\bullet\big)$ becomes
singular with respect to
the resulting new rigged configuration $(\mu', J')_{L+1}$.
\end{enumerate}

The algorithm is known to be well-defined and
the resulting object gives the sought rigged configuration
$(\mu', J')_{L+1} =\phi(i_1\ldots i_L d)$.

\smallskip\noindent
{\mathversion{bold} {\bf Algorithm of $\phi^{-1}$ for general $n$}}.

Given a rigged configuration $(\mu, J)_L$,
we construct a highest path
$i_1\ldots i_L = \phi^{-1}((\mu, J)_L)$
inductively with respect to $L$.
We are to determine $d (=i_L) \in \{1,\ldots, n+1\}$
and $(\mu', J')_{L-1}$ such that
$\phi^{-1}((\mu, J)_L) =
\phi^{-1}((\mu', J')_{L-1}) \,d$.
\begin{enumerate}\itemsep=0pt

\item[$(i)$]
Set $\ell^{(0)}=1$.
For $c=1,2, \ldots, n$ in this order, proceed as follows until stopped.
Find the color $c$ singular string whose length $\ell^{(c)}$
is smallest within the condition $\ell^{(c-1)}\le \ell^{(c)}$.
If there are more than one such strings, pick any one of them.
If there is no such string with color $c$,  set $d=c$ and stop.
If $c=n$ and such a color $n$ string still exists, set $d=n+1$
and stop.
Denote these selected strings by
$\big(c, \ell^{(c)}, J^{(c)}_\ast\big)$ with $c=1,2, \ldots, d-1$.

\item[$(ii)$]
Replace the selected string
$\big(c, \ell^{(c)}, J^{(c)}_\ast\big)$ by
$\big(c, \ell^{(c)}-1, J^{(c)}_\bullet\big)$
for all $c=1,2, \ldots, d-1$ leaving the other strings unchanged.
When $\ell^{(c)}=1$, this means that the length one string
is to be eliminated.
The new rigging $J^{(c)}_\bullet$ is to be chosen so that
the shortened string $\big(c, \ell^{(c)}-1, J^{(c)}_\bullet\big)$ becomes
singular in the new data $(\mu', J')_{L-1}$.
\end{enumerate}
For an empty rigged configuration, we understand that
$\phi^{-1}((\varnothing, \varnothing)_L) =
\phi^{-1}((\varnothing, \varnothing)_{L-1}) \,1
= \cdots = \overbrace{11\ldots 1}^L$.
The algorithm is known to be well-defined and ends up with
the empty rigged configuration at $L=0$.
The resulting sequence gives the sought highest path
$i_1\ldots i_L = \phi^{-1}((\mu, J)_{L})$.

\begin{Example}
The algorithm of $\phi^{-1}$ 
for the rigged configurations in Example \ref{k:ex2}.
For convenience the vacancy $p^{(a)}_j$ is shown to the left of 
each block $(a,j)$.
The rightmost cell in the singular strings to be shorten are indicated by $\times$.

{\small
\begin{center}
\unitlength 10pt
\begin{picture}(18,3)
\put(-4,1.15){$\overset{3}{\longrightarrow}$}
\put(0,2.25){$(1^{13})$}
\multiput(4,0)(1,0){3}{\line(0,1){3}}
\multiput(4,0)(0,1){2}{\line(1,0){2}}
\multiput(4,2)(0,1){2}{\line(1,0){4}}
\multiput(7,2)(1,0){2}{\line(0,1){1}}
\put(3.2,2.15){\rm 0}
\put(3.2,1.15){\rm 4}
\put(3.2,0.15){\rm 4}
\put(8.3,2.15){\rm 0}
\put(6.3,1.15){\rm 4}
\put(6.3,0.15){\rm 3}
\put(5.1,1.25){$\times$}
\multiput(11,1)(1,0){2}{\line(0,1){2}}
\put(11,1){\line(1,0){1}}
\multiput(11,2)(0,1){2}{\line(1,0){2}}
\put(13,2){\line(0,1){1}}
\put(10.2,2.15){\rm 1}
\put(10.2,1.15){\rm 0}
\put(12.3,1.15){\rm 0}
\put(13.3,2.15){\rm 1}
\put(12.1,2.25){$\times$}
\multiput(16,2)(1,0){2}{\line(0,1){1}}
\multiput(16,2)(0,1){2}{\line(1,0){1}}
\put(15.2,2.15){\rm 0}
\put(17.3,2.15){\rm 0}
\end{picture}
\end{center}
\begin{center}
\unitlength 10pt
\begin{picture}(18,3)
\put(-4,1.15){$\overset{3}{\longrightarrow}$}
\put(0,2.25){$(1^{12})$}
\multiput(4,0)(1,0){2}{\line(0,1){3}}
\multiput(4,0)(0,1){2}{\line(1,0){1}}
\put(6,1){\line(0,1){1}}\put(5,1){\line(1,0){1}}
\multiput(4,2)(0,1){2}{\line(1,0){4}}
\multiput(7,2)(1,0){2}{\line(0,1){1}}
\put(6,2){\line(0,1){1}}
\put(3.2,2.15){\rm 0}
\put(3.2,0.15){\rm 8}
\put(3.2,1.15){\rm 4}
\put(8.3,2.15){\rm 0}
\put(5.3,0.15){\rm 8}
\put(6.3,1.15){\rm 3}
\put(4.1,0.25){$\times$}
\multiput(11,1)(1,0){2}{\line(0,1){2}}
\multiput(11,1)(0,1){3}{\line(1,0){1}}
\put(10.2,2.15){\rm 0}
\put(10.2,1.15){\rm 0}
\put(12.3,1.15){\rm 0}
\put(12.3,2.15){\rm 0}
\put(11.1,2.25){$\times$}
\multiput(16,2)(1,0){2}{\line(0,1){1}}
\multiput(16,2)(0,1){2}{\line(1,0){1}}
\put(15.2,2.1){\rm 0}
\put(17.3,2.1){\rm 0}
\put(16.1,2.25){$\times$}
\end{picture}
\end{center}
\begin{center}
\unitlength 10pt
\begin{picture}(18,2)
\put(-4,1.15){$\overset{4}{\longrightarrow}$}
\put(0,1.25){$(1^{11})$}
\multiput(4,0)(1,0){3}{\line(0,1){2}}
\put(4,0){\line(1,0){2}}
\multiput(4,1)(0,1){2}{\line(1,0){4}}
\multiput(7,1)(1,0){2}{\line(0,1){1}}
\put(3.2,1.15){\rm 0}
\put(3.2,0.15){\rm 4}
\put(8.3,1.15){\rm 0}
\put(6.3,0.15){\rm 3}
\put(7.1,1.25){$\times$}
\multiput(11,1)(1,0){2}{\line(0,1){1}}
\multiput(11,1)(0,1){2}{\line(1,0){1}}
\put(10.2,1.15){\rm 0}
\put(12.3,1.15){\rm 0}
\put(16.4,1.2){$\emptyset$}
\end{picture}
\end{center}
\begin{center}
\unitlength 10pt
\begin{picture}(18,2)
\put(-4,1.15){$\overset{2}{\longrightarrow}$}
\put(0,1.25){$(1^{10})$}
\multiput(4,0)(1,0){3}{\line(0,1){2}}
\put(4,0){\line(1,0){2}}
\multiput(4,1)(0,1){2}{\line(1,0){3}}
\multiput(7,1)(1,0){1}{\line(0,1){1}}
\put(3.2,1.15){\rm 1}
\put(3.2,0.15){\rm 3}
\put(7.3,1.15){\rm 1}
\put(6.3,0.15){\rm 3}
\put(5.1,0.25){$\times$}
\multiput(11,1)(1,0){2}{\line(0,1){1}}
\multiput(11,1)(0,1){2}{\line(1,0){1}}
\put(10.2,1.15){\rm 0}
\put(12.3,1.15){\rm 0}
\put(16.4,1.2){$\emptyset$}
\end{picture}
\end{center}
\begin{center}
\unitlength 10pt
\begin{picture}(18,2)
\put(-4,1.15){$\overset{2}{\longrightarrow}$}
\put(0,1.25){$(1^{9})$}
\multiput(4,0)(1,0){2}{\line(0,1){2}}
\put(4,0){\line(1,0){1}}
\multiput(4,1)(0,1){2}{\line(1,0){3}}
\multiput(6,1)(1,0){2}{\line(0,1){1}}
\put(3.2,1.15){\rm 2}
\put(3.2,0.15){\rm 6}
\put(7.3,1.15){\rm 1}
\put(5.3,0.15){\rm 6}
\put(4.1,0.25){$\times$}
\multiput(11,1)(1,0){2}{\line(0,1){1}}
\multiput(11,1)(0,1){2}{\line(1,0){1}}
\put(10.2,1.15){\rm 0}
\put(12.3,1.15){\rm 0}
\put(11.1,1.25){$\times$}
\put(16.4,1.2){$\emptyset$}
\end{picture}
\end{center}
\begin{center}
\unitlength 10pt
\begin{picture}(18,1)
\put(-4,0.15){$\overset{3}{\longrightarrow}$}
\put(0,0.25){$(1^{8})$}
\multiput(4,0)(1,0){4}{\line(0,1){1}}
\multiput(4,0)(0,1){2}{\line(1,0){3}}
\put(3.2,0.15){\rm 2}
\put(7.3,0.15){\rm 1}
\put(11.4,0.2){$\emptyset$}
\put(16.4,0.2){$\emptyset$}
\end{picture}
\end{center}
\begin{center}
\unitlength 10pt
\begin{picture}(18,1)
\put(-4,0.15){$\overset{1}{\longrightarrow}$}
\put(0,0.25){$(1^{7})$}
\multiput(4,0)(1,0){4}{\line(0,1){1}}
\multiput(4,0)(0,1){2}{\line(1,0){3}}
\put(3.2,0.15){\rm 1}
\put(7.3,0.15){\rm 1}
\multiput(4.12,0.25)(1,0){3}{$\times$}
\put(11.4,0.2){$\emptyset$}
\put(16.4,0.2){$\emptyset$}
\end{picture}
\end{center}
\begin{center}
\unitlength 10pt
\begin{picture}(18,1)
\put(-5.5,0.15){$\overset{2}{\rightarrow}
\overset{2}{\rightarrow} \overset{2}{\rightarrow}$}
\put(0,0.25){$(1^{4})$}
\put(5.4,0.2){$\emptyset$}
\put(11.4,0.2){$\emptyset$}
\put(16.4,0.2){$\emptyset$}
\end{picture}
\end{center}
\begin{center}
\unitlength 10pt
\begin{picture}(18,1)
\put(-6,0.15){$\overset{1}{\rightarrow}\overset{1}{\rightarrow}
\overset{1}{\rightarrow} \overset{1}{\rightarrow}$}
\put(0.4,0.2){$\emptyset$}
\put(5.4,0.2){$\emptyset$}
\put(11.4,0.2){$\emptyset$}
\put(16.4,0.2){$\emptyset$.}
\end{picture}
\end{center}}

Thus the image is the highest path $11112221322433
\in B^{\otimes 14}_1$.
The algorithm of $\phi$ proceeds backward.
\end{Example}

\begin{Remark}\label{k:nonh}
Let ${\mathcal P}(L,\lambda)=
\{p \in B^{\otimes L}_1 | \mathrm{wt}\, p = \lambda\}$ 
be the set of all the weight $\lambda$ paths including non highest paths.
It is known that the algorithms for $\phi$ and $\phi^{-1}$ actually work 
in a wider setting so that (\ref{k:rcp}) is generalized to 
$\phi\bigl({\mathcal P}(L,\lambda)\bigr)
\underset{\phi}{\overset{\phi^{-1}}{\rightleftarrows}}
{\mathcal P}(L,\lambda)$.
The set 
$\phi\bigl({\mathcal P}(L,\lambda)\bigr)$
of extended rigged configurations is 
characterized by (\ref{k:rp}) with a non-trivial lower bound \cite{DS}.
\end{Remark}

\subsection{Inverse scattering method}\label{subsec:ism}

Let  $p=b_1\otimes \cdots \otimes b_L \in B^{\otimes L}_1$
be a state of the BBS satisfying 
the boundary condition (\ref{t:aug1c}).
Suppose that $p$ is highest and of weight $\lambda$, 
i.e. $p \in {\mathcal P}_+(L,\lambda)$.
Then the state $T_l(p)$ after the time evolution 
also belongs to ${\mathcal P}_+(L,\lambda)$.
Thus, via the KKR bijection (\ref{k:rcp}), 
$T_l$ on BBS states induces a 
time evolution of rigged configurations. 
The following theorem presents its explicit form.

\begin{Theorem}{\rm \cite[Prop.~2.6]{KOSTY06}}\label{k:th:lin}
For the subset of paths that undergo time evolutions without boundary effects,
the commutative diagram 
\begin{equation}\label{k:cd}
\begin{CD}
{\mathcal P}_+(L,\lambda) @>{\phi}>> 
\mathrm{RC}(L,\lambda)\\
@V{T_l}VV @VV{T_l}V\\
{\mathcal P}_+(L,\lambda) @>{\phi}>> 
\mathrm{RC}(L,\lambda)
\end{CD}
\end{equation}
holds with the following time evolution 
$T_l$ on rigged configurations:
\begin{align}
&T_l: (\mu, J)_L \mapsto (\mu, J')_L,\\
&J'= (J^{\prime (1)}, J^{(2)},\ldots, J^{(n)}),\quad
J^{\prime (1)}_{j,\alpha} = J^{(1)}_{j,\alpha}+\min(l,j).\label{k:rr}
\end{align} 
\end{Theorem} 

Namely, the KKR bijection {\em linearizes} the dynamics.
More concretely, we find 
\begin{align}
&(\mu^{(1)}, \ldots, \mu^{(n)}), (J^{(2)}, \ldots, J^{(n)})\;
\text{ are conserved (action variable)},\label{k:av}\\
&J^{(1)}_{j,\alpha}\;
\text{ changes linearly (angle variables).}
\end{align}
Let us write (\ref{k:rr}) as $J' = J+h_l$,
where $h_l=(\delta_{a1} \min(l,j))_{a,j,\alpha}$ plays the role of the ($l$th) velocity vector. 
The commutative diagram (\ref{k:cd}) provides a solution of the 
initial value problem in BBS.
For a state $p$, 
it is given as $T^N_l(p) = \phi^{-1}\circ T^N_l \circ \phi(p)$,
where the $T^N_l$ in the RHS is just to change the rigging as 
$J \mapsto J+Nh_l$.
The variety of time evolutions $T_1, T_2,\ldots$ is 
reflected in the velocity vectors $h_1, h_2, \ldots$.

\begin{Example}\label{k:rcd}
The time evolution of the rigged configurations 
under $T_\infty$ corresponding to Example \ref{t:exp2}. 
\begin{equation*}
\begin{picture}(180,39)(15,5)
\setlength{\unitlength}{0.32cm}
\put(0,2){$(1^{58})$}
\put(0.5,3.5){$\mu^{(0)}$}
\multiput(5,0)(1,0){3}{\line(0,1){3}}
\put(5,0){\line(1,0){2}}
\put(5,1){\line(1,0){3}}
\put(5,2){\line(1,0){4}}
\put(5,3){\line(1,0){4}}
\put(8,1){\line(0,1){2}}
\put(9,2){\line(0,1){1}}
\put(9.3,2.15){$4+4t$}
\put(8.3,1){$10+3t$}
\put(7.3,-0.1){$15+2t$}
\put(6.5,3.5){$\mu^{(1)}$}
\multiput(14,1)(1,0){2}{\line(0,1){2}}
\put(14,1){\line(1,0){1}}
\multiput(14,2)(0,1){2}{\line(1,0){3}}
\multiput(16,2)(1,0){2}{\line(0,1){1}}
\put(15,3.5){$\mu^{(2)}$}
\put(17.3,2.15){\rm 1}
\put(15.3,0.9){\rm 0}
\multiput(20,2)(1,0){2}{\line(0,1){1}}
\multiput(20,2)(0,1){2}{\line(1,0){1}}
\put(20,3.5){$\mu^{(3)}$}
\put(21.3,2.15){\rm 0}
\end{picture}
\end{equation*}
\end{Example}

In Example \ref{k:rcd}, 
one notices that $\mu^{(1)}=(4,3,2)$ 
gives the list of amplitudes of solitons.  
This fact holds in general, which is 
a manifestation of the Bethe ansatz structure in BBS: 
\begin{equation}\label{k:ss}
\mu^{(1)} = \text{list of amplitudes of solitons}.
\end{equation}
We call it {\em soliton/string correspondence}.
In fact, $\mu^{(1)}$ 
is related to the earlier introduced conserved quantity 
$E_l$ (\ref{t:e}) as 
\begin{equation}\label{k:elm}
E_l = \text{ number of cells in the left $l$ columns of $\mu^{(1)}$}.
\end{equation}
There are still more conserved quantities in (\ref{k:av}) than $\mu^{(1)}$.
They are responsible for the {\em internal labels} 
of colliding solitons\footnote{The data (\ref{k:av}) is regarded 
as a rigged configuration for $\widehat{\mathfrak{sl}}_n$ 
(instead of $\widehat{\mathfrak{sl}}_{n+1}$) and the 
solitons are determined as the image of it under the KKR map $\phi^{-1}$.
See \cite{KOSTY06, KSY} for detail.}.

The inverse scattering scheme explained so far
is naturally extended to not necessarily highest states
by Remark \ref{k:nonh} as long as the boundary effect is absent.
For $n=1$, the solution of the initial value problem 
in the same spirit has also been obtained in \cite{Takagi1}. 
 
It was an essential insight of the quantum inverse scattering method \cite{STF}
that Bethe ansatz can be viewed as a quantization of the classical 
inverse scattering method \cite{AS,GGKM}.
It is gratifying to realize that 
the combinatorial version of the 
Bethe ansatz here provides the inverse scattering scheme
of the BBS which is a crystalline quantum integrable system.
In this respect, 
the KKR maps $\phi$ and $\phi^{-1}$ are
the direct and inverse scattering transforms 
and the rigged configurations play the role of scattering data \cite{KOSTY06}.


\section{Ultradiscretization --- min-plus algebra}

\subsection{Tropicalization and ultradiscretization}
\label{i:subsec:4-1}

Define $\mathbb{T} = \R \cup \{ \infty \}$
where $\infty$ is the infinity which satisfies 
$ a < \infty$ and $\infty + a = \infty$ for any $a \in \R$. 
The algebra $(\mathbb{T}, \oplus, \odot)$ is called the {\it min-plus algebra} 
(or the {\it tropical semifield}) \cite{SS04},
where the two operations ``$\oplus$'' and ``$\odot$'' in $\mathbb{T}$
are respectively called
{\it tropical addition} and {\it tropical multiplication} defined by
$$
a \oplus b := \min(a, b), 
\qquad 
a \odot b := a + b.
$$
The additive identity is $\infty$, and the multiplicative identity
is $0$, i.e. 
$$
a \oplus \infty = a,
\qquad 
a \odot 0 = a
$$
hold for any $a \in \mathbb{T}$. 
We have the inverse of $\odot$ as $a \odot (-a) = 0$, 
but not the inverse of $\oplus$.
In the following we also write $(\mathbb{T}, \min, +)$ 
for $(\mathbb{T}, \oplus, \odot)$.

We are to introduce a limiting procedure called the {\it tropicalization},
which links the subtraction-free algebra
$(\R_{>0},+,\times)$ to the min-plus algebra. 
We define a map $\mathrm{Log}_\ve : \R_{>0} \to \R$ with an infinitesimal 
parameter $\ve > 0$ by
\begin{align}
  \label{i:loge-map}
  \mathrm{Log}_\ve : a \mapsto - \ve \log a.
\end{align}
For $a > 0$, define $A \in \R$ by $a = \e^{-\frac{A}{\ve}}$.
Then we have $\mathrm{Log}_\ve (a) = A$. 
Moreover, for $a, b > 0$ define $A,B \in \R$ by $a = \e^{-\frac{A}{\ve}}$ and 
$b = \e^{-\frac{B}{\ve}}$. 
Then we have  
$$
  \mathrm{Log}_\ve (a + b) = 
  -\ve \log (\e^{-\frac{A}{\ve}} + \e^{-\frac{B}{\ve}}),
  \quad
  \mathrm{Log}_\ve (a \times b) = A + B.
$$
In the limit $\ve \to 0$, $\mathrm{Log}_\ve (a + b)$ becomes $\min(A,B)$.
In this manner, the algebra $(\R_{>0},+,\times)$ reduces to 
the min-plus algebra,
and the procedure $\lim_{\ve \to 0} \mathrm{Log}_\ve$ 
with the transformation as $a = \e^{-\frac{A}{\ve}}$
is called the {\it tropicalization}.
 
Through the tropicalization,
subtraction-free rational equations on $\R_{>0}$ reduce to 
piecewise-linear equations on $\R$ described by min-plus algebra,
which is summarized as follows:
for $A,B,C\in \R$ set 
$$a=e^{-\frac{A}{\ve}},\ b=e^{-\frac{B}{\ve}}, 
\ c=e^{-\frac{C}{\ve}}$$
and take the limit $\ve \to 0$ of the image $\mathrm{Log}_\ve$ of 
the equations
$$
({\rm i})\ a+b=c, \quad ({\rm ii})\ ab=c, 
\quad ({\rm iii})\ \frac{a}{b}=c.
$$
Then we obtain 
$$
({\rm i})\ \min (A,B)=C, \quad ({\rm ii})\ A+B=C, 
\quad ({\rm iii}) \ A-B=C.
$$
We remark that the distributive law of the algebra $(\R_{>0},+,\times)$,
$a(b+c) = ab + ac$, reduces to that of the min-plus algebra,
$A + \min(B,C) = \min(A+B, A+C)$.

Let us show an example.
\begin{Example}
The discrete Lotka-Volterra equation for the variables
$\{v_j^m ~|~ (j,m) \in \Z^2 \}$ is given by
\begin{align}\label{i:eq:d-LV}
\frac{v_j^{m+1}}{v_j^m} 
= 
\frac{1+\delta v_{j-1}^m}{1+\delta v_{j+1}^{m+1}},
\end{align}
where $\delta$ is a positive parameter.
We restrict $v_j^m \in \R_{>0}$, and take 
transformations $\delta = \mathrm{e}^{-\frac{1}{\ve}}$
and $v_j^m = \mathrm{e}^{-\frac{V_j^m}{\ve}}$.
Then the tropicalization of \eqref{i:eq:d-LV} is calculated as
\begin{align}\label{i:eq:vv}
\begin{split}
  V_j^{m+1} - V_j^m 
  &= - \lim_{\ve \to 0} \ve \Bigl( 
  \log( 1+\mathrm{e}^{\frac{-1- V_{j-1}^m}{\ve}})
  - \log( 1+\mathrm{e}^{\frac{-1-V_{j+1}^{m+1}}{\ve}})
  \Bigr) 
  \\
  &= \min( 0, V_{j-1}^m+1) - \min( 0, V_{j+1}^{m+1}+1).
\end{split}
\end{align}
\end{Example}

By construction, the tropicalization of a discrete equation is defined on $\R$,
i.e. the dependent variables of the tropicalization are in $\R$.
When the tropicalization is defined on $\Z$,
we call it the {\it ultradiscretization} of the discrete equation.  
In the above example, \eqref{i:eq:vv} allows the ultradiscretization, 
since $V_j^{m+1}$ is determined as an integer
if $V_j^m$, $V_{j-1}^m$ and $V_{j+1}^{m+1}$ are integers.  
 
\begin{Remark}\label{i:rem:dLV}
The original Lotka-Volterra equation
$\overset{.}{v}_j = v_j (v_{j+1} - v_{j-1})$
is the continuous limit 
$\delta \to 0$ of \eqref{i:eq:d-LV} with $v_j^{m} = v_j(-\delta m)$.
Here $\delta$ is a unit of the discrete time and 
$\overset{.}{v}_j$ is a derivation of $v_j = v_j(t)$ by the time $t$.
\end{Remark}

\subsection{Evolution equations of BBS}
\label{i:subsec:4-2}

The original BBS in \S \ref{i:sec-intro}
corresponds to the time evolution $T_\infty$ 
in the formalism of \S \ref{t:2-3}, which is the only case 
that admits the algorithms (albeit non-local) without carrier.
One can set up two kinds of evolution equations for it:
\begin{enumerate}
\item[(i)]
the equation for the number $u_k^t$ of balls 
in the $k$-th box at time $t$ \cite{TTMS96} (the spatial description),

\item[(ii)]
the equation for the number $Q_j^t$ of balls in the $j$-th soliton 
(from the left) and 
the number $W_j^t$ of empty boxes between the $j$-th and 
the $j+1$-th solitons at time $t$ \cite{TNS99} (the soliton description).  
\end{enumerate}
These descriptions are respectively related to the ultradiscretization
of famous integrable difference equations, 
the discrete Lotka-Volterra equation (\S \ref{i:subsubsec-LV}) and 
the discrete Toda lattice equation (\S \ref{i:sec:TodaBBS}).


\subsubsection{Lotka-Volterra equation and infinite BBS.}
\label{i:subsubsec-LV}

Let $u_k^t \in \{0,1\}$ be the number of balls in the 
$k$-th box at time $t$.
The evolution equation for $u_k^t$ is described by a
piecewise-linear equation \cite{TTMS96}:
\begin{align}\label{i:eq:BBS-u}
u_k^{t+1} = \min \bigl(1-u_k^t, ~ 
            \sum_{j=-\infty}^{k-1}(u_j^t - u_j^{t+1}) \bigr).
\end{align}
This equation has a piecewise-linear version of the bilinear form
in the following sense:
assume that the variables $\{\rho_k^t ~|~ k,t \in \Z\}$ satisfy 
\begin{align}\label{i:eq:tau-BBS}
\rho_{k+1}^{t+1} + \rho_k^{t-1} 
=
\max\bigl(\rho_{k+1}^t + \rho_{k}^t, 
~ \rho_{k+1}^{t-1} + \rho_{k}^{t+1} -1 \bigr).
\end{align}
Then the variables $\{ u_k^t ~|~ k,t \in \Z \}$ defined by 
\begin{align}\label{i:u-T}
u_k^t = \rho_k^t + \rho_{k-1}^{t+1} - \rho_k^{t+1} - \rho_{k-1}^t
\end{align}
satisfy \eqref{i:eq:BBS-u}.

On the other hand, the discrete Lotka-Volterra equation 
\eqref{i:eq:d-LV} has a bilinear form:
\begin{align}\label{i:eq:tau-LV}
(1 + \delta) \tau^{m}_{j+1}\tau^{m+1}_j
=
\delta \tau^{m+1}_{j+2}\tau^{m}_{j-1}
+
\tau^{m+1}_{j+1}\tau^{m}_{j},
\end{align}
i.e. if the variables $\{ \tau_j^{m} ~|~ j,m \in \Z \}$ satisfy
the bilinear difference equation \eqref{i:eq:tau-LV}, then 
the variables $\{v_j^{m} ~|~ j,m \in \Z \}$ defined by 
\begin{align}\label{i:eq:v-tau}
v_j^{m} 
= 
\frac{\tau^{m+1}_{j+2}\tau^{m}_{j-1}}{\tau^{m+1}_{j+1}\tau^{m}_{j}}
\end{align}
satisfy \eqref{i:eq:d-LV}.

\begin{Prop}{\rm \cite{TTMS96}}
Eq. \eqref{i:eq:tau-BBS} is 
the ultradiscretization of the bilinear form \eqref{i:eq:tau-LV}  
with the transformations $\delta = \mathrm{e}^{-\frac{1}{\ve}}$
and $\sigma_k^t = \mathrm{e}^{\frac{\rho_k^t}{\ve}}$
under a coordinate transformation $\sigma^t_k := \tau^k_{k-t}$.
\end{Prop}

\proof
It is obvious that \eqref{i:eq:tau-BBS} can be defined on $\Z$.
Via the coordinate transformation, \eqref{i:eq:tau-LV} becomes
\begin{align}\label{i:eq:tau-KdV}
(1 + \delta) \sigma^{t-1}_k\sigma^{t+1}_{k+1}
=
\delta \sigma^{t-1}_{k+1}\sigma^{t+1}_k
+
\sigma^{t}_{k}\sigma^{t}_{k+1}.
\end{align}
By applying the tropicalization with the transformation, 
we have
\begin{align*}
- \lim_{\ve \to 0} \ve \log(1+\mathrm{e}^{-\frac{1}{\ve}}) 
- \rho_k^{t-1} - \rho_{k+1}^{t+1} 
= 
- \lim_{\ve \to 0} \ve \log
\bigl(\mathrm{e}^{\frac{\rho_{k+1}^{t-1} + \rho_{k}^{t+1} -1}{\ve}}
+ \mathrm{e}^{\frac{\rho^{t}_{k} + \rho^{t}_{k+1}}{\ve}} \bigr), 
\end{align*}
which yields \eqref{i:eq:tau-BBS}.
Here we use 
$$
\lim_{\ve \to 0} \ve \log(1+\mathrm{e}^{-\frac{1}{\ve}}) = 0,
\qquad 
\lim_{\ve \to 0} \ve \log(\mathrm{e}^\frac{A}{\ve}+\mathrm{e}^\frac{B}{\ve}) 
= \max(A,B).   
$$
\qed

\begin{Remark} 
At \eqref{i:eq:BBS-u},
we can regard $v_{k-1}^t := \sum_{j=-\infty}^{k-1}(u_j^t - u_j^{t+1})$
as ``the number of balls in the carrier'',
which is identified with $v_{k-1}$ at \eqref{t:apr20}.
Then \eqref{i:eq:BBS-u} is rewritten as
\begin{align}\label{i:eq:uv}
  u_k^{t+1} = \min(1-u_k^t, ~v_{k-1}^t),
  \qquad v_k^t = u_k^t + v_{k-1}^t - u_k^{t+1}.
\end{align}
One sees that these correspond to the description of BBS with 
the combinatorial $R$ of $\widehat{\mathfrak{sl}}_2$ crystal in \S \ref{t:2-2}.
In fact, for $n=1$, \eqref{t:udRPP} simply reads
\begin{equation}
{\tilde x}_i -x_i = -{\tilde y}_i + y_i = 
\min(x_{i+1},y_i)-\min(x_i,y_{i+1}).
\end{equation}
Thus by setting $y=(1-u_k^t,u_k^t), ~\tilde{y}=(1-u_k^{t+1},u_k^{t+1}) \in B_1$
and $x = (\theta - v_{k-1}^t,v_{k-1}^t), 
~\tilde{x} = (\theta - v_{k}^{t},v_{k}^{t})
\in B_\theta$,
\eqref{t:udRPP} reduces to
\eqref{i:eq:uv} in the limit $\theta \to \infty$.
\end{Remark}

The $\widehat{\mathfrak{sl}}_{n+1}$ BBS also has  
the bilinear form as \eqref{i:eq:tau-BBS}.
Given a state at time $t=0$ as 
$p= \cdots \otimes x^0_k \otimes x^0_{k+1} \otimes \cdots$
with $x^0_k \in B_1$,
we consider its time evolution
$T^t_\infty(p)  = \cdots \otimes x^t_k \otimes x^t_{k+1} \otimes \cdots$
for $t \ge 0$.
Here $x^t_k \in B_1$ specifies the local state of the $k$-th box at time $t$.
According to \eqref{t:jun24a}, we express it as
$x^t_k = (x^t_{k,1},\ldots, x^t_{k,n+1})$.
Define $\rho_{k,i}^t \in \Z$ 
($k \in \Z, i = 0,1,\ldots, n+1, t \in \Z_{\geq 0}$) by
\begin{align}\label{i:rho}
\begin{split}
&\rho_{k,i}^t 
=
\sum_{j=-\infty}^k (x_{j,2}^t + x_{j,3}^t + \cdots + x_{j,i}^t)
+
\sum_{t' \geq t+1} \sum_{j=-\infty}^k 
(x_{j,2}^{t'} + x_{j,3}^{t'} + \cdots + x_{j,n+1}^{t'})
\\
&\hspace{10cm} i=1,\cdots,n+1,
\\
&\rho_{k,0}^t = \rho_{k,n+1}^t - k.
\end{split}
\end{align}
This counts the number of balls in the SW quadrant of the time evolution profile as in Example \ref{t:exp2}. 
The variables $(k,t)$ specify the position of the top 
right corner of the quadrant.
The first term in \eqref{i:rho} means that only those balls 
with color $\le i$ are counted on the top row of the quadrant.
The quantities $\rho^t_{k,i}$ are finite due to the boundary condition and 
the BBS time evolution rule.
To see this concretely, note that each state at time $t$ has a finite number of balls,
therefore we have $x^t_{j,i}=0 ~(j<k_0, ~i=2,\ldots,n+1)$ for some $k_0 \le k$.
Then,  the time evolution rule (Proposition \ref{t:jun15c}) implies that
the nonzero contribution to the double infinite sum in 
\eqref{i:rho}  actually comes 
from the finite region depicted as 
\begin{align}\label{i:rho-sum}
\begin{matrix}
x_{k_0}^t & x_{k_0+1 }^t & \cdots & x_{k-1}^t & x_k^t &
\\
& x_{k_0+1}^{t+1} & \cdots & \cdots & x_{k}^{t+1} &  
\\
& & x_{k_0+2}^{t+2} & \cdots & x_{k}^{t+2} &  
\\
& & & \ddots & \vdots & 
\\
& & & & x_k^{t+k-k_0} &
\end{matrix}.
\end{align}
By the definition, we have $\rho_{k,n+1}^{t+1} = \rho_{k,1}^{t}$ and 
\begin{align}\label{i:rho-x}
x_{k,i}^t 
= 
\rho_{k,i}^t - \rho_{k-1,i}^t - \rho_{k,i-1}^t + \rho_{k-1,i-1}^t
& & i=1,\ldots,n+1.
\end{align}

\begin{Prop}{\rm \cite[Prop. 4.2]{KSY}} 
The following relation holds:
\begin{align}\label{i:rho-bilinear}
\rho_{k,i-1}^{t+1} + \rho_{k-1,i}^{t} 
=
\max(\rho_{k,i}^{t+1} + \rho_{k-1,i-1}^{t},
\rho_{k-1,i-1}^{t+1} + \rho_{k,i}^{t} - 1) & & 
i=2,\ldots, n+1.
\end{align}  
\end{Prop}
We note that the similar fact is studied in \cite[IV]{HHIKTT01}.
When $n=1$, we recover \eqref{i:eq:tau-BBS} and \eqref{i:u-T}
via 
$\rho_k^t = \rho_{k,2}^t$, $u_k^t = x_{k,2}^t$.
The variables $\rho_{k,i}^t$ will play an important role to 
solve the BBS in \S 4.4.

\subsubsection{Toda lattice and infinite BBS.}
\label{i:sec:TodaBBS}

Consider a state of the $\widehat{\mathfrak{sl}}_2$ BBS with $N$ solitons,
and let $Q_j^t$ be the number of balls in the $j$-th soliton and 
$W_j^t$ be the number of empty boxes between the $j$-th soliton and 
the $j+1$-th soliton at time $t$ as follows:
\begin{align*}
....1111111\overbrace{222..22}^{Q_1^t}\overbrace{11...1}^{W_1^t}
\overbrace{2...22}^{Q_2^t}\overbrace{1...1}^{W_2^t}
........\overbrace{11...1}^{W_{N-1}^t}\overbrace{22...2}^{Q_N^t}11111......
\end{align*}
We have positive integers $Q_j^t$ for $j=1,\ldots,N$
and $W_j^t$ for $j=1,\ldots,N-1$,
and have $W_0^t = W_N^t = \infty$. 

The evolution equations for $Q_j^t$ and $W_j^t$ are written as \cite{TNS99}
\begin{align}\label{i:eq:Toda-Q}
&Q_j^{t+1} 
= 
\min\bigl(\sum_{k=1}^{j} Q_k^t - \sum_{k=1}^{j-1} Q_k^{t+1}, ~ W_j^t \bigr)
& & j=1,\ldots,N,
\\
\label{i:eq:Toda-W}
&W_j^{t+1} = Q_{j+1}^t + W_j^t - Q_j^{t+1}
& & j=1,\ldots,N-1.
\end{align}

\begin{Example}
Let us consider the case of $N=3$. 
In the following, the evolution of a $3$-soliton state
at the left is written in terms of $(Q_1^t,W_1^t,Q_2^t,W_2^t,Q_3^t)$
at the right:
\begin{verbatim}                                              
     t=0   ...2222...222...2.....................   (4,3,3,3,1)
     t=1   .......222...222.22...................   (3,3,3,1,2)
     t=2   ..........222...2..2222...............   (3,3,1,2,4)
     t=3   .............222.2.....2222...........   (3,1,1,5,4)
     t=4   ................2.222......2222.......   (1,1,3,6,4)
     t=5   .................2...222.......2222...   (1,3,3,7,4)
\end{verbatim}
One sees that the variables $(Q_1^t,W_1^t,Q_2^t,W_2^t,Q_3^t)$
satisfy \eqref{i:eq:Toda-Q} and \eqref{i:eq:Toda-W}.
\end{Example}

On the other hand, the discrete Toda lattice equation is given by
\begin{align}\label{i:eq:Toda-q}
&q_j^{t+1} = q_j^t + w_j^t - w_{j-1}^{t+1} 
\\
\label{i:eq:Toda-w}
&w_j^{t+1} = \frac{q_{j+1}^t w_j^t}{q_{j}^{t+1}}
\end{align}
for $j,t \in \Z$.
Now we only consider \eqref{i:eq:Toda-q} for $j=1,\ldots,N$ and 
\eqref{i:eq:Toda-w} for $j=1,\ldots,N-1$
with the boundary condition $w_0^t = w_N^t = 0$.
This is what is called the {\it discrete Toda molecule equation}.
\begin{Prop}{\rm \cite[\S 3]{TNS99}}
Eqs. \eqref{i:eq:Toda-Q} and \eqref{i:eq:Toda-W} are
the ultradiscretization of the discrete Toda molecule equation 
with $q_j^t = \mathrm{e}^{-\frac{Q_j^t}{\ve}}$ and 
$w_j^t = \mathrm{e}^{-\frac{W_j^t}{\ve}}$.
\end{Prop}
\proof
By using \eqref{i:eq:Toda-w} iteratively,
\eqref{i:eq:Toda-q} becomes subtraction-free:  
\begin{align}\label{i:eq:qq}
q_j^{t+1} 
= 
\frac{\prod_{k=1}^j q_k^t}{\prod_{k=1}^{j-1} q_k^{t+1}} + w_j^t. 
\end{align}
We apply the tropicalization and obtain the claim.
Note that the boundary condition for $w_0^t$ and $w_N^t$
is consistent with that for $W_0^t$ and $W_N^t$. 
It is clear that \eqref{i:eq:Toda-Q} and \eqref{i:eq:Toda-W}
are defined on $\Z$.
\qed

\begin{Remark}
The original Toda lattice equation 
$\overset{..}{x}_j 
=  \mathrm{e}^{x_{j+1} - x_j} - \mathrm{e}^{x_{j} - x_{j-1}}$
is the continuous limit $\delta \to 0$ of \eqref{i:eq:Toda-q}
and \eqref{i:eq:Toda-w}
with $w_j^t = \delta^2 \mathrm{e}^{x_{j+1} - x_j}$
and $q_j^t = 1 + \delta \overset{.}{x}_i$,
in the same manner as the Lotka-Volterra equation
at Remark \ref{i:rem:dLV}.
\end{Remark}

\begin{Remark}
The description \eqref{i:eq:Toda-Q}, \eqref{i:eq:Toda-W} can be generalized to 
the infinite BBS of type $\widehat{\mathfrak{sl}}_{n+1}$ \cite[\S 3]{TNS99}.
Again we consider a state with $N$ solitons,
where we regard a non-increasing sequence of $2,3,\ldots,n+1$
as a soliton (\S \ref{t:2-3-2}).
Let $Q_{j,i}^t$ be the number of $i$-balls in the $j$-th soliton and 
$W_j^t$ be the number of empty boxes between the $j$-th soliton and 
the $j+1$-th soliton at time $t$.
Then we have non-negative integers $Q_{j,i}^t$ 
($j=1,\ldots,N,~ i=2,\ldots,n+1$) and $W_j^t$ ($j=0,\ldots,N$)
satisfying  $\sum_{i=2}^{n+1} Q_{j,i}^t >0$ and
$W_0^t = W_N^t = \infty$. 
The other $W_j^t$ can be zero only if 
the color of the rightmost ball in the 
$j$-th soliton is strictly smaller than that of the leftmost 
ball in the $j+1$-th soliton.

We define $W_j^t$ for $t \in \Z/n$, and 
regard $W_j^t$ with $t \notin \Z$ as the intermediate states.
The evolution equations are written as
\begin{align}\label{i:eq:g-udToda}
&Q_{j,i}^{t+1} 
= 
\min\bigl(\sum_{k=1}^{j} Q_{k,i}^t - \sum_{k=1}^{j-1} Q_{k,i}^{t+1}, 
~ W_j^{t+\frac{n+1-i}{n}} \bigr)
& & j=1,\ldots,N,
\\ \displaybreak[0]
&W_j^{t+\frac{n+2-i}{n}} = Q_{j+1,i}^t + W_j^{t+\frac{n+1-i}{n}} 
- Q_{j,i}^{t+1}
& & j=1,\ldots,N-1,
\end{align}
where we run these equations from $i=n+1$ to $i=2$.
These piecewise-linear equations correspond to the ultradiscretization 
of the generalized (or hungry) Toda molecule equation.
\end{Remark}

\begin{Remark}
In this description
only the information of relative coordinates of solitons survive,
and the information of the absolute coordinates are lost.
However, it is sufficient to study the basic features of BBS such as 
the soliton scattering and the conserved quantities.
See \cite{TNS99} for the detail. 
\end{Remark}

\subsection{Birational $R$ and Geometric crystal}\label{t:4-3}

The purpose of this subsection is to introduce
{\em birational} $R$ and {\em geometric crystal} for $\widehat{\mathfrak{sl}}_{n+1}$ \cite{KOTY03, KOTY04}.
Besides their conceptual importance,
they are useful
to describe local evolution rules of
discrete integrable systems related to BBS.
They consist of birational maps and many other relations between certain sets of variables.
The combinatorial $R$ and the crystal for $\widehat{\mathfrak{sl}}_{n+1}$ in \S \ref{t:2-2} are obtained from them
by ultradiscretization.

\subsubsection{Birational $R$.}
Let ${\mathcal B} = \{x = (x_1,\ldots, x_{n+1})  \} \subset (\mathbb{C}^{\times})^{n+1}$ be a set of variables.
The birational $R$
(introduced under the name of {\em tropical} $R$ in \cite{KOTY03, KOTY04})
for $\widehat{\mathfrak{sl}}_{n+1}$ is the birational map 
$R: {\mathcal B} \times {\mathcal B} \rightarrow {\mathcal B} \times {\mathcal B}$ specified by
$R(x,y) = (\tilde{y},\tilde{x})$ in which
\begin{equation}\label{t:eq:RPP}
\begin{split}
&\tilde{x}_i=x_i \frac{P_{i-1}(x,y)}{P_i(x,y)}, \quad \tilde{y}_i=y_i \frac{P_i(x,y)}{P_{i-1}(x,y)},\\
&P_i(x,y)=\sum_{k=1}^{n+1} \left( \prod_{j=k}^{n+1} x_{i+j} \prod_{j=1}^k y_{i+j} \right),
\end{split}
\end{equation}
where all the indices are considered to be in $\Z_{n+1}$.
It satisfies the inversion relation $R^2=id$ on ${\mathcal B} \times {\mathcal B}$ and the 
Yang-Baxter equation 
\begin{equation}\label{t:YBtropR}
R_1R_2R_1 = R_2R_1R_2,
\end{equation}
on ${\mathcal B} \times {\mathcal B} \times {\mathcal B}$,
where $R_1(x,y,z) = (R(x,y),z)$ and 
$R_2(x,y,z) = (x,R(y,z))$. 
A proof will be given later (Proposition \ref{t:jun29b}).

The birational $R$ 
is characterized as the unique solution to a version of discrete Toda lattice equation.
\begin{Prop}{\rm \cite[Th.2.2]{Y01}}\label{t:jun29a} 
Given $(x,y)$,
the birational $R$ 
is the unique solution to the equations on $(\tilde{y},\tilde{x})$:
\begin{equation}\label{t:eq:toda}
x_iy_i = \tilde{y}_i \tilde{x}_i,\qquad 
\frac{1}{x_i}+\frac{1}{y_{i+1}} = 
\frac{1}{\tilde{y}_i}+\frac{1}{\tilde{x}_{i+1}},
\end{equation}
with an extra constraint
$\prod_{i=1}^{n+1}(x_i/\tilde{x}_i) = 
\prod_{i=1}^{n+1}(y_i/\tilde{y}_i)= 1$.
\end{Prop}
\proof
We prove that the $(\tilde{y},\tilde{x})$ given by (\ref{t:eq:RPP}) satisfies 
(\ref{t:eq:toda}).
The uniqueness of the solution will be discussed later (Proposition \ref{t:jun15b}).
The former equation is clearly satisfied by (\ref{t:eq:RPP}).
Let us check the latter one.
It is equivalent to 
$P_{i+1}(x,y)/x_{i+1} + P_{i-1}(x,y)/y_{i} =  P_{i}(x,y)/x_{i} + P_{i}(x,y)/y_{i+1}$
which is verified as
\begin{align*}
P_{i+1}(x,y)/x_{i+1} - P_{i}(x,y)/x_{i} &= \sum_{k=1}^{n} \left( \prod_{j=k}^{n} x_{i+j+1} \prod_{j=1}^k y_{i+j+1} -
 \prod_{j=k}^{n} x_{i+j} \prod_{j=1}^k y_{i+j} \right) \\
&= \sum_{k=2}^{n+1} \left( \prod_{j=k}^{n+1} x_{i+j} \prod_{j=2}^k y_{i+j} -
 \prod_{j=k}^{n+1} x_{i+j-1} \prod_{j=2}^k y_{i+j-1} \right) \\
&= P_{i}(x,y)/y_{i+1} - P_{i-1}(x,y)/y_{i}.
\end{align*}
\qed

In order to relate the birational $R$ to the combinatorial $R$,
we introduce the {\em max-plus version of the tropicalization}.
It is a slight modification of the tropicalization in \S \ref{i:subsec:4-1}.
For $a > 0$, define $A \in \R$ by $a = \e^{\frac{A}{\ve}}$.
Then we have $- \mathrm{Log}_\ve (a) = A$. 
Moreover, for $a, b > 0$ define $A,B \in \R$ by $a = \e^{\frac{A}{\ve}}$ and 
$b = \e^{\frac{B}{\ve}}$. 
Then we have  
$$
  - \mathrm{Log}_\ve (a + b) = 
  \ve \log (\e^{\frac{A}{\ve}} + \e^{\frac{B}{\ve}}),
  \quad
  -\mathrm{Log}_\ve (a \times b) = A + B.
$$
In the limit $\ve \to 0$, $-\mathrm{Log}_\ve (a + b)$ becomes $\max(A,B)$.
In this manner, the algebra $(\R_{>0},+,\times)$ reduces to 
the ``max-plus'' algebra,
and the procedure $- \lim_{\ve \to 0} \mathrm{Log}_\ve$ 
with the transformation as $a = \e^{\frac{A}{\ve}}$
is also called the tropicalization.
As in \S \ref{i:subsec:4-1} it is called ultradiscretization when defined on $\Z$.
We note that this version of the ultradiscretization of (\ref{t:eq:RPP}) is (\ref{t:udRPP}),
and that of (\ref{t:eq:toda}) is (\ref{t:udtoda}),
when we take ${\mathcal B} = \{x = (x_1,\ldots, x_{n+1}) \vert x_i \in {\mathbb R}_{>0} \, \mbox{for all} \, i \}$.

%

\subsubsection{Geometric crystal.}
A representation theoretical background for the birational $R$ is 
provided by the geometric crystals \cite{BK00} 
and their natural extrapolation into the affine setting \cite[\S 1]{KOTY03}.
We explain this notion
for $\widehat{\mathfrak{sl}}_{n+1}$.

To give an overview of the basic idea,
first we show a few relations in the case of up to 2-fold tensor products.
Let us begin with the crystal.
As a result of \eqref{t:jun24e}-\eqref{t:jun24i} and by interpreting $\ft{i}=\et{i}^{-1}$,
the action of the Kashiwara operator 
${\tilde e}_i$ with a parameter $c \in \Z$ is given, unless they vanish, by
\begin{align}
& {\tilde e}^c_i(x) = (\ldots, x_{i-1}, x_i+c, x_{i+1}-c, x_{i+2},\ldots),\\
\begin{split}
& {\tilde e}_i^c(x \otimes y) 
= {\tilde e}^{c_1}_i(x) \otimes {\tilde e}_i^{c_2}(y),\\
&c_1 = \max(x_i+c,y_{i+1}) - \max(x_i,y_{i+1}), \\
&c_2 = \max(x_i,y_{i+1}) - \max(x_i,y_{i+1}-c).
\end{split}\label{t:aug26i}
\end{align}
In the geometric crystal, one still has the coordinates 
$x=(x_1,\ldots, x_{n+1}) \in {\mathcal B}$ and the corresponding 
structure looks as $(c \in \mathbb{C}^{\times})$
\begin{align}
&e^c_i(x) = (\ldots, x_{i-1}, cx_i, c^{-1}x_{i+1}, x_{i+2},\ldots),\label{t:aug25c}\\
\begin{split}
&e_i^c(x, y) 
= (e^{c_1}_i(x), e_i^{c_2}(y)),\\
&c_1 = \frac{cx_i+y_{i+1}}{x_i+y_{i+1}}, \;\;  
 c_2 = \frac{x_i+y_{i+1}}{x_i+c^{-1}y_{i+1}}.
\end{split}\label{t:aug25d}
\end{align}
We call $e_i^c$ the {\em geometric Kashiwara operator}.
Note that the $c_1, c_2$ in (\ref{t:aug26i}) are piecewise-linear 
and obtained from (\ref{t:aug25d}) by the ultradiscretization, i.e.~replacing 
$+,\times, /$ with $\max, +, -$, respectively.
%

Now we define the geometric crystal for $\widehat{\mathfrak{sl}}_{n+1}$ in more general setting.
In what follows, let $c \in \mathbb{C}^{\times}$ be a parameter which takes generic values, $e_i^c$ be a 
rational transformation
on a variable set ${\mathcal V} \subset (\mathbb{C}^{\times})^N$ where $N \in \mathbb{Z}_{>0}$,
and
$\varepsilon_i, \gamma_i$ are rational functions on ${\mathcal V}$. 

\begin{Definition} \label{t:def:cry}
A geometric crystal for $\widehat{\mathfrak{sl}}_{n+1}$
is a family $\{ {\mathcal V} ,\varepsilon_i,\gamma_i, e_i^c\}$ 
which satisfies the following relations.
For any $x \in  {\mathcal V}$, $c, c' \in \mathbb{C}^{\times}$, and 
$i,j \in I = \{0,1,\ldots,n \}$,
\begin{itemize}
\item[(i)] $e_i^{c}e_i^{c'}(x)=e_i^{c c'}(x), ~~e_i^1(x)=x$,
\item[(ii)] 
$\varepsilon_i(e_i^c(x))=c^{-1}\varepsilon_i(x)$,
\item[(iii)] 
$\gamma_i(e_j^c(x))=c^2 \gamma_i(x) \, (i=j), \quad =c^{-1} \gamma_i(x) 
\; (i-j \equiv \pm 1), \quad = \gamma_i(x) \; (\text{otherwise})$,
\item[(iv)]
$e_i^{c}e_j^{c'} (x) = e_j^{c'}e_i^{c} (x)$
if $i-j \not\equiv \pm 1$,
\item[(v)]
$e_i^{c}e_j^{c c'}e_i^{c'} (x)=e_j^{c'}e_i^{c c'}e_j^{c} (x)$
if $i-j \equiv \pm 1$.
\end{itemize}
Here $i \equiv j$ means $i - j \in (n+1) \Z$. 
\end{Definition}
In what follows, we introduce the function $\varphi_i$ by 
$\gamma_i  = \varphi_i /\varepsilon_i $.
\begin{Example}\label{t:aug26a}
For $x \in {\mathcal V} = {\mathcal B}$, 
define $e_i^c$ by (\ref{t:aug25c}) and let $\varepsilon_i (x) = x_{i+1}, 
~\varphi_i (x) = x_{i}$.
\end{Example}
\begin{Example}\label{t:aug26b}
For $(x,y) \in {\mathcal V} = {\mathcal B} \times {\mathcal B}$, 
define $e_i^c$ by (\ref{t:aug25d}) and let 
$\varepsilon_i(x,y) = x_{i+1}(1+y_{i+1}/x_i), ~
\varphi_i(x,y) = y_{i}(1+x_{i}/y_{i+1})$.
\end{Example}

\begin{Example}\label{t:aug26c}
For $\boldsymbol{x} = (x^{(1)}, \ldots, x^{(L)}) \in {\mathcal V} = {\mathcal B}^L$, 
define $\varepsilon_i, \varphi_i, e_i^c$ by 
\begin{align}
\label{t:aug26d}
&\varepsilon_i(\boldsymbol{x}) =
\frac{\sum_{k=1}^L \left( \prod_{j=1}^k \varepsilon_i(x^{(j)}) \right)
\left( \prod_{j=k}^{L-1} \varphi_i(x^{(j)}) \right)}
{\prod_{j=1}^{L-1} \varphi_i(x^{(j)})}, \\
\label{t:aug26e}
&\varphi_i(\boldsymbol{x}) =
\frac{\sum_{k=1}^L \left( \prod_{j=2}^k \varepsilon_i(x^{(j)}) \right)
\left( \prod_{j=k}^{L} \varphi_i(x^{(j)}) \right)}
{\prod_{j=2}^{L} \varepsilon_i(x^{(j)})},\\
\begin{split}
&e^c_i(\boldsymbol{x}) = (e^{c_1}_i(x^{(1)}),\ldots, e^{c_L}_i(x^{(L)})),\\
&\mbox{with}\quad c_l = \frac{\sum_{k=1}^L c^{\theta(k \leq l)}
\left( \prod_{j=2}^k \varepsilon_i(x^{(j)}) \right)
\left( \prod_{j=k}^{L-1} \varphi_i(x^{(j)}) \right)}
{\sum_{k=1}^L c^{\theta(k \leq l-1)}
\left( \prod_{j=2}^k \varepsilon_i(x^{(j)}) \right)
\left( \prod_{j=k}^{L-1} \varphi_i(x^{(j)}) \right)}.
\end{split}\label{t:aug26f}
\end{align}
Here $\theta(s)=1$ if $s$ is true and $=0$ otherwise.
The $\varepsilon_i, \varphi_i$ in the right hand sides are those defined in Example \ref{t:aug26a}.
\end{Example}
When $c, c' \in \mathbb{R}_{>0}$,
all the above relations in the geometric crystal for $\widehat{\mathfrak{sl}}_{n+1}$
reduces to the corresponding relations in the crystal for $\widehat{\mathfrak{sl}}_{n+1}$ in \S \ref{t:2-2-1},
via the max-plus version of the ultradiscretization.
\subsubsection{Matrix realization.}
There is a matrix realization
of the geometric crystal for $\widehat{\mathfrak{sl}}_{n+1}$, where 
each element $x \in {\mathcal B}$ is associated with the matrix 
\begin{equation}\label{t:aug26g}
M(x,\zeta) = \begin{pmatrix}
x^{-1}_1 &  & & & -\zeta \\
-1 & x^{-1}_2 & & & \\
& -1 & \ddots & & \\
& &\ddots &x^{-1}_{n} & \\
& & & -1 & x^{-1}_{n+1}
\end{pmatrix}^{-1}
\end{equation}
involving the spectral parameter $\zeta$.
The structure of $\widehat{\mathfrak{sl}}_{n+1}$ geometric crystal 
is realized as simple matrix operations.
The action of the geometric Kashiwara operator
is induced by a multiplication of (product of) $M$ with certain unipotent matrices.
For simplicity we assume $i \ne 0$ in what follows.
Let $G_i(a) = E + a E_{i,i+1}$ where $E$ is the identity matrix.
Then we have
\begin{equation}\label{t:aug26h}
G_i \left(\frac{c-1}{\varepsilon_i(x)}\right) M(x,\zeta ) 
G_i \left(\frac{c^{-1}-1}{ \varphi_i(x)} \right) = M(e^c_i(x), \zeta ),
\end{equation}
for $e^c_i(x) \in {\mathcal B}$ in (\ref{t:aug25c}).
In the same way the action of $e^c_i$ on $(x,y) \in {\mathcal B}^2$
in (\ref{t:aug25d}) is represented by
\begin{equation}
G_i \left(\frac{c-1}{\varepsilon_i(x,y)}\right) M(x,\zeta ) M(y,\zeta ) 
G_i \left(\frac{c^{-1}-1}{ \varphi_i(x,y)} \right) = M(e^{c_1}_i(x), \zeta ) M(e^{c_2}_i(y), \zeta ).
\end{equation}
%
By using the formulas in Example \ref{t:aug26c},
one can also define the action of $e^c_i$ for general multiple product case.
Through these examples, we observe that
the product of matrices 
$M(x^{(1)},\zeta) \cdots M(x^{(L)},\zeta)$ 
corresponds to the product of geometric crystals 
$(x^{(1)},\ldots, x^{(L)}) \in 
{\mathcal B}^L$.

It is easy to see that
equation (\ref{t:eq:toda}) is equivalent to the matrix equation
\begin{equation}\label{t:eq:MM}
M(x,\zeta)M(y,\zeta) = M(\tilde{y},\zeta)M(\tilde{x},\zeta).
\end{equation}
Due to the presence of the spectral parameter $\zeta$, 
its non-trivial solution is unique.
This characterizes the birational $R$ as the unique intertwiner 
(i.e. the operator that interchanges the order of product)
of the geometric crystals.
For $x = (x_1, \ldots , x_{n+1}) \in {\mathcal B}$ we set $\ell (x) = x_1 \cdots x_{n+1}$.
The uniqueness of the solution to (\ref{t:eq:MM}) can be verified as a consequence of the following:
\begin{Prop}\label{t:jun15b}
Suppose
$M(x,\zeta)M(y,\zeta) = M(x',\zeta)M(y',\zeta)$
for $\ell(x) = \ell(x') \ne  \ell(y) = \ell(y')$.
Then $x=x', y=y'$.
\end{Prop}
\proof
We define $\overline{M}(x,\zeta) = (1 - \zeta \ell(x)) M(x,\zeta)$ to avoid a singularity of $M(x,\zeta)$ at $\zeta = \ell (x) ^{-1}$.
Now the relation $\overline{M}(x,\zeta)\overline{M}(y,\zeta) = \overline{M}(x',\zeta)\overline{M}(y',\zeta)$ is 
supposed to be satisfied
under the condition $\ell(x) = \ell(x') \ne  \ell(y) = \ell(y')$. 
It is easy to see that the matrix elements of $\overline{M}(x,\zeta)$ are given by
\begin{equation}
\overline{M}(x,\zeta)_{ij} =
\begin{cases}
\prod_{k=j}^i x_k & \mbox{for $i \geq j$,} \\
\zeta \left( \prod_{k=1}^{i} x_k \right) \left( \prod_{k=j}^{n+1} x_k \right)  & \mbox{for $i < j$.}
\end{cases}
\end{equation}
Note that the rank of the matrix $\overline{M}(x,\zeta)$ reduces to one when $\zeta = \ell (x)^{-1}$ or more precisely
$\overline{M}(x,\ell (x)^{-1})_{ij} = \ell (x)^{-1} \left( \prod_{k=1}^{i} x_k \right) \left( \prod_{k=j}^{n+1} x_k \right)$
for any $i,j$.
Thus the relation $\overline{M}(x,\zeta)\overline{M}(y,\zeta) = \overline{M}(x',\zeta)\overline{M}(y',\zeta)$
at $\zeta = \ell (x)^{-1}$ yields the condition
$\left( \prod_{k=1}^{i} x_k \right) = \alpha \left( \prod_{k=1}^{i} x'_k \right)$
for any $i$ and with some constant $\alpha$, forcing $x_i = x'_i$ for any $i$.
In the same way we obtain $y=y'$ by taking $\zeta = \ell (y)^{-1}$ in the relation.
\qed

Now we show that the birational $R$ is the intertwiner 
of the geometric crystals.
\begin{Prop}\label{t:aug16b}
$R e^c_i = e^c_i R$.
\end{Prop}
\proof
Let $R(x,y) = (\tilde{y},\tilde{x})$, $e^c_i R(x,y) = ((\tilde{y})',(\tilde{x})')$,
$e^c_i(x,y) =(x',y')$ and $R e^c_i(x,y) = (\widetilde{y'},\widetilde{x'})$.
Then we have
\begin{eqnarray*}
M(\widetilde{y'}, \zeta ) M(\widetilde{x'}, \zeta ) &=&
M(x', \zeta ) M(y', \zeta ) \\
&=& G_i \left(\frac{c-1}{\varepsilon_i(x,y)}\right) M(x,\zeta ) M(y,\zeta ) 
G_i \left(\frac{c^{-1}-1}{ \varphi_i(x,y)} \right) \\
&=& G_i \left(\frac{c-1}{\varepsilon_i(\tilde{y},\tilde{x})}\right) M(\tilde{y},\zeta ) M(\tilde{x},\zeta ) 
G_i \left(\frac{c^{-1}-1}{ \varphi_i(\tilde{y},\tilde{x})} \right) \\
&=& M((\tilde{y})', \zeta ) M((\tilde{x})', \zeta ).
\end{eqnarray*}
Here we used $\varepsilon_i(\tilde{y},\tilde{x}) = \varepsilon_i(x,y)$ and
$\varphi_i(\tilde{y},\tilde{x}) = \varphi_i(x,y)$ which are verified by (\ref{t:eq:toda}).
By Proposition \ref{t:jun15b} we have
$\widetilde{y'} = (\tilde{y})', \widetilde{x'} = (\tilde{x})'$.
\qed

We show that the birational $R$ satisfies the Yang-Baxter equation.
\begin{Prop}{\rm \cite[Th.2.2]{Y01}}\label{t:jun29b}
The birational $R$ satisfies (\ref{t:YBtropR}).
\end{Prop}
\proof
Let $R_1 R_2 R_1 (x,y,z)=(z',y',x')$ and $R_2 R_1 R_2 (x,y,z)=(z'',y'',x'')$.
By Proposition \ref{t:jun29a} and
since (\ref{t:eq:toda}) is equivalent to (\ref{t:eq:MM}) we have
\begin{equation}\label{t:jun29c}
M(z',\zeta) M(y',\zeta) M(x',\zeta) = M(x,\zeta) M(y,\zeta) M(z,\zeta) = M(z'',\zeta) M(y'',\zeta) M(x'',\zeta).
\end{equation}
By an obvious extension of Proposition \ref{t:jun15b} this leads to $x'=x'', y'=y'', z'=z''$.
\qed

\subsubsection{Bilinearization.}
The birational $R$ is equivalent to a system of 
bilinear difference equations of Hirota type \cite{KOTY04}.
To see this, introduce the functions 
$\tau^J_i \, (1 \le J \le 4, i \in \Z_{n+1})$ and 
the parameters $\lambda_i, \kappa_i$, and make the change of variables
\begin{equation}\label{t:aug5b}
\begin{split}
&x^{-1}_i = \lambda_i\delta\tau^3_i/\delta\tau^2_i,\quad
y^{-1}_i = \kappa_i\delta\tau^2_i/\delta\tau^1_i,\\
&{\tilde{y}}^{-1}_i = \kappa_i\delta\tau^3_i/\delta\tau^4_i,\quad
{\tilde{x}}^{-1}_i = \lambda_i\delta\tau^4_i/\delta\tau^1_i,
\end{split}
\end{equation}
with $\delta\tau^J_i = \tau^J_i/\tau^J_{i-1}$.
In order to memorize the relations (\ref{t:aug5b})
it is useful to draw the following vertex diagram 
and regard the tau functions as residing in the quadrants.
\small
\begin{equation}\label{t:aug5a}
\setlength{\unitlength}{3mm}
\begin{picture}(9,9)
\put(1,4){\line(1,0){6}}
\put(4,1){\line(0,1){6}}
\put(0.2,3.8){$x$}
\put(7.3,3.8){$\tilde{x}$}
\put(3.8,7.5){$y$}
\put(3.8,0){$\tilde{y}$}
\put(5,5){$\tau^1$}
\put(2,5){$\tau^2$}
\put(2,2){$\tau^3$}
\put(5,2){$\tau^4$}
\end{picture}
\end{equation}
\normalsize
Then the former relation in (\ref{t:eq:toda}) is automatically satisfied
and the latter is translated into 
\begin{equation}\label{t:eq:abl}
\lambda_i\tau^2_{i-1}\tau^4_i - \kappa_i\tau^2_i\tau^4_{i-1} 
= \alpha\tau^1_i\tau^3_{i-1}
\end{equation}
for any nonzero parameter $\alpha$ independent of $i$.
The birational map $R:(x,y) \mapsto (\tilde{y},\tilde{x})$ is induced by 
an automorphism  $\tau^2_i \leftrightarrow \tau^4_i$, 
$\lambda_i \leftrightarrow \kappa_i$, $\alpha \rightarrow -\alpha$
of (\ref{t:eq:abl}).
Eq. (\ref{t:eq:abl}) is a version of so-called 
Hirota-Miwa (non-autonomous discrete KP) equation.

\subsection{General solution}\label{subsec:gsol}

Recall that the KKR map $\phi^{-1}$ (\S \ref{subsec:kkr})
transforms a rigged configuration into a highest path.
It turns out that its image allows 
an explicit formula in terms of {\em ultradiscrete tau functions}.
In view of the remarks after \eqref{k:elm},  
this yields the {\em general} solution of the BBS time evolution 
equation (\ref{i:rho-bilinear}) corresponding to an arbitrary initial condition.

To formulate the ultradiscrete tau function, it  is convenient 
to regard a rigged configuration (\S \ref{subsec:kkr}) as 
a {\em multiset} i.e. a set with multiplicity of each element taken into account
\begin{equation}\label{k:S}
S = \{(a_i,l_i,J_i)\in \{1,2,\ldots, n\}\times 
\Z_{\ge 1}\times \Z_{\ge 0}
\mid i=1,2,\ldots, N\},
\end{equation}
where $N\ge 0$ is arbitrary and 
each triplet $s=(a,l,r)$ signifies a string having  
color $a$,  length $l$ and rigging $r$.
This fact will be 
denoted by ${\rm cl}(s)=a, {\rm lg}(s)=l$
and ${\rm rg}(s)=r$\footnote{Colors 
$1,2,\ldots, n$ of strings in rigged configurations should not be confused with 
the colors of balls in BBS.}.
$S$ is a rigged configuration if 
${\rm rg}(s) \le p^{({\rm cl}(s))}_{{\rm lg}(s)}$
is satisfied for all $s\in S$,
where $p^{(a)}_j$ is defined in (\ref{k:paj}). 

For a rigged configuration $S$ (\ref{k:S}), 
let $T \subseteq S$ be a (possibly empty) subset of $S$.
We allow the fact that $T$ is no longer a rigged configuration in general.
Introduce the piecewise-linear functions $c_{k,a}(T)$
($0 \le k \le L$ and $1 \le a \le n+1$) by
\begin{align}
c(T) &= \frac{1}{2}\sum_{s,t \in T}C_{{\rm cl}(s),{\rm cl}(t)}
\min({\rm lg}(s), {\rm lg}(t)) 
+ \sum_{s\in T}{\rm rg}(s), \label{k:ct}\\
c_{k,a}(T) &= c(T)
+\sum_{s\in T, {\rm cl}(s)=a}{\rm lg}(s)
-k\sum_{s\in T, {\rm cl}(s)=1}1,\label{k:cka}
\end{align}
where $C_{a,b}=2\delta_{a,b}-\delta_{|a-b|,1}$
is an element of Cartan matrix $(C_{a,b})_{1 \le a,b\le n}$ of $\mathfrak{sl}_{n+1}$. 
($L$ will be the length of the corresponding path.)
By the definition, the second term in $c_{k,a}(T)$  is 
$0$ when $a=n+1$.
Obviously we have 
$c(\emptyset) = c_{k,a}(\emptyset) = 0$.
The quantity $c(S)$ is known as the {\em cocharge} of
the rigged configuration $S$ \cite{KR}.
The ultradiscrete tau function is a 
$\Z_{\ge 0}$-valued piecewise-linear function 
$\tau_{k,a}=\tau_{k,a}(S)$ on $S$
defined by ($0 \le k \le L$)
\begin{equation}\label{k:ttau}
\begin{split}
\tau_{k,a}&= -\min_{T \subseteq S}
\left(c_{k,a}(T)\right)
\quad (1 \le a \le n+1),\\
\tau_{k,0}&= \tau_{k,n+1} -k.
\end{split}
\end{equation}
\begin{Example}\label{k:ex:tau}
$\tau_{0,n+1}=-\min_{T \subseteq S}(c(T))$ 
for $S$ (\ref{k:S}) with $N=1,2,3$ is given by
\begin{align*}
\tau_{0,n+1} &= -\min(0,\xi_1),\\
\tau_{0,n+1} &= -\min(0,\xi_1, \xi_2, \xi_1+\xi_2+A_{1,2}),\\
\tau_{0,n+1} &= -\min(0,\xi_1, \xi_2, \xi_3, \xi_1+\xi_2+A_{1,2},
\xi_1+\xi_3+A_{1,3,} \xi_2+\xi_3+A_{2,3},\\
&\qquad \qquad \quad \xi_1+\xi_2+\xi_3+A_{1,2}+A_{1,3}+A_{2,3}),
\end{align*}
where we have used the shorthand
$\xi_i = l_i+J_i$ and $A_{i,i'}=C_{a_i,a_{i'}}\min(l_i, l_{i'})$.
\end{Example}

In general, 
the minimum (\ref{k:ttau}) 
for $S$ (\ref{k:S}) extends over $2^N$ 
candidates and reminds us of the structure of 
tau functions in the theory of solitons \cite{MJD00}.
In fact, (\ref{k:ttau}) can be deduced from the tau functions in 
the discrete KP hierarchy by ultradiscretization with 
an elaborate turning of parameters between 
KP solitons and rigged configurations \cite[sec.~5]{KSY}.

In \S \ref{subsec:ism}, we have seen that 
rigged configurations undergo linear time evolution (\ref{k:rr}).
In the present notation, it is rephrased as 
\begin{equation}\label{k:rte}
S=\{(a_i,l_i, J_i)\} \overset{T_l}{\longmapsto}
T_l(S):=
\{(a_i, l_i, J_i+\delta_{1, a_i}\min(l,l_i))\}.
\end{equation}

\begin{Theorem}\label{k:th:ksy}
Let $b_1\otimes \cdots \otimes b_L =\phi^{-1}(S)$ 
be the image (highest path) of a rigged configuration $S$
under the KKR map $\phi^{-1}$.
\\
(i) {\rm \cite[Th.~2.1]{KSY}}
\;$b_k = (x_{k,1},\ldots, x_{k,n+1}) \in B_1$ $(\ref{t:jun24a})|_{l=1}$  is 
expressed as
\begin{equation}\label{k:xta}
x_{k,a} = \tau_{k,a}- \tau_{k-1,a}- \tau_{k,a-1}+ \tau_{k-1,a-1}.
\end{equation}

\noindent
(ii) {\rm \cite[Prop.~5.1]{KSY}} 
Denote by $\overline{\tau}_{k,a}$ the ultradiscrete tau function 
associated with $T_\infty(S)$ defined by (\ref{k:rte}).
(Thus $\tau_{k,1} = \overline{\tau}_{k,n+1}$.)
Then the following ultradiscrete Hirota-Miwa equation is satisfied. 
\begin{equation}\label{k:bi}
\overline{\tau}_{k,a-1}+\tau_{k-1,a} = 
\max(\overline{\tau}_{k,d}+\tau_{k-1,a-1},\;
\overline{\tau}_{k-1,a-1}+\tau_{k,a}-1)\quad
(2 \le a \le n+1).
\end{equation}

\noindent
(iii) {\rm \cite[Th.~4.9]{KSY}} 
Define $p=\cdots \otimes x^0_k \otimes x^0_{k+1} \otimes \cdots$ 
by $x^0_k = b_k$ if $1 \le k \le L$ 
and $x^0_k = \fbox{\rm 1}$ otherwise.
Let  $\rho^0_{k,a}$ 
be the number of balls specified from $p$ as in (\ref{i:rho}).
Then 
$\tau_{k,a} = \rho^0_{k,a}$ holds for $1\le k \le L, 1\le a \le n+1$.
\end{Theorem}

Theorem \ref{k:th:ksy} is known to hold also for 
extended rigged configurations (Remark \ref{k:nonh}) 
and non highest paths \cite[sec.7]{KSY}.
In view of the inverse scattering method (\S \ref{subsec:ism}),
it provides the explicit piecewise-linear formula 
describing BBS under any time evolution. 

\begin{Remark}
Let $N_a$ be the number of strings in a rigged configuration 
$S$ having color $a$.
The soliton/string correspondence (\ref{k:ss})
tells that $S$ describes the $N_1$-soliton states of BBS.
On the other hand,  $\tau_{k,a}(S)$ is an ultradiscretization of an 
$N = N_1+\cdots + N_n$-soliton solution of 
the discrete KP equation \cite{KSY}.
For $n>1$, the ``extra" $N_2+\cdots + N_n$ solitons in KP
specify the internal labels of the 
BBS solitons. 
\end{Remark}

The cocharge mentioned under (\ref{k:cka})
is related to the {\em energy} of a path, which involves 
the energy function $H$ (\ref{t:aug1a}) as a building block.
See for example \cite{NY97, O07, S07}.
In this context, ultradiscrete tau functions are 
combinatorial analogues of corner transfer matrices 
in solvable lattice models \cite{B},
and (\ref{k:xta}) is regarded as the formula for a ``one-point function".

These features and the insights gained in 
\S \ref{subsec:ism} are summarized in the following table.
One can compare the format of the solutions of BBS
coming from the two basic tools in quantum integrable systems, 
Bethe ansatz and corner transfer matrices.

\begin{table}[h]
\begin{center}
\begin{tabular}{c|c|c}
\hfil  & Bethe roots &  Corner transfer matrix \\
\hline 
&&\vspace{-0.5cm}\\
Combinatorial analogue
& rigged configuration & 
energy in crystal \\
&& \vspace{-0.5cm} \\
\hline
&& \vspace{-0.5cm}\\
Role in BBS
& action-angle variable 
& tau function\\
&& \vspace{-0.5cm} \\
\hline
&& \vspace{-0.5cm}\\
Dynamics & linear & bilinear
\end{tabular}
\end{center}
\end{table}


\section{Periodic BBS}

\subsection{Basic features}
\label{t:5-1}

In this section we restrict ourselves to $\widehat{\mathfrak{sl}}_2$ case
and consider the box-ball system with the periodic boundary condition, 
which we call periodic BBS for short \cite{KTT06, MIT06, YYT, YT02}.
An example of the time evolutions of this system appeared in \S 1.
Compared with the infinite system, there are many interesting features in periodic BBS
which come from the finiteness of its phase space.  
For an attempt to generalize our formalism to the case of
periodic $\widehat{\mathfrak{sl}}_{n+1}$ BBS, see \cite{KT10}.

Let us recall the formalism in \S \ref{t:2-3} which is based on the crystal base theory.
In the case of $\widehat{\mathfrak{sl}}_2$, 
the vertex diagrams for combinatorial $R$ (\ref{t:july15a}) 
on $B_l \otimes B_1$ look like those in (\ref{k:wtm1}).  
Let $L$ be the system size and $M \, (\leq \! L/2)$ be the number of balls.
Consider the diagram (\ref{t:apr20}) with not necessarily large $L$.
Let
\begin{equation}\label{t:july20g}
{\mathcal P}_L = \{ b_1 \otimes \cdots \otimes b_L \in (B_1)^{\otimes L}
 | \# \{ i | b_i = \onebox{2}\,\} \leq L/2 \}.
\end{equation}
To attain the periodic boundary condition we want to find $v' \in B_l$ such that $v' = v$.
See (\ref{k:hatten}) for an example.
In general $v'$ is a function of $v \in B_l$ and $p:=b_1 \otimes \cdots \otimes b_L \in (B_1)^{\otimes L}$.
Hence we can denote it by $v' = v'(v,p)$.
\begin{Prop}{\rm \cite[Proposition 2.1]{KTT06}}\label{t:july15d}
For any $p \in {\mathcal P}_L$ under the condition $\# \{ i | b_i = 2\} <L/2$,
the solution $v \in B_l$ to the equation $v'(v,p) = v$ is unique and is given by
$v = v'(u_l,p)$,
where $u_l = \threeboxes{1}{\scriptstyle \cdots}{1} \in B_l$.
\end{Prop}
Let $v_l (p) = v'(u_l,p)$.
By setting $v = v' = v_l(p)$ in (\ref{t:apr20}) we define the time evolution operator $T_l$ by the relation (\ref{t:t}). 
That is, we have
\begin{equation}\label{t:july25a}
v_l (p) \otimes p \simeq T_l (p) \otimes v_l (p),
\end{equation}
as elements of $B_l \otimes (B_1)^{\otimes L} \simeq (B_1)^{\otimes L} \otimes B_l$.
We note that (\ref{t:july25a}) is a periodic version of the Lax equation (\ref{t:aug31a}).
In particular $T_1$ yields a cyclic shift by one unit cell to the right.
The evolution by $T_\infty$ admits the description without carrier 
given in \S 1,
which is  also equivalent to the ``arc rule" in \cite{YYT}.
\begin{Remark}\label{t:aug3e}
In \S \ref{t:5-1} we restrict ourselves to the case $M < L/2$ for simplicity.
However, our formalism of the periodic BBS
based on the crystal base theory also enables one to treat the case $M \geq L/2$ \cite{KTT06}.
\end{Remark}

The energy associated with  $T_l$ is defined by (\ref{t:e}) with $v_0 = v_l (p)$,
where the values of the energy function are given by $H=0$ for the bottom right diagram in (\ref{k:wtm1})
and $H=1$ otherwise.

In what follows, we write 
for example $\onebox{1}\otimes \onebox{2}\otimes \onebox{1}\otimes \onebox{2}$
simply as $1212$.

\begin{Example}\label{t:july15c}
The time evolutions of $p=222111211111$ by $T_l$ with 
$l \geq 3, T_2$ and $T_1$:
\par\noindent
\begin{verbatim}
      t=0  222...2.....  |  222...2.....  |  222...2.....
      t=1  ...222.2....  |  ..222..2....  |  .222...2....
      t=2  ......2.222.  |  ....222.2...  |  ..222...2...
      t=3  22.....2...2  |  ......22.22.  |  ...222...2..
      t=4  ..222...2...  |  2.......2.22  |  ....222...2.
      t=5  .....222.2..  |  222......2..  |  .....222...2
      t=6  2.......2.22  |  ..222.....2.  |  2.....222...
\end{verbatim}
\end{Example}

\begin{Theorem}{\rm \cite[Th.2.2]{KTT06}}\label{t:july15f}
The commutativity $T_lT_k(p) = T_kT_l(p)$, and the 
conservation of the energy $E_l (T_k(p)) = E_l(p)$ hold.
\end{Theorem}
\proof
Let $R ({v}_{k}(T_l(p)) \otimes {v}_{l}(p)) = \overline{v_{l}(p)} \otimes \overline{v_{k}(T_l(p))}$. 
See the following diagram

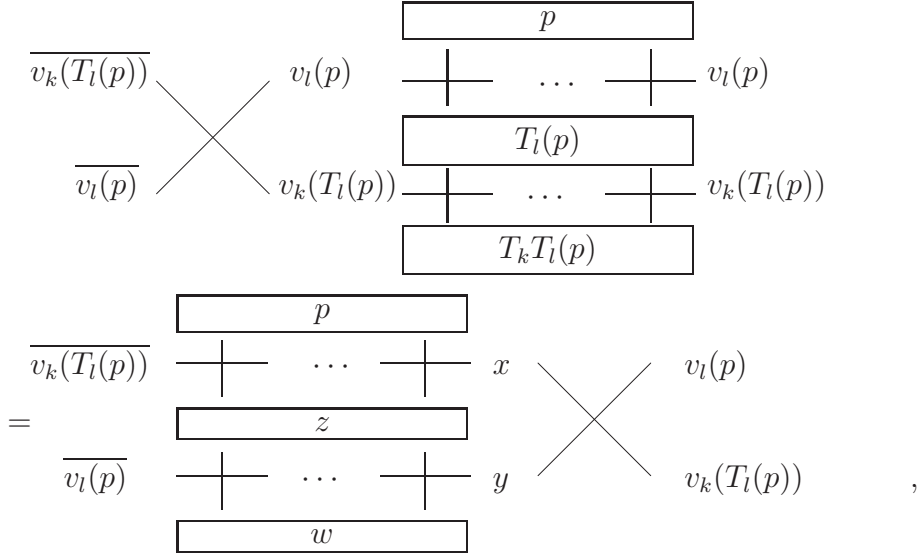
\setlength{\unitlength}{1.5mm}
\begin{picture}(80,55)(-20,0)

\put(0,10){\line(1,0){8}}
\put(0,20){\line(1,0){8}}

\put(11,9){$\cdots$} 
\put(12,19){$\cdots$}

\put(18,10){\line(1,0){8}}
\put(18,20){\line(1,0){8}}

\put(4,7){\line(0,1){5}}
\put(22,7){\line(0,1){5}}

\put(4,17){\line(0,1){5}}
\put(22,17){\line(0,1){5}}

\put(-13,19){$\overline{v_{k}(T_l(p))}$}
\put(-10,9){$\overline{v_{l}(p)}$}

\put(0,24){\framebox[110pt][c]{$p$}}
\put(0,14){\framebox[110pt][c]{$z$}}
\put(0,4){\framebox[110pt][c]{$w$}}

\put(28,9){$y$}
\put(28,19){$x$}

\put(32,10){\line(1,1){10}}
\put(32,20){\line(1,-1){10}}

\put(45,9){${v}_{k}(T_l(p))$}
\put(45,19){${v}_{l}(p)$}

\put(-15,14){$=$}

\put(-1.8,35){\line(1,1){10}}
\put(-1.8,45){\line(1,-1){10}}

\put(-9,35){$\overline{v_{l}(p)}$}
\put(-13,45){$\overline{v_{k}(T_l(p))}$}

\put(9,35){${v}_{k}(T_l(p))$}
\put(10,45){${v}_{l}(p)$}

\put(20,35){\line(1,0){8}}
\put(20,45){\line(1,0){8}}

\put(31,34){$\cdots$} 
\put(32,44){$\cdots$}

\put(38,35){\line(1,0){8}}
\put(38,45){\line(1,0){8}}

\put(24,32.5){\line(0,1){4.7}}
\put(42,32.5){\line(0,1){4.7}}

\put(24,42.8){\line(0,1){5}}
\put(42,42.8){\line(0,1){5}}

\put(20,50){\framebox[110pt][c]{$p$}}
\put(20,39){\framebox[110pt][c]{$T_{l}(p)$}}
\put(20,29.3){\framebox[110pt][c]{$T_{k}T_{l}(p)$}}

\put(47,35){${v}_{k}(T_l(p))$}
\put(47,45){${v}_{l}(p)$}

\put(65,9){,}

\end{picture}

\noindent
where the Yang-Baxter relation is used
to move the symbol ``$\times$'' for the $R$ from the left to the right.
Consider what $x,y,z$ and $w$ should be.
Since $R ({v}_{k}(T_l(p)) \otimes {v}_{l}(p)) = \overline{v_{l}(p)} \otimes \overline{v_{k}(T_l(p))}$
we have $x = \overline{v_{k}(T_l(p))}$ and $y = \overline{v_{l}(p)}$.
Hence by Proposition \ref{t:july15d} the equality $\overline{v_{k}(T_l(p))} = {v}_{k}(p)$ holds.
Thus $z = T_k (p)$.
Then again by Proposition \ref{t:july15d} the equality $\overline{v_{l}(p) }= {v}_{l}(T_k(p))$ holds, and we have
$w = T_l (T_k(p))$.
Hence $T_k (T_l(p)) = T_l (T_k(p))$.
For a proof of the conservation of the energy, see \cite[Th.2.2]{KTT06}.
\qed

\subsection{Linearization and general solution}
\label{subsec:general-sol}

Here we construct action-angle variables of the periodic BBS,
solve the initial value problem and present an explicit formula 
for $N$-soliton solutions in terms of tropical Riemann theta function.
These results were firstly obtained in \cite{KS06,KTT06}.

\subsubsection{Action variable.}\label{k:ss:act}
We are going to introduce 
the action variable of a state.
It is equivalent to the list of 
amplitudes of solitons contained in a state, which is the conserved quantity.
Recall that a state 
$p= b_1 b_2 \ldots  b_L\,(b_j =1,2)$  
is highest if the condition (\ref{k:hp1}) is satisfied.
It is elementary to show that any state in ${\mathcal P}_L$
can be made highest by a cyclic shift. 
Namely, there exist a highest state $p_+ \in {\mathcal P}_L$ 
and $d\in \Z$ such that
$p = T_1^d(p_+)$.
Given $p$, such a pair $(d, p_+)$ is not necessarily unique.
Nevertheless one can show 
\begin{Prop}{\rm \cite[Proposition 3.3]{KTT06}}
(i) Let $\mu$ be the configuration of the rigged configuration
$\phi(p_+)$. 
(Namely, $\mu$ is the Young diagram denoted by $\mu^{(1)}$ in \S \ref{subsec:kkr}.)
Then $\mu$ is independent of the not necessarily unique choice  
of $(d, p_+)$.
\\
(ii) The energy $E_l$ of $p$ is related to $\mu$ via
\begin{equation}\label{k:mue}
E_l(p) = \text{number of cells in the left $l$ columns of } \mu.
\end{equation} 
\end{Prop}

Due to (i), the Young diagram $\mu$ 
is uniquely determined from a state $p$.
We denote it by $\mu_\ast(p)$ and call it {\it action variable} of $p$.
Due to (ii), it is a conserved quantity, namely, 
$\mu_\ast(T_l(p)) = \mu_\ast(p)$ holds for any $l$.
Let $m_k=m_k(p)$ be 
the number of length $k$ rows in the Young diagram $\mu_\ast(p)$. 
Then (\ref{k:mue}) is rephrased by the same formula as (\ref{t:jun10a}):
\begin{equation}\label{k:ejp}
E_l(p) = \sum_{k\ge 1}\min(l,k)m_k.
\end{equation}
In the context of the KKR bijection, $m_k$ is the number of 
length $k$ strings.
On the other hand, $m_k$ is the number of 
amplitude $k$ solitons contained in $p$.
This is another manifestation of the {\em soliton/string correspondence},
which was observed earlier also in BBS on infinite lattice in (\ref{k:ss}).
We introduce the {\it isolevel set} of states characterized by 
the action variable
\begin{equation}\label{k:ils}
{\mathcal P}_L(\mu) = \{p \in {\mathcal P}_L \mid \mu_\ast(p) = \mu\}
\end{equation}
for any Young diagram $\mu$ such that 
$|\mu| \le L/2$.

\begin{Example}\label{k:ex:pp}
Take 
$p=2211221112122111221 \in {\mathcal P}_{19}$.
It can be expressed as cyclic shifts of highest paths as 
$p=T^2_1(p_+)=T^6_1(p'_+)=T^{13}_1(p''_+)$, where 
$p_+ = 1122111212211122122$,
$p'_+=1112122111221221122$ and 
$p''_+ = 1112212211221112122$.
Their image by the KKR map $\phi$ is given by
\begin{equation*}
\begin{picture}(250,11)(-50,-11)
\unitlength 0.34mm
\multiput(-120,0)(0,0){1}{
\put(-42,-25){$p_+ \,\overset{\phi}{\longmapsto}$}
\put(0,0){\line(0,-1){50}}
\put(10,0){\line(0,-1){50}}
\put(20,0){\line(0,-1){30}}
\put(30,0){\line(0,-1){10}}

\put(0,0){\line(1,0){30}}\put(33,-9){\rm 1}
\put(0,-10){\line(1,0){30}}\put(23,-19.7){\rm 1}
\put(0,-20){\line(1,0){20}}\put(23,-29.7){\rm 0}
\put(0,-30){\line(1,0){20}}\put(13,-39.7){\rm 8}
\put(0,-40){\line(1,0){10}}\put(13,-49.7){\rm 4}
\put(0,-50){\line(1,0){10}}}

\multiput(-10,0)(0,0){1}{
\put(-45,-25){$p'_+ \,\overset{\phi}{\longmapsto}$}
\put(0,0){\line(0,-1){50}}
\put(10,0){\line(0,-1){50}}
\put(20,0){\line(0,-1){30}}
\put(30,0){\line(0,-1){10}}

\put(0,0){\line(1,0){30}}\put(33,-9){\rm 1}
\put(0,-10){\line(1,0){30}}\put(23,-19.7){\rm 3}
\put(0,-20){\line(1,0){20}}\put(23,-29.7){\rm 1}
\put(0,-30){\line(1,0){20}}\put(13,-39.7){\rm 6}
\put(0,-40){\line(1,0){10}}\put(13,-49.7){\rm 2}
\put(0,-50){\line(1,0){10}}}

\multiput(92,0)(0,0){1}{
\put(-45,-25){$p''_+ \,\overset{\phi}{\longmapsto}$}
\put(0,0){\line(0,-1){50}}
\put(10,0){\line(0,-1){50}}
\put(20,0){\line(0,-1){30}}
\put(30,0){\line(0,-1){10}}

\put(0,0){\line(1,0){30}}\put(33,-9){\rm 0}
\put(0,-10){\line(1,0){30}}\put(23,-19.7){\rm 3}
\put(0,-20){\line(1,0){20}}\put(23,-29.7){\rm 2}
\put(0,-30){\line(1,0){20}}\put(13,-39.7){\rm 8}
\put(0,-40){\line(1,0){10}}\put(13,-49.7){\rm 3}
\put(0,-50){\line(1,0){10}}
\put(40,-50){.}}

\end{picture}
\end{equation*} 
They all lead to $\mu_\ast(p)=(3,2,2,1,1)$.
\end{Example}
\begin{Example}\label{k:ex:ils} The isolevel sets ${\mathcal P}_6(\mu)$ with 
$|\mu |= 3$ are given by
\begin{equation}\label{k:pex}
\begin{split}
{\mathcal P}_6\bigl((1,1,1)\bigr)&=\{121212, 212121\},\\
{\mathcal P}_6((3))
&=\{111222, 211122, 221112, 222111, 122211, 112221\},\\
{\mathcal P}_6\bigl((2,1)\bigr)&=
\{121122, 212112, 221211, 122121, 112212, 211221,\\
&\quad\;\;\, 112122, 211212, 221121, 122112, 212211, 121221\}.
\end{split}
\end{equation}
\end{Example}

\subsubsection{Angle variable.}\label{k:ss:ang}
Let us observe Example \ref{k:ex:pp}.
Recall that in the infinite system, the time evolution $T_1$ is 
the uniform shift of all the riggings by $1$. See (\ref{k:rr}).
If we adopt the same feature also in the periodic BBS,
the following identification should be made for the 
``periodic version" of the rigged configuration:

\begin{equation}\label{k:mit}
\begin{picture}(250,11)(-50,-11)
\unitlength 0.34mm
\multiput(-100,0)(0,0){1}{

\put(-9,-8){1}\put(-9,-25){3}\put(-9,-45){9}
\put(0,0){\line(0,-1){50}}
\put(10,0){\line(0,-1){50}}
\put(20,0){\line(0,-1){30}}
\put(30,0){\line(0,-1){10}}

\put(0,0){\line(1,0){30}}\put(33,-9){3}
\put(0,-10){\line(1,0){30}}\put(23,-19.7){3}
\put(0,-20){\line(1,0){20}}\put(23,-29.7){2}
\put(0,-30){\line(1,0){20}}\put(13,-39.7){10}
\put(0,-40){\line(1,0){10}}\put(13,-49.7){6}
\put(0,-50){\line(1,0){10}}}

\put(-45,-25){$\equiv$}

\multiput(-10,0)(0,0){1}{
\put(-9,-8){1}\put(-9,-25){3}\put(-9,-45){9}
\put(0,0){\line(0,-1){50}}
\put(10,0){\line(0,-1){50}}
\put(20,0){\line(0,-1){30}}
\put(30,0){\line(0,-1){10}}

\put(0,0){\line(1,0){30}}\put(33,-9){7}
\put(0,-10){\line(1,0){30}}\put(23,-19.7){9}
\put(0,-20){\line(1,0){20}}\put(23,-29.7){7}
\put(0,-30){\line(1,0){20}}\put(13,-39.7){12}
\put(0,-40){\line(1,0){10}}\put(13,-49.7){8}
\put(0,-50){\line(1,0){10}}}

\put(45,-25){$\equiv$}

\multiput(80,0)(0,0){1}{
\put(-9,-8){1}\put(-9,-25){3}\put(-9,-45){9}
\put(0,0){\line(0,-1){50}}
\put(10,0){\line(0,-1){50}}
\put(20,0){\line(0,-1){30}}
\put(30,0){\line(0,-1){10}}

\put(0,0){\line(1,0){30}}\put(33,-9){13}
\put(0,-10){\line(1,0){30}}\put(23,-19.7){16}
\put(0,-20){\line(1,0){20}}\put(23,-29.7){15}
\put(0,-30){\line(1,0){20}}\put(13,-39.7){21}
\put(0,-40){\line(1,0){10}}\put(13,-49.7){16}
\put(0,-50){\line(1,0){10}}
\put(50,-50){.}}

\end{picture}
\end{equation} 
Here we have attached the vacancy $p_i$ on the left of the block of 
length $i$ strings.
The riggings are no longer bounded by it.
The basic idea in constructing angle variables is to introduce an appropriate 
equivalence relation among such extended riggings.

We proceed to the precise definition.
Consider the isolevel set ${\mathcal P}_L(\mu)$ with 
$\mu = (i_g^{m_{i_g}} \ldots i_1^{m_{i_1}})$.
Here  $i_1<\cdots < i_g$ are the length of the rows in $\mu$ and 
$m_{i_j}$ is the multiplicity of $i_j$\footnote{
The $i_1,\ldots, i_g$ here and in \S \ref{i:pBBS-Toda}
will denote the amplitude of solitons.
They should not be confused with the ones in \S \ref{sec:baa} like
$p=i_1\cdots i_L$ or 
in (\ref{k:hp1}) and (\ref{k:hp3}). }.
For instance $\mu=(3^12^21^2)$  in (\ref{k:mit}).
We set 
\begin{equation}\label{k:ipj}
{\mathcal I}=\{i_1<\cdots < i_g\},\quad 
p_j = L-2\sum_{i \in {\mathcal I}}\min(j,i)m_i\quad (j \in \Z_{\ge 0}),
\end{equation}
where the latter is the vacancy $p^{(1)}_j$ (\ref{k:paj}) with $n=1$.
Recall that a rigged configuration 
$(\mu, J)_L$ is the data in which the vicinity of the block of length $i$ strings looks as
\begin{equation}\label{k:zurg}
\begin{picture}(50,20)(0,1)
\unitlength 0.38mm
\put(-30,5){\line(0,1){70}}

\put(-13,40){\vector(-1,0){15}} 
\put(-7,37){$i$}
\put(2,40){\vector(1,0){15}}

\put(-50,37){$p_i$}

\put(40,60){\line(0,1){15}}
\put(44,70){$\cdot$}\put(44,66){$\cdot$}\put(44,62){$\cdot$}
\put(20,60){\line(1,0){20}}
\put(20,20){\line(0,1){40}}
\put(23,50){$\scriptstyle{J_{i, m_i}}$}
\put(27,40){$\cdot$}\put(27,36){$\cdot$}\put(27,32){$\cdot$}
\put(24,23){$\scriptstyle{J_{i,1}}$}
\put(0,20){\line(1,0){20}}
\put(4,12){$\cdot$}\put(4,8){$\cdot$}\put(4,4){$\cdot$}
\put(0,5){\line(0,1){15}}

\put(-5,3){
\put(92,42){$J=(J_{i,\alpha}),\;\;\; i \in {\mathcal I},\;\;
1 \le \alpha \le m_i,$}
\put(90,22){$0\le J_{i,1}\le \cdots \le J_{i,m_i}\le p_i$.}
}
\end{picture}
\end{equation}
We extend the integer sequence 
$J_{i,\alpha}$ from $1 \le \alpha \le m_i$ to ${\alpha \in \Z}$ 
by imposing the {\it quasi-periodicity} as
\begin{equation}\label{k:quasi}
J_{i,\alpha+m_i}= J_{i,\alpha}+p_i\quad (\alpha \in \Z).
\end{equation}
The resulting sequence will be denoted by 
${\bf J} = (J_{i,\alpha})_{(i\alpha)\in {\mathcal I}\times \Z}$
and called a {\it quasi-periodic extension} of the rigging $J$.
(Indices will be suppressed as ${\bf J} = (J_{i,\alpha})$.)
By the definition, ${\bf J}$ ranges over the set
\begin{align}
\tilde{\mathcal J}_L(\mu) &=
\prod_{i \in {\mathcal I}}{\tilde \Lambda}(m_i,p_i),\label{k:jtm0}\\
\tilde{\Lambda}(m, p) &=
\{ (\lambda_\alpha)_{\alpha \in \Z} \mid 
\lambda_\alpha \in \Z, \,
\lambda_\alpha \le \lambda_{\alpha+1}, \, 
\lambda_{\alpha+m} = \lambda_\alpha + p \;
(\forall\alpha) \},\label{k:lamti}
\end{align}
where $L$-dependence enter (\ref{k:jtm0}) via $p_i$ (\ref{k:ipj}).

Now we introduce the equivalence relation on 
$\tilde{\mathcal J}_L(\mu)$.
For $k \in {\mathcal I}$, define $\sigma_k$ by
\begin{equation}\label{k:slide}
\sigma_k: \tilde{\mathcal J}_L(\mu)\rightarrow \tilde{\mathcal J}_L(\mu);
\quad
(J_{i,\alpha})\mapsto (J_{i, \alpha+\delta_{i k}}+2\min(i,k)).
\end{equation}
Let ${\mathcal A}$ be the abelian multiplicative group generated by 
$\sigma_{i_1},\ldots, \sigma_{i_g}$.
Define
\begin{equation}\label{k:Jde}
{\mathcal J}_L(\mu)=\tilde{\mathcal J}_L(\mu)/{\mathcal A},
\end{equation}
which is the set of equivalence classes of $\tilde{\mathcal J}_L(\mu)$
under ${\mathcal A}$.
The image $[{\bf J}]\in {\mathcal J}_L(\mu)$ 
of ${\bf J} \in \tilde{\mathcal J}_L(\mu)$ 
will also be written as
${\bf J}$ for simplicity unless emphasis is preferable.
Elements of ${\mathcal J}_L(\mu)$ are called {\em angle variables}.

Angle variables are also depicted as (\ref{k:zurg}).
Actually, infinitely many such diagrams that are transformable by ${\mathcal A}$ 
all correspond to a single angle variable. 
For instance in (\ref{k:mit}), 
if the leftmost one is $[{\bf J}]$, then the middle and the rightmost ones
are $[\sigma_2({\bf J})]$ and 
$[\sigma_1\sigma^2_2({\bf J})]$, respectively.

We introduce the time evolution $T_l\,(l \ge 1)$ 
on $\tilde{\mathcal J}_L(\mu)$ by
\begin{equation}\label{k:tmjt}
T_l: \tilde{\mathcal J}_L(\mu)\rightarrow \tilde{\mathcal J}_L(\mu);
\quad
(J_{i,\alpha}) \mapsto (J_{i, \alpha}+\min(i,l)),
\end{equation}
and denote its induced action on ${\mathcal J}_L(\mu)$ also by $T_l$. 
Obviously, $T_l$ is linear and commutative.
In particular we use the abbreviation 
$T_1^d({\bf J}) = {\bf J} + d$ for the uniform shift.
Readers are highly recommended to check that 
$\left(\prod_{i \in {\mathcal I}}\sigma^{m_i}_i\right)({\bf J})
={\bf J}+L \in \tilde{\mathcal J}_L(\mu)$,
which implies that any angle variable is invariant 
under $T_1^L$ as it should.

\subsubsection{Linearization of time evolution.}\label{k:ss:lin}
Let us assign an angle variable to each state in the isolevel set 
${\mathcal P}_L(\mu)$.
Namely, we are going to construct a 
{\em direct scattering map} 
$\Phi: {\mathcal P}_L(\mu)\rightarrow {\mathcal J}_L(\mu)$.
We do this by suitably adapting the KKR map $\phi$ to the periodic setting.
Let 
${\mathcal P}^+_L(\mu) = \{p \in {\mathcal P}_L(\mu)\mid 
p: \,\text{highest}\}$ be the subset of ${\mathcal P}_L(\mu)$ 
consisting of highest paths.
We consider the following scheme:
\begin{equation}\label{k:eq:pj}
\begin{split}
\Phi: \quad {\mathcal P}_L(\mu) &\longrightarrow \;\,
\Z \times {\mathcal P}_L^+(\mu)\;\;
\longrightarrow \;\;\quad \tilde{\mathcal J}_L(\mu) 
\;\;\;\;\longrightarrow 
\;\; \;{\mathcal J}_L(\mu)\\
p \;\; &\longmapsto \;\;\;\;(d, p_+) \;\;\;\quad\longmapsto
\quad \;{\bf J} +d \quad\;\;\longmapsto \;\;[{\bf J}+d].
\end{split}
\end{equation}
First arrow: Pick any $(d, p_+)$ such that $p=T^d_1(p_+)$.
Second arrow: Apply the KKR map $\phi(p_+)=(\mu,J)_L$ and 
quasi-periodically extend the so obtained rigging $J$ to ${\bf J}$ 
followed by a uniform shift by $d$. 
Third arrow: Take the image in ${\mathcal J}_L(\mu)$
(identify by ${\mathcal A})$.
In order to make sense of the scheme (\ref{k:eq:pj}) 
as a definition of the map $\Phi$,
the non-uniqueness in the first arrow must be canceled 
in the identification in the third arrow.
It was indeed the case in the example (\ref{k:mit}).
Here comes the main result of this section.
\begin{Theorem}{\rm \cite[Th.3.11]{KTT06}}\label{k:th:pmain}
$\Phi$ is well-defined, bijective and makes the following diagram
commutative.
\begin{equation}\label{k:pbbscd}
\begin{CD}
{\mathcal P}_L(\mu) @>{\Phi}>> {\mathcal J}_L(\mu) \\
@V{T_l}VV @VV{T_l}V\\
{\mathcal P}_L(\mu) @>{\Phi}>> {\mathcal J}_L(\mu). 
\end{CD}
\end{equation}
Here $T_l$ on the left and right sides are defined by (\ref{t:july25a}) and 
(\ref{k:tmjt}), respectively. 
\end{Theorem}
The commutative diagram (\ref{k:pbbscd})
is the periodic version of (\ref{k:cd}). 
According to Theorem \ref{k:th:pmain},
the nonlinear time evolution on ${\mathcal P}_L(\mu)$
is transformed to a straight motion on ${\mathcal J}_L(\mu)$.
This is a characteristic feature in finite dimensional integrable systems,
where the dynamics is linearized, via what is called the eigenvector map,
on Jacobian variety (or more precisely, abelian variety).
In our setting here, the modified KKR map $\Phi$
and the set of angle variables ${\mathcal J}_L(\mu)$
play the analogous roles to them.
In \S 6.3 we will study this feature with tropical geometry,
and see that ${\mathcal J}_L(\mu)$ corresponds to the lattice points
of some tropical abelian variety.

Theorem \ref{k:th:pmain} led to the first complete solution 
of the initial value problem of the periodic BBS.
It is obtained by going along the commutative diagram
(\ref{k:pbbscd}) as
$T_l^{\mathcal N}=\Phi^{-1}\circ T_l^{\mathcal N} \circ \Phi$.
The variety of time evolutions $T_1, T_2, \ldots$ 
is reflected in the corresponding velocity vectors in (\ref{k:tmjt}).

\begin{Example}\label{k:ex:sks}
Let us derive a time evolution of the length $19$ path 
$p$ in Example \ref{k:ex:pp}:
\begin{equation}\label{k:kotae}
T^{5}_3(p) = 1221112211211221122.
\end{equation}
The angle variable of $p$ has been obtained in (\ref{k:mit}).
Using the leftmost representation, we find
\begin{equation*}
\begin{picture}(15,14)(-5,-12)
\setlength{\unitlength}{0.35mm}
\put(0,0){\line(0,-1){50}}
\put(10,0){\line(0,-1){50}}
\put(20,0){\line(0,-1){30}}
\put(30,0){\line(0,-1){10}}

\put(-30,-18){$\Phi$}
\put(-45,-25){$p \;\longmapsto$}

\put(0,0){\line(1,0){30}}
\put(33,-9){\rm 3}
\put(0,-10){\line(1,0){30}}\put(-8,-9){\rm 1}
\put(23,-19.7){\rm 3}
\put(0,-20){\line(1,0){20}}\put(-8,-24){\rm 3}
\put(23,-29.7){\rm 2}
\put(0,-30){\line(1,0){20}}
\put(13,-39.7){\rm 10}
\put(0,-40){\line(1,0){10}}\put(-8,-44){\rm 9}
\put(13,-49.7){\rm 6}
\put(0,-50){\line(1,0){10}}
\end{picture}
\qquad
\begin{picture}(15,14)(-5,-12)
\setlength{\unitlength}{0.35mm}
\put(0,0){\line(0,-1){50}}
\put(10,0){\line(0,-1){50}}
\put(20,0){\line(0,-1){30}}
\put(30,0){\line(0,-1){10}}

\put(-32,-14){$T^5_3$}
\put(-36,-25){$\longmapsto$}

\put(0,0){\line(1,0){30}}
\put(33,-9){\rm 18}
\put(0,-10){\line(1,0){30}}
\put(23,-19.7){\rm 13}
\put(0,-20){\line(1,0){20}}
\put(23,-29.7){\rm 12}
\put(0,-30){\line(1,0){20}}
\put(13,-39.7){\rm 15}
\put(0,-40){\line(1,0){10}}
\put(13,-49.7){\rm 11}
\put(0,-50){\line(1,0){10}}
\end{picture}
\begin{picture}(20,14)(-23,-12)
\setlength{\unitlength}{0.35mm}
\put(0,0){\line(0,-1){50}}
\put(10,0){\line(0,-1){50}}
\put(20,0){\line(0,-1){30}}
\put(30,0){\line(0,-1){10}}

\put(-72,-13){$\sigma_2\sigma^{-2}_3$}
\put(-70,-25){$\longmapsto\;\; \;\;\;\; 8 \; + $}

\put(0,0){\line(1,0){30}}
\put(33,-9){\rm 0}
\put(0,-10){\line(1,0){30}}
\put(23,-19.7){\rm 3}
\put(0,-20){\line(1,0){20}}
\put(23,-29.7){\rm 1}
\put(0,-30){\line(1,0){20}}
\put(13,-39.7){\rm 5}
\put(0,-40){\line(1,0){10}}
\put(13,-49.7){\rm 1}
\put(0,-50){\line(1,0){10}}
\put(40,-50){.}
\end{picture}
\end{equation*} 
The vacancies are exhibited only in the leftmost diagram.
The rightmost diagram is a rigged configuration and 
corresponds to the highest path
$p'= 1121122112212211122$.
Therefore the image of the rightmost 
angle variable by $\Phi^{-1}$ is 
$T^8_1(p')$  giving the RHS of 
 (\ref{k:kotae}). 
\end{Example}
As this example indicates, to compute the inverse image $\Phi^{-1}({\bf J})$,
one first finds a representative of the angle variable ${\bf J}$
of the form (rigged configuration)$+ \,e$ with some $e \in \Z$.
Namely, one transforms ${\bf J}$ into 
$\sigma({\bf J}) = {\bf J'}+e$ with an appropriate 
element $\sigma \in {\mathcal A}$ so that 
${\bf J'}$ becomes the quasi-periodic extention of some rigged configuration 
$(\mu,J')_L$.
Then one applies the KKR map $\phi^{-1}$ 
to get a highest path $p'_+$ as $p'_+ = \phi^{-1}((\mu,J')_L)$.
Finally, the inverse image is obtained as
$\Phi^{-1}({\bf J}) = T_1^e(p'_+)$.
The fact that these procedures are always possible and the result is unique is 
guaranteed by Theorem \ref{k:th:pmain}.
We note that the solution of the initial value problem 
based on the procedure called 10-elimination \cite{MIT06} 
is equivalent \cite{KirS09} to the preceding solution \cite{KTT06}
explained here.

\subsubsection{$N$-soliton solution.}\label{k:ss:nsol}
Let us present an explicit formula of the path 
$p = \Phi^{-1}({\bf J}) \in B^{\otimes L}_1$ that
corresponds to the given angle variable 
${\bf J} \in {\mathcal J}_L(\mu)$.
This is a combinatorial analogue of the Jacobi inversion problem, and 
the result is indeed expressed in terms of tropical Riemann theta function.
For simplicity 
we restrict ourselves to the case $m_i=1$ for all 
$i \in {\mathcal I}$.
(See \cite{KS2} for the general case, where the 
tropical Riemann theta function
with rational characteristics is involved.)
We retain the notation (\ref{k:ipj}), where the latter 
reduces to $p_j = L-2\sum_{i \in {\mathcal I}}\min(j,i)$.
The angle variable 
${\bf J}= (J_{i,\alpha})_{(i \alpha) \in {\mathcal I}\times \Z}$ 
can be presented 
simply as a $g$-dimensional vector ${\bf J} = (J_i)_{i \in {\mathcal I}}$
with $J_i = J_{i,1}$ since the other components are
specified as $J_{i,\alpha} = J_i+(\alpha-1)p_i$ by 
the quasi-periodicity.
It is easy to translate the identification by (\ref{k:slide})
into this representation of ${\bf J}$, leading to the simple description
\begin{equation}\label{k:jfred}
{\mathcal J}_L(\mu)= \Z^g/F\Z^g,\qquad
F=\bigl(\delta_{ij}p_i+2\min(i,j)\bigr)_{i,j \in {\mathcal I}}.
\end{equation}

The $g\times g$ integer matrix $F$ is positive definite
and has the origin in the study of Bethe equation at $q=0$ \cite{KuN}.
We introduce the tropical Riemann theta function
(Definition \ref{i:def:theta})
\begin{equation}\label{k:udt}
\Theta({\bf Z}; F) = \min_{{\bf n} \in \Z^g}
\left\{{\bf n}\cdot \bigl(\tfrac{1}{2}\, F{\bf n}+ {\bf Z}\bigr)\right\}
\quad({\bf Z} \in {\mathbb R}^g)
\end{equation}
in which the $F$ is built in as the period matrix.
We further introduce the $g$-dimensional vectors
\begin{equation}\label{k:phvec}
{\bf p} = (p_i)_{i \in {\mathcal I}},\quad 
{\bf h}_l = \bigl(\min(i,l)\bigr)_{i \in {\mathcal I}},
\end{equation}
where the latter is the velocity corresponding to 
$T_l$ (\ref{k:tmjt}).

\begin{Theorem}{\rm \cite[Th.3.3]{KS06}}\label{k:th:udt}
The state $\Phi^{-1}({\bf J}) = b_1b_2\ldots b_L\;(b_j =1,2)$ 
corresponding to the angle variable ${\bf J} \in {\mathcal J}(\mu)$ 
is expressed as $(\Theta({\bf Z})= \Theta({\bf Z}; F)) $
\begin{equation}\label{k:xt4}
\begin{split}
b_{k}&= 1-\Theta\left({\bf J}-\frac{\bf p}{2}
-k{\bf h}_1\right)
+ \Theta\left({\bf J}-\frac{\bf p}{2}
-(k-1){\bf h}_1\right)\\
&+ \Theta\left({\bf J}-\frac{\bf p}{2}
-k{\bf h}_1+{\bf h}_\infty\right)
- \Theta\left({\bf J}-\frac{\bf p}{2}
-(k-1){\bf h}_1+{\bf h}_\infty\right).
\end{split}
\end{equation}
\end{Theorem}
The time evolution by $T_l$ is attained just by replacing 
${\bf J}$ with ${\bf J} + {\bf h}_l$.
Thanks to the quasi-periodicity
(\ref{i:theta-quasi}), $b_k$ is invariant under the change 
${\bf J} \mapsto {\bf J} + F\Z^g$.
The invariance $b_k =b_{k + L}$ is 
due to $F{\bf h}_1 = L{\bf h}_1$.

\begin{Remark}
The matrix $F$ \eqref{k:jfred} will be related to the period matrix $\Omega$
\eqref{i:omega}
of the tropical spectral curve of the tropical periodic Toda lattice
at Proposition \ref{i:prop:Omega-F}. 
\end{Remark}

\subsection{Decomposition into Torus}
\label{t:5-3}

\subsubsection{Introduction.}\label{t:5-3int} 
Let us discuss the structure of the isolevel set of the periodic BBS.
Recall the definition of the isolevel set ${\mathcal P}_L (\mu)$ in (\ref{k:ils}).
We use the notations 
$\mu = (i_g^{m_{i_g}}\ldots i_1^{m_{i_1}}), ~{\mathcal I}, 
~p_j$ in \S \ref{subsec:general-sol}.
In contrast to the tropicalization of the periodic discrete Toda lattice in \S \ref{i:sec:pToda}
where the isolevel set $\mathcal{T}_C $ given by
(\ref{i:TC}) is isomorphic to a real torus $\R^g / \Omega \Z^g$,
the set ${\mathcal P}_L (\mu)$ is a finite set.
As we have shown in Theorem \ref{k:th:pmain} the set ${\mathcal P}_L (\mu)$ is identified with the set
${\mathcal J}_L (\mu)$ which was defined as a quotient set (\ref{k:Jde}).

The set ${\mathcal P}_L (\mu)$ can be regarded as a graph in the sense of graph theory.
Let ${\mathcal T}$ be the abelian multiplicative group generated by $T_1, T_2, \ldots$.
Then ${\mathcal T}$ acts on the isolevel set ${\mathcal P}_L (\mu)$.
If one represents the elements of ${\mathcal P}_L (\mu)$ by nodes and the actions of the time evolutions by arrows,
then one has a colored oriented graph. 
The graph for ${\mathcal P}_L (\mu)$ has usually several connected components.
From the viewpoint of the group action we say that
the isolevel set ${\mathcal P}_L (\mu)$ decomposes into ${\mathcal T}$-orbits.

To begin with let us illustrate a few simple examples.
First let us assume $m_j =1$ for all $j \in {\mathcal I}$.
Then we have ${\mathcal J}_L(\mu)= \Z^g/F\Z^g$ (\ref{k:jfred}) which is
the set of all integer points on the torus ${\mathbb R}^g/F\Z^g$.
%
We shall call $\Z^g/F\Z^g$ itself a ($g$-dimensional) torus for short.
\begin{Example}\label{t:july25b}
The isolevel set
${\mathcal P}_5((2))$ is depicted as follows.
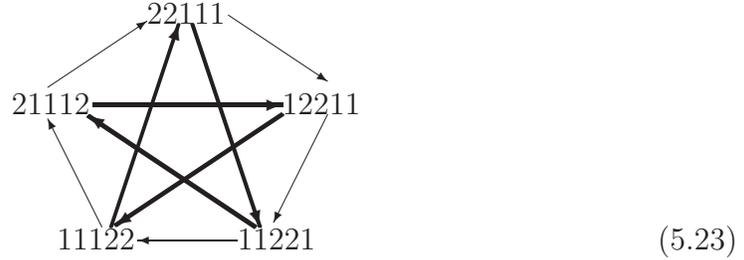
\begin{equation}
\setlength{\unitlength}{6mm}
\begin{picture}(6,6)
\put(3,5){$22111$}
\put(6,3){$12211$}
\put(5,0){$11221$}
\put(1,0){$11122$}
\put(0,3){$21112$}
\put(4.8,5.2){\vector(3,-2){2.2}}
\put(7,3){\vector(-1,-2){1.2}}
\put(5,0.2){\vector(-1,0){2.2}}
\put(2,0.5){\vector(-1,2){1.2}}
\put(0.8,3.6){\vector(3,2){2.2}}
\put(4,5){\thicklines\vector(1,-3){1.5}}
\put(4.02,5){\thicklines\line(1,-3){1.5}}
\put(3.98,5){\thicklines\line(1,-3){1.5}}
\put(5.4,0.5){\thicklines\vector(-3,2){3.7}}
\put(5.42,0.52){\thicklines\line(-3,2){3.7}}
\put(5.38,0.48){\thicklines\line(-3,2){3.7}}
\put(1.8,3.2){\thicklines\vector(1,0){4.2}}
\put(1.8,3.22){\thicklines\line(1,0){4.2}}
\put(1.8,3.18){\thicklines\line(1,0){4.2}}
\put(6,3){\thicklines\vector(-3,-2){3.7}}
\put(5.98,3.02){\thicklines\line(-3,-2){3.7}}
\put(6.02,2.98){\thicklines\line(-3,-2){3.7}}
\put(2.2,0.5){\thicklines\vector(1,3){1.5}}
\put(2.22,0.5){\thicklines\line(1,3){1.5}}
\put(2.18,0.5){\thicklines\line(1,3){1.5}}
\end{picture}
\end{equation}
Here the actions of $T_1$ and $T_2$ are represented by thin and thick arrows respectively.
Note that ${\mathcal P}_5((2)) \simeq {\mathcal J}_5((2)) = \Z / 5 \Z$.
The velocity vectors (\ref{k:phvec}) are given by ${\bf h}_1 = (1)$ and ${\bf h}_2 = (2)$
which reflects the relation $T_2 = (T_1)^2$ on this isolevel set.
\end{Example}

Now we consider the case with $m_i >1$ where the graph for ${\mathcal P}_L (\mu)$ has indeed several connected components.
Let
\begin{equation}
\Sigma (p) = \{ p' \in {\mathcal P}_L (\mu) \vert p' = g p \quad \mbox{for some} \quad g \in {\mathcal T} \} \subset {\mathcal P}_L (\mu)
\end{equation}
be the ${\mathcal T}$-orbit of $p \in {\mathcal P}_L (\mu)$.
\begin{Example}\label{t:aug4a}
The graph for the isolevel set
${\mathcal P}_6((1,1))$ is depicted as follows.
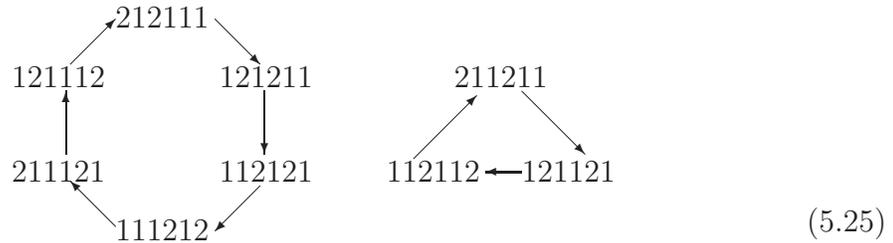
\begin{equation}
\setlength{\unitlength}{6mm}
\begin{picture}(12,5)
\put(2,4.5){$212111$}
\put(4.3,3.2){$121211$}
\put(4.3,1.1){$112121$}
\put(2,-0.2){$111212$}
\put(-0.3,1.1){$211121$}
\put(-0.3,3.2){$121112$}
\put(9.5,3.2){$211211$}
\put(11,1.1){$121121$}
\put(8,1.1){$112112$}
\put(4.2,4.7){\vector(1,-1){1}}
\put(5.3,3.1){\vector(0,-1){1.4}}
\put(5.2,1){\vector(-1,-1){1}}
\put(2,0.1){\vector(-1,1){1}}
\put(0.9,1.7){\vector(0,1){1.4}}
\put(1,3.7){\vector(1,1){1}}
\put(11,3.1){\vector(1,-1){1.4}}
\put(11,1.3){\vector(-1,0){0.8}}
\put(8.6,1.6){\vector(1,1){1.4}}
\end{picture}
\end{equation}
Here the actions of $T_1$ are represented by arrows.
It is not necessary to consider $T_{l \geq 2}$ because they effectively coincide with $T_1$ in the present case.
The graph has two connected components,
as ${\mathcal P}_6((1,1)) = \Sigma (212111) \sqcup \Sigma (211211)$.
Each connected component has the structure of a one-dimensional torus,
$\Sigma (212111) \simeq (\Z / 6 \Z)$ and $\Sigma (211211) \simeq (\Z / 3 \Z)$.
\end{Example}

\subsubsection{Internal symmetry.}\label{t:5-3is}
The isolevel set in Example \ref{t:aug4a} has two connected components with different sizes.
The difference reflects the {\em internal symmetry}.
If the state has a larger internal symmetry, then it belongs to a smaller connected component.
Let us briefly discuss this notion here.

The internal symmetry of a BBS state is represented by an integer vector
$\boldsymbol{\gamma} = (\gamma_{i_1}, \ldots, \gamma_{i_g}) \in (\Z_{>0})^g$.
We demonstrate how one can read off $\boldsymbol{\gamma}$ from $p$.
Given $p \in {\mathcal P}_L (\mu)$ there exist a highest path $p_+$ and $d \in \Z$ such that $T_1^d (p_+) = p$.
By the KKR map in \S \ref{subsec:kkr} a rigged configuration is constructed from $p_+$ as
$\phi (p_+) = (\mu, J)_L$
where $J = (J_{i,\alpha})_{i \in {\mathcal I}, 1 \leq \alpha \leq m_i}$.
We adopt the quasi-periodic extension for the angle variables (\ref{k:quasi}).
Now the integer $\gamma_i \, (i \in {\mathcal I} ~ \eqref{k:ipj})$ 
is defined as the largest common divisor of $m_i$ and $p_i$ such that
\begin{equation}\label{t:july27d}
J_{i,\alpha+\frac{m_i}{\gamma_i}}= J_{i,\alpha}+ \frac{p_i}{\gamma_i} \quad (\alpha \in \Z).
\end{equation}
Neither the action of $\sigma_k$ in (\ref{k:slide}) nor
that of $T_l$ in (\ref{k:tmjt}) changes the relation (\ref{t:july27d}). 
Therefore Theorem \ref{k:th:pmain}
ensures that
the internal symmetry of $p$ is uniquely determined from the above procedure
even if there is non-unique choice of $p_+$.

\begin{Example}\label{t:july27a}
Take $p = 1212111222, \tilde{p} = 1211121222 \in {\mathcal P}_{10}$ in Example \ref{t:july27b}.
They are already highest.
By the KKR map we obtain the following rigged configurations.
\begin{equation}
\setlength{\unitlength}{3.6mm}
\begin{picture}(20,3)
\put(-1,1){$p \, \stackrel{\phi}{\mapsto}$}
\multiput(3,0)(1,0){2}{\line(0,1){3}}
\multiput(5,2)(1,0){2}{\line(0,1){1}}
\multiput(3,0)(0,1){2}{\line(1,0){1}}
\multiput(3,2)(0,1){2}{\line(1,0){3}}
\put(6.3,2){$0$}
\put(4.3,1){$0$}
\put(4.3,0){$0$}
\put(2,2){$0$}
\put(2,0.5){$4$}
\put(12,1){$\tilde{p} \, \stackrel{\phi}{\mapsto}$}
\multiput(16,0)(1,0){2}{\line(0,1){3}}
\multiput(18,2)(1,0){2}{\line(0,1){1}}
\multiput(16,0)(0,1){2}{\line(1,0){1}}
\multiput(16,2)(0,1){2}{\line(1,0){3}}
\put(19.3,2){$0$}
\put(17.3,1){$2$}
\put(17.3,0){$0$}
\put(15,2){$0$}
\put(15,0.5){$4$}
\end{picture}
\end{equation}
In this example one has $g=2, ~{\mathcal I} = \{1,3 \}$, hence ${\boldsymbol \gamma} = ( \gamma_1, \gamma_3)$.
Note that the configuration implies $m_3 = 1$ which imposes $\gamma_3 = 1$.
While
$ J_{1, \alpha + \frac22} = J_{1,\alpha} + \frac42$
for $\tilde{p}$ , no such symmetry exists for $p$.
Hence one has ${\boldsymbol \gamma} = (1,1)$ for $p$, and
${\boldsymbol \gamma} = (2,1)$ for $\tilde{p}$.
\end{Example}

\begin{Example}\label{t:july27e}
Take $p=121122111212211222121111$ with $L = 24$ which is already highest.
By the KKR map we obtain the following rigged configuration.
\begin{equation}
\setlength{\unitlength}{3.6mm}
\begin{picture}(9,6)
\put(-1,3){$p \, \stackrel{\phi}{\mapsto}$}
\multiput(3,0)(1,0){2}{\line(0,1){6}}
\put(5,3){\line(0,1){3}}
\put(6,5){\line(0,1){1}}
\multiput(3,0)(0,1){3}{\line(1,0){1}}
\multiput(3,3)(0,1){2}{\line(1,0){2}}
\multiput(3,5)(0,1){2}{\line(1,0){3}}
\put(6.5,5){$0$}
\put(5.5,4){$1$}
\put(5.5,3){$0$}
\put(4.5,2){$8$}
\put(4.5,1){$4$}
\put(4.5,0){$0$}
\put(1.8,5){$4$}
\put(1.8,3.5){$6$}
\put(1.3,1){$12$}
\end{picture}
\end{equation}
Here we have $g=3$ and ${\mathcal I} = \{1,2,3 \}$, 
hence ${\boldsymbol \gamma} = ( \gamma_1, \gamma_2, \gamma_3)$.
Observe that
$ J_{1, \alpha + \frac33} = J_{1,\alpha} + \frac{12}{3} $
which implies $\gamma_1 = 3$.
No such symmetry exists for $J_{2,\alpha}$ (and $J_{3,\alpha}$).
Hence one has ${\boldsymbol \gamma} = (3,1,1)$.
\end{Example}

Let
\begin{equation}
{\mathcal P}_{L, {\boldsymbol \gamma}} (\mu) = \{ 
p \in {\mathcal P}_{L} (\mu) \vert \mbox{the internal symmetry of $p$ is ${\boldsymbol \gamma}$}
\}.
\end{equation}
Then we have ${\mathcal P}_{L} (\mu) = \bigsqcup_{{\boldsymbol \gamma}} {\mathcal P}_{L, {\boldsymbol \gamma}} (\mu)$.
\subsubsection{Connected component as torus.}\label{t:5-3cct}
So far the states of periodic BBS are classified as
\begin{equation}\label{t:aug15a}
{\mathcal P}_L \supset {\mathcal P}_L(\mu) \supseteq {\mathcal P}_{L,{\boldsymbol \gamma}}(\mu) \supseteq \Sigma(p).
\end{equation}
Now we study the structure of a single connected component $\Sigma(p)$.
Let $F_{\boldsymbol{\gamma}}$ be a $g \times g$ matrix
defined as
\begin{align}\label{t:july26b}
F_{\boldsymbol{\gamma}} &= F \, \cdot \, {\rm diag} (\gamma_{i_1}^{-1}, \ldots \gamma_{i_g}^{-1}), \\
F &=\bigl( \delta_{ij}p_i+2\min(i,j) m_j \bigr)_{i,j \in {\mathcal I}}. \label{t:july26bx} 
\end{align}
This matrix $F_{\boldsymbol{\gamma}}$ is a generalization of $F$ in (\ref{k:jfred}).
Then the structure of a connected component is stated as follows.
\begin{Prop}{\rm \cite[Th.2]{T10}}\label{t:july28e}
Every connected component of the isolevel set ${\mathcal P}_L (\mu)$
has the structure of a $g$-dimensional torus $\Z^g / F_{\boldsymbol{\gamma}} \Z^g$.
The time evolution $T_l$ is realized as the straight motion with 
a constant velocity vector ${\bf h}_l = \bigl(\min(i,l)\bigr)_{i \in {\mathcal I}}$
on the torus.
\end{Prop}

This result may be viewed as an ultradiscrete analogue of the 
classical Arnold-Liouville theorem \cite{Ar}. 

\begin{Example}\label{t:july27b}
Take $p = 1212111222, \tilde{p} = 1211121222 \in {\mathcal P}_{10}$.
They belong to two distinct connected components of the level set ${\mathcal P}_{10} ((3,1,1))$.
The ${\mathcal T}$-orbits $\Sigma (p)$ and $\Sigma (\tilde{p})$ have the structure of
two-dimensional tori $\Z^2 / F_{(1,1)} \Z^2$ and $\Z^2 / F_{(2,1)} \Z^2$ respectively,
where $F_{(1,1)} =$
\small
$ 
\begin{pmatrix}
8 & 2 \\
4 & 6 
\end{pmatrix}$
\normalsize
and
$F_{(2,1)} =$
\small
$ \begin{pmatrix}
4 & 2 \\
2 & 6 
\end{pmatrix}$.
\normalsize
They are depicted as follows.
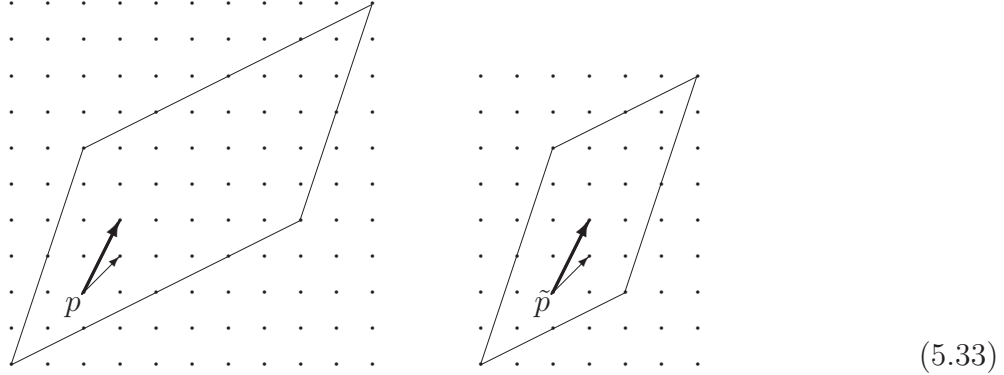
\begin{equation}\label{t:aug4g}
\setlength{\unitlength}{4.8mm}
\begin{picture}(10,10)(6,0)
\multiput(-0.2,-0.2)(0,1){11}{\multiput(0.1,0)(1,0){11}{$\cdot$}}
\multiput(0,0)(8,4){2}{\line(1,3){2}}
\multiput(0,0)(2,6){2}{\line(2,1){8}}
\put(2,2){\vector(1,1){1}}
\put(2,2){\thicklines\vector(1,2){1}}
\put(2.03,2.03){\thicklines\line(1,2){0.98}}
\put(1.97,1.97){\thicklines\line(1,2){0.98}}
\put(1.5,1.5){$p$}
\multiput(12.8,-0.2)(0,1){9}{\multiput(0.1,0)(1,0){7}{$\cdot$}}
\multiput(13,0)(4,2){2}{\line(1,3){2}}
\multiput(13,0)(2,6){2}{\line(2,1){4}}
\put(15,2){\vector(1,1){1}}
\put(15,2){\thicklines\vector(1,2){1}}
\put(15.03,2.03){\thicklines\line(1,2){0.98}}
\put(14.97,1.97){\thicklines\line(1,2){0.98}}
\put(14.5,1.5){$\tilde{p}$}
\end{picture}
\end{equation}
The nodes within and on the edges of the parallelograms represent the states of periodic BBS
which belong to each of the connected components.
Every pair of the parallel edges of each parallelogram should be identified.
Thin and thick arrows represent the velocity vectors ${\bf h}_1$ and ${\bf h}_2$ respectively.
\end{Example}
\subsubsection{Fundamental period.}\label{t:5-3fp}
The time evolution $T_l$ is invertible
and the isolevel set ${\mathcal P}_L (\mu)$ is a finite set.
Therefore every path $p \in {\mathcal P}_L (\mu)$ possesses the 
property $T_l^N(p)=p$ for some integer $N\ge 1$.
We say any such integer a {\em period} of $p$.
The minimum period is called the 
{\em fundamental period} of $p$ and denoted by 
${\mathcal N}_l(p)$.
Every period is a multiple of the fundamental period.
A formula for 
the fundamental period under any $T_l$ was established in \cite{KTT06}.
Here we show a derivation of the formula
based on Proposition \ref{t:july28e}.
Note that ${\mathcal N}_l(p)$ is common to all the states in one connected component.
Hence it is determined by the action variable $\mu$ and the internal symmetry $\boldsymbol{\gamma}$.

To avoid double indices 
we denote by $\boldsymbol{f}_j$ the $i_j$-th column of the matrix $F$ (\ref{t:july26bx}), hence
$F = (\boldsymbol{f}_1, \ldots , \boldsymbol{f}_g)$.
For any $\boldsymbol{b} \in \Z^g$ we define
\begin{equation}\label{t:aug4c}
F_{i}[\boldsymbol{b}] = (\boldsymbol{f}_1, \ldots , \boldsymbol{f}_{i-1},\boldsymbol{b} ,\boldsymbol{f}_{i+1}, \ldots , \boldsymbol{f}_g).
\end{equation}
Let $p \in {\mathcal P}_{L, {\boldsymbol{\gamma}}} (\mu)$ be a state of the periodic BBS, and
$\Sigma (p) \simeq \Z^g/F_{\boldsymbol{\gamma}} \Z^g $.
We define the least common multiple of nonzero rational numbers $r_1,\ldots, r_n$ 
as 
\begin{equation}\label{t:aug11a}
{\rm LCM} ( r_1, \ldots, r_n ) = \min \big| \Z \cap \Z r_1 \cap \cdots \cap \Z r_n \setminus \{0\} \big|. 
\end{equation}
For instance we let ${\rm LCM} (1/3,2/3)$ be $2$ rather than $2/3$.
Now we have
\begin{Prop}{\rm \cite[Th.4.9]{KTT06}}\label{t:aug4b}
\begin{equation}\label{t:july28d}
{\mathcal N}_l(p) = {\rm LCM}\!
\left(
\frac{\det F}{  {\gamma}_{i_1} \det F_{1}[\boldsymbol{h}_l] },
\ldots,
\frac{\det F}{  {\gamma}_{i_g} \det F_{g}[\boldsymbol{h}_l] }
\right),
\end{equation}
where we exclude any entries of LCM such that 
$\det F_{i}[\boldsymbol{h}_l] = 0$.
\end{Prop}

\proof
Proposition \ref{t:july28e} implies that
${\mathcal N}_l(p)$ is defined as the smallest positive integer $N$ such that
the following linear equation
has an integer solution $\boldsymbol{n}  = (n_1, \ldots ,n_g )\in \Z^g$:
\begin{equation}\label{t:aug11b}
N \boldsymbol{h}_l = F_{\boldsymbol{\gamma}} \boldsymbol{n}.
\end{equation}
By demanding all these conditions on the expression
$n_j = N {\gamma}_{i_j} \det F_{j}[{\boldsymbol{h}_l}]/\det F$ of the solution
to (\ref{t:aug11b}), 
one obtains the formula (\ref{t:july28d}).
\qed

\begin{Example}
Take $p = 1212111222, \tilde{p} = 1211121222 \in {\mathcal P}_{10}$ in Example \ref{t:july27b}.
Then
$\det F= 
\det$
\small
$
\begin{pmatrix}
8 & 2 \\
4 & 6 
\end{pmatrix} = 40$ 
\normalsize
and
\begin{align}
\det F_1 [\boldsymbol{h}_1] &= 
\det
\begin{pmatrix}
1 & 2 \\
1 & 6 
\end{pmatrix} = 4,
\quad
\det F_2 [\boldsymbol{h}_1] = 
\det
\begin{pmatrix}
8 & 1 \\
4 & 1 
\end{pmatrix} = 4, \nonumber\\
\det F_1 [\boldsymbol{h}_2] &= 
\det
\begin{pmatrix}
1 & 2 \\
2 & 6 
\end{pmatrix} = 2,
\quad
\det F_2 [\boldsymbol{h}_2] = 
\det
\begin{pmatrix}
8 & 1 \\
4 & 2 
\end{pmatrix} = 12, \nonumber\\
\det F_1 [\boldsymbol{h}_3] &= 
\det
\begin{pmatrix}
1 & 2 \\
3 & 6 
\end{pmatrix} = 0,
\quad
\det F_2 [\boldsymbol{h}_3] = 
\det
\begin{pmatrix}
8 & 1 \\
4 & 3 
\end{pmatrix} = 20.
\end{align}
Hence
\begin{align}\label{t:aug4d}
{\mathcal N}_1(p) &= {\rm LCM} ( 40/4 , 40/4 ) = 10, \quad
{\mathcal N}_2(p) = {\rm LCM} ( 40/2 , 40/12 ) = 20, \nonumber\\
{\mathcal N}_3(p) &= {\rm LCM} ( 40/20 ) = 2.
\end{align}
For $\tilde{p}$ one has ${\gamma}_1 = 2, {\gamma}_3 = 1$.
Hence
\begin{align}\label{t:aug4f}
{\mathcal N}_1(\tilde{p}) &= {\rm LCM} ( 40/8 , 40/4 ) = 10, \quad
{\mathcal N}_2(\tilde{p}) = {\rm LCM} ( 40/4 , 40/12 ) = 10, \nonumber\\
{\mathcal N}_3(\tilde{p}) &= {\rm LCM} ( 40/20 ) = 2.
\end{align}
For instance, one can deduce $T_2^{10}(\tilde{p}) = \tilde{p}$ and $T_2^{10}(p) \ne p$.
One can easily check these results by using the figures in (\ref{t:aug4g}).
\end{Example}
When the internal symmetry is trivial, i.e. 
$\forall \gamma_{i_j}=1$, 
Proposition \ref{t:aug4b}
reduces to the result in \cite{YYT}.
\subsubsection{Multiplicity of torus and structure of isolevel set.}\label{t:5-3mult}
The number of the elements of the isolevel set ${\mathcal P}_L (\mu)$ is expressed by
the formula \cite{KTT06}
\begin{align}\label{t:july26a}
|{\mathcal P}_L (\mu)| &= \det F \prod_{i \in {\mathcal I}} \frac{1}{m_i} {p_i + m_i -1 \choose m_i -1} \\
&= \frac{L}{p_{i_g}} \prod_{i \in {\mathcal I}} {p_i + m_i -1 \choose m_i},
\label{t:july26c}
\end{align}
where the $g \times g$ matrix $F$ is defined as (\ref{t:july26bx}).
While the former expression (\ref{t:july26a}) was first obtained in the context of Bethe ansatz \cite[Th.3.5]{KuN},
the latter one (\ref{t:july26c}) was originally found as a formula for the cardinality $|{\mathcal P}_L (\mu)|$ and
proved by elementary combinatorial arguments \cite[Prop.2.2]{YYT}.
Their equivalence is due to the relation $\det F = p_{i_1} \cdots p_{i_{g-1}} L$.

We demonstrate the decomposition of the isolevel set ${\mathcal P}_L (\mu)$ into connected components
from the viewpoint of their cardinality.
First we consider the case $m_i = 1$ for all $i \in {\mathcal I}$.
Then by (\ref{t:july26a}) one has $|{\mathcal P}_L (\mu)| = \det F$.
Actually this is a consequence of ${\mathcal P}_L (\mu) \simeq {\mathcal J}_L (\mu)$ and (\ref{k:jfred}).
In this case ${\mathcal P}_L (\mu)$ itself is a connected graph.
Next we consider general cases with $m_i \geq 1$.
Let $m,p$ be a pair of positive integers and $\gamma$ any common divisor of $m$ and $p$.
We define 
\begin{equation}\label{t:july26d}
C_{\gamma}(m, p) =
\sum_\beta \mu\Bigl(\frac{\beta}{\gamma}\Bigr)
{\frac{p+m}{\beta} -1 \choose \frac{m}{\beta} -1},
\end{equation}
where $\beta$ runs over  
all the common divisors of $m$ and $p$ that is a multiple of $\gamma$.
Here $\mu$ is the 
M{\"o}bius function in number theory defined by
\begin{equation}\label{t:july26e}
\mu (k) = 
\begin{cases}
1 & \mbox{if $k=1$}, \\
(-1)^j & \mbox{if $k$ is the product of $j$ distinct primes},\\
0 & \mbox{otherwise.}
\end{cases}
\end{equation}
(This $\mu$ should not be confused with 
the Young diagram.)
By the M{\"o}bius inversion formula we have
\begin{equation}\label{t:july26f}
{ p+m-1 \choose m-1 } = \sum_\gamma C_\gamma(m,p),
\end{equation}
where $\gamma$ runs over  
all the common divisors of $m$ and $p$.
By (\ref{t:july26a}) and (\ref{t:july26f}) we obtain
\begin{align}
|{\mathcal P}_L (\mu)| &= 
\sum_{\boldsymbol{\gamma}}
{\rm mult} (\boldsymbol{\gamma}) \, \det F_{\boldsymbol{\gamma}}, \label{t:july26g} \\
{\rm mult} (\boldsymbol{\gamma}) &=
\prod_{i \in {\mathcal I}} \frac{\gamma_i C_{\gamma_i} (m_i,p_i)}{m_i},
\label{t:july26h}
\end{align}
where the numbers $(\gamma_i/m_i) C_{\gamma_i} (m_i,p_i)$ turn out to be
integers as a result of certain cyclic group actions \cite{T10}.
This formula is a consequence of the following fact.
\begin{Prop}{\rm \cite{T10}}\label{t:july26i}
The ${\rm mult} (\boldsymbol{\gamma})$ in (\ref{t:july26h}) is the multiplicity of the tori in ${\mathcal P}_L (\mu)$.
That is, the following relation is satisfied:
\begin{equation*}
{\mathcal P}_L (\mu) \simeq  
\bigsqcup_{\boldsymbol{\gamma}}
{\rm mult} (\boldsymbol{\gamma}) \,\, \left( \Z^g/F_{\boldsymbol{\gamma}} \Z^g \right).
\end{equation*}
\end{Prop}

\begin{Example}\label{t:july26j}
Take ${\mathcal P}_{24} (\mu)$ with
\begin{equation}
\setlength{\unitlength}{3.6mm}
\begin{picture}(9,6)
\put(0.5,3){$\mu =$}
\multiput(3,0)(1,0){2}{\line(0,1){6}}
\put(5,3){\line(0,1){3}}
\put(6,5){\line(0,1){1}}
\multiput(3,0)(0,1){3}{\line(1,0){1}}
\multiput(3,3)(0,1){2}{\line(1,0){2}}
\multiput(3,5)(0,1){2}{\line(1,0){3}}
\end{picture}.
\end{equation}
Then we have $(m_1, m_2, m_3) = (3,2,1)$ and $(p_1, p_2, p_3) = (12,6,4)$.
The matrix $F$ is given by
\begin{equation}
F =
\begin{pmatrix}
18 & 4 & 2 \\
6 & 14 & 4 \\
6 & 8 & 10
\end{pmatrix}.
\end{equation}
Possible internal symmetries are $\boldsymbol{\gamma} = (1,1,1), (1,2,1),(3,1,1)$, and $(3,2,1)$.
By using (\ref{t:july26d}) and (\ref{t:july26h}) one has
\begin{equation}\label{t:july26k}
{\mathcal P}_{24} (\mu) \simeq 90 \,\, (\Z^3/F_{(1,1,1)} \Z^3)
\sqcup 30 \,\, (\Z^3/F_{(1,2,1)} \Z^3)
\sqcup 3 \,\, (\Z^3/F_{(3,1,1)} \Z^3)
\sqcup (\Z^3/F_{(3,2,1)} \Z^3).
\end{equation}
The configuration $\mu$ is common to that in Example \ref{t:july27e}.
Hence we can deduce that $\Sigma (p) \simeq (\Z^3/ F_{(3,1,1)} \Z^3)$ for $p=121122111212211222121111$.
\end{Example}


\section{Approach by tropical geometry}

\subsection{Preliminary of tropical geometry} 

Tropical geometry is the algebraic geometry of the min-plus algebra 
$(\mathbb{T}, \min, +)$. 
We introduce the basic notion of the theory of tropical curves,
following \cite{MZ06}.

\subsubsection{Tropical curve.}

A {\it tropical polynomial} $F(X)$ of one variable $X$ is written as
$$
F(X) = \min_{i \in I}(n_i X + C_i) 
\qquad n_i \in \Z_{\geq 0},~ C_i \in \R,
$$
where $I$ is a finite subset of $\Z$. 
One can regard $F(X)$ as the tropicalization of a polynomial 
$f(x) = \sum_{i \in I} c_i x^{n_i} \in \R_{>0}[x]$ 
with the transformation
$c_i = \mathrm{e}^{-\frac{C_i}{\ve}}$ and $x = \mathrm{e}^{-\frac{X}{\ve}}$.
In the same manner, a tropical polynomial $F(X,Y)$ of two variables 
$X$ and $Y$ is written as
$$
F(X,Y) = \min_{i \in I} (n_i X + m_i Y + C_i)
\qquad n_i, m_i \in \Z_{\geq 0},~ C_i \in \R.
$$


Very roughly speaking, a {\it tropical curve} is a finite graph 
(i.e. a graph with a finite number of vertices and edges) 
with a {\it metric structure}.
In the following we only consider affine tropical curves in $\R^2$
given by a tropical polynomial of two variables.
Fix a finite set $I$ and a tropical polynomial $F(X,Y)$
of two variables. 
The tropical curve $\Gamma$ given by $F(X,Y)$ is defined as
$$
\Gamma = \{(X,Y) \in \R^2 ~|~ 
F(X,Y) \text{ is {\it indifferentiable}} \}.
$$

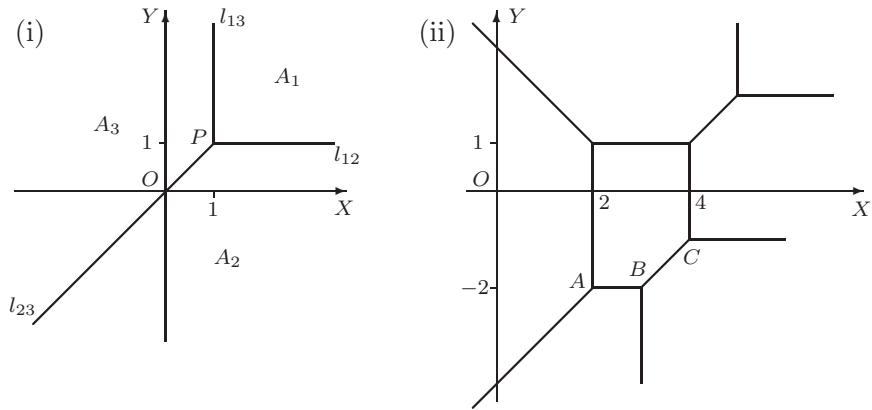
\begin{figure}
\unitlength=0.8mm

\begin{picture}(100,65)(0,-5)

\put(5,55){\small (i)}

\put(5,30){\vector(1,0){55}}
\put(58,26){\scriptsize$X$}
\put(30,5){\vector(0,1){55}}
\put(26,58){\scriptsize$Y$}
\put(26,31){\scriptsize$O$}

\put(38,30){\line(0,-1){1}}
\put(37,26){\scriptsize$1$}
\put(30,38){\line(-1,0){1}}
\put(26,37){\scriptsize$1$}

\thicklines

\put(8,8){\line(1,1){30}}
\put(38,38){\line(1,0){20}}
\put(38,38){\line(0,1){20}}

\put(48,48){\scriptsize$A_1$}
\put(38,18){\scriptsize$A_2$}
\put(18,40){\scriptsize$A_3$}
\put(34,38){\scriptsize$P$}
\put(39,58){\scriptsize$l_{13}$}
\put(58,35){\scriptsize$l_{12}$}
\put(4,10){\scriptsize$l_{23}$}


\thinlines

\put(72,55){\small (ii)}

\put(80,30){\vector(1,0){66}}
\put(85,-5){\vector(0,1){65}}
\put(144,26){\scriptsize$X$}
\put(87,58){\scriptsize$Y$}
\put(81,31){\scriptsize$O$}

\put(118,27){\scriptsize$4$}
\put(102,27){\scriptsize$2$}
\put(85,38){\line(-1,0){1}}
\put(81,37){\scriptsize$1$}
\put(85,14){\line(-1,0){1}}
\put(79,13){\scriptsize$-2$}

\thicklines
\put(101,14){\line(-1,-1){20}}
\put(101,14){\line(1,0){8}}
\put(109,14){\line(1,1){8}}
\put(109,14){\line(0,-1){16}}
\put(117,22){\line(1,0){16}}
\put(117,22){\line(0,1){16}}
\put(117,38){\line(1,1){8}}
\put(125,46){\line(1,0){16}}
\put(125,46){\line(0,1){12}}
\put(117,38){\line(-1,0){16}}
\put(101,38){\line(-1,1){20}}
\put(101,38){\line(0,-1){24}}

\put(97,14){\scriptsize$A$}
\put(107,16){\scriptsize$B$}
\put(116,18){\scriptsize$C$}
\end{picture} 

\caption{Tropical curves: (i) a line, (ii) an elliptic curve}
\label{i:fig:trop-examples}
\end{figure}

\begin{Example}
See Figure \ref{i:fig:trop-examples} for examples of tropical curves,
where (i) is a ``{\it tropical line}'' given by 
$F(X,Y) = \min(X, \ Y, \ 1)$ and (ii) is a ``{\it tropical elliptic curve}''
given by $F(X,Y) = \min(2Y+7,\ 2Y+X+4,\ Y+2X,\ Y+X+2,\ Y+6,\ X+3,\ 8)$.

The meaning of ``indifferentiable'' is seen at (i) as follows:
let $A_1$, $A_2$ and $A_3$ be three open domains divided 
by the tropical line. 
We write $l_{12}$, $l_{23}$ and $l_{13}$ for the three boundaries,
and $P$ for their intersection point $l_{12} \cap l_{23} \cap l_{13}$. 
The function $F(X,Y) = \min(X,\ Y,\ 1)$ is ``differentiable'' at 
$(X,Y) \in A_1 \cup A_2 \cup A_3$, 
namely we have $F(X,Y) = 1$ in $A_1$,  $F(X,Y) = Y$ in $A_2$ and 
$F(X,Y) = X$ in $A_3$. 
The function $F(X,Y)$ is ``indifferentiable''
at $(X,Y) \in l_{12} \cup l_{23} \cup l_{13}$
where at least two of $X$, $Y$ and $1$ become the minimum. 
For instance, 
$F(X,Y) = Y = 1$ on $l_{12} \setminus \{P\}$, 
$F(X,Y) = X = Y$ on $l_{23} \setminus \{P\}$,
and $F(X,Y) = X = Y = 1$ at $P$.
\end{Example}

The edges in tropical curves have rational slopes,
and we associate each vertex with a {\it primitive tangent vector}
which is a tangent vector given by a pair of coprime integers.
(If one of the integers is zero, then let another be $\pm 1$.)
The primitive tangent vector is uniquely determined up to sign.
For two vectors $\xi = (n_1,n_2)$ and $\xi' = (n_1',n_2')$,
we set $\xi \wedge \xi'=n_1 n_2' - n_2 n_1'$. 
The following is the definitions of a {\it smooth} tropical curve and 
its {\it genus}.

\begin{Definition}\label{i:def:smoothness}
The tropical curve $\Gamma \subset \R^2$ is smooth if
the following two conditions hold:
\begin{itemize}
\item[(a)]
all vertices in $\Gamma$ are $3$-valent. 
\item[(b)]
For each $3$-valent vertex $v$, let $\xi_1, \xi_2, \xi_3$ be the
primitive tangent vectors which are outgoing from $v$.
Then these vectors satisfy
$\xi_1+\xi_2+\xi_3 = (0,0)$ and $|\xi_i \wedge \xi_j| = 1$ for
$i,j \in \{1,2,3\}, ~i \neq j$.  
\end{itemize}
When a tropical curve $\Gamma$ is smooth, 
the {\it genus} of $\Gamma$ is dim$ \ H_1(\Gamma, \Z)$.
\end{Definition}

A smooth tropical curve is equipped with 
the {\it metric structure} as follows:
\begin{Definition}
Assume $\Gamma$ is a smooth tropical curve.
Let $E(\Gamma)$ be the set of edges in $\Gamma$,
and let $\xi_e$ be the primitive tangent vector of $e \in E(\Gamma)$.
We define the length of edges $l: ~E(\Gamma) \to \R_{\geq 0}$ by
$$
  e \mapsto l(e) = \frac{\parallel e \parallel}
  {\parallel \xi_e \parallel},
$$
where $\parallel~~\parallel$ is any norm in $\R^2$.
\end{Definition}

\begin{Example}
Both of the two tropical curves at Figure \ref{i:fig:trop-examples} are smooth.
For instance, in (ii), 
the $3$-valent vertex $A$ has outgoing primitive tangent 
vectors $(1,0)$, $(0,1)$ and $(-1,-1)$, and the conditions (a) and (b) 
in Definition \ref{i:def:smoothness} are satisfied.
The genera of the curves (i) and (ii) are respectively $0$ and $1$.
The lengths of the edges are $l(AB) = l(BC) = 1$ in (ii), for example. 
\end{Example}

We omit the metric structure for non-smooth tropical curves
for simplicity. See \cite{MZ06}.

\subsubsection{Abelian integral and tropical Jacobian variety.}

Let $\Gamma$ be a smooth tropical curve whose genus $g$ is not zero.
We fix $g$ generators $B_1,\cdots,B_g$ of the fundamental group of 
$\Gamma$ (i.e. generators of the cycles in $\Gamma$).
 
To describe the abelian integral on $\Gamma$, we need some preparations:
for each $e \in E(\Gamma)$, 
we fix a linear map $\alpha_e: e \to [0,1]$
(where we have only two choices).
For $p_1, p_2 \in e$,
we define a {\it fundamental path} $p$ by 
$p = (e; p_1, p_2) \in E(\Gamma) \times \Gamma \times \Gamma$.
For a fundamental path $p = (e; p_1, p_2)$,
we define $[p_1,p_2;\alpha_e] \subset [0,1]$ by
$$
[p_1,p_2;\alpha_e] 
= 
\begin{cases}
[\alpha_e(p_1), \alpha_e(p_2)] & \text{ if } \alpha_e(p_1) \leq \alpha_e(p_2)
\\
[\alpha_e(p_2), \alpha_e(p_1)] & \text{ if } \alpha_e(p_1) > \alpha_e(p_2)
\end{cases}.
$$
For two fundamental paths $p = (e;p_1,p_2)$ and $p' = (e';p_1',p_2')$, 
we define the addition rule 
only when $e=e'$ and $p_2=p_1'$ or $p_1=p_2'$, by
$$
p + p'= 
\begin{cases}
(e;p_1,p_2') \qquad \text{ if } p_2 = p_1' 
\\
(e;p_1',p_2) \qquad \text{ if } p_1 = p_2'.
\end{cases}
$$ 
We define a {\it set of paths} $\mathcal{P}$ on $\Gamma$ by
$$
\mathcal{P} 
= 
\bigl( \underset{p: \, \text{fundamental path}}{\oplus} \Z p \bigr) ~/~ 
\text{\small the addition rule}.
$$
Then $\mathcal{P}$ is an infinite dimensional vector space.
For two fundamental paths $p = (e;p_1,p_2)$ and $p' = (e';p_1',p_2')$,
we define 
\begin{align}\label{i:eq:sgn}
\begin{split}
&\mathrm{sgn}(p,p') 
=
\begin{cases} 
0 & \text{ if } e \neq e'
\\
\mathrm{sgn}\bigl[
(\alpha_e(p_1)-\alpha_e(p_2))(\alpha_e(p_1')-\alpha_e(p_2')) \bigr]
& \text{ if } e=e'
\end{cases},
\\
&l(p \cap p') = 
\begin{cases} 
0 & \text{ if } e \neq e'
\\
\bigl|[p_1,p_2; \alpha_e] \cap[p_1',p_2';\alpha_e]\bigr| 
\cdot 
l(e)
& \text{ if } e=e'
\end{cases},
\end{split}
\end{align}
and define a {\it bilinear form} of fundamental paths by
\begin{align}\label{i:eq:bilinear}
  \langle ~,~ \rangle 
  ~: (p, p') \mapsto \langle p,p'\rangle 
  = \mathrm{sgn}(p,p') \cdot l(p \cap p'). 
\end{align} 
This naturally gives the bilinear form 
$\langle ~,~ \rangle:~ \mathcal{P} \times \mathcal{P} \to \R$ 
on $\mathcal{P}$.
Briefly speaking, the bilinear form of two paths $p$ and $p'$ 
on $\Gamma$ is ``the length of the common part of $p$ and $p'$ 
with the sign depending on the directions of the two paths''.

\begin{Example}
See Figure \ref{i:fig:trop-example2} for 
the smooth tropical curve $\Gamma$ given by
\begin{align*}
F(X,Y) = \min(2Y, \ Y+3 X, \ Y+2X, \ Y+X+1, \ Y+4, 11).
\end{align*}
The genus of $\Gamma$ is $2$, and we fix the basis $B_1$ and $B_2$ 
of the fundamental group of $\Gamma$ as depicted.
The bilinear forms of $B_1$ and $B_2$ take the values as
$$
\langle B_1, B_1 \rangle = 20, 
\qquad 
\langle B_1, B_2 \rangle = -7,
\qquad 
\langle B_2, B_2 \rangle = 14.
$$
Let us demonstrate how to compute $\langle B_1, B_2 \rangle$:
the common part of $B_1$ and $B_2$ is 
the edge $PQ$. We set $\overset{\curvearrowright}{QP} \subset B_1$
and $\overset{\curvearrowright}{PQ} \subset B_2$
which are fundamental paths on $\Gamma$.
Then we have $l(\overset{\curvearrowright}{QP} \cap 
\overset{\curvearrowright}{PQ}) = l(PQ) = 7$ and 
$\mathrm{sgn}(\overset{\curvearrowright}{QP},\overset{\curvearrowright}{PQ}) 
= -1$, and the result follows.
\end{Example}

\begin{figure}
\unitlength=0.8mm

\begin{picture}(100,70)(-20,0)

\put(10,10){\vector(1,0){50}}
\put(58,6){\scriptsize$X$}
\put(20,0){\vector(0,1){66}}
\put(16,64){\scriptsize$Y$}

\thicklines

\put(20,10){\line(-2,-3){8}}
\put(20,10){\line(1,1){8}}
\put(28,18){\line(2,1){16}}
\put(44,26){\line(1,0){15}}
\put(20,54){\line(-2,3){8}}
\put(20,54){\line(1,-1){8}}
\put(28,46){\line(2,-1){16}}
\put(44,38){\line(1,0){15}}
\put(20,10){\line(0,1){44}}
\put(28,18){\line(0,1){28}}
\put(44,26){\line(0,1){12}}

\thinlines

\put(24,32){\oval(4,25)}
\put(25.1,36){\scriptsize$\wedge$}
\put(22,31){\scriptsize${B_{\!1}}$}
\put(36,32){\oval(12,15)}
\put(40.7,31){\scriptsize$\wedge$}
\put(34,31){\scriptsize$B_{\!2}$}

\put(15,53){\scriptsize$11$}
\put(16,11){\scriptsize$O$}
\put(28,10){\line(0,-1){1}}
\put(27.5,6){\scriptsize$1$}
\put(44,10){\line(0,-1){1}}
\put(43.5,6){\scriptsize$3$}
\put(20,18){\line(-1,0){1}}
\put(16,17){\scriptsize$2$}
\put(20,46){\line(-1,0){1}}
\put(16,45){\scriptsize$9$}

\put(27.5,47){\scriptsize$P$}
\put(27.5,15){\scriptsize$Q$}
\end{picture}
\caption{Tropical curve of genus $2$}
\label{i:fig:trop-example2}
\end{figure}

Now we introduce the abelian integral and the tropical 
Jacobian variety for $\Gamma$:
\begin{Definition}
Fix $P_0 \in \Gamma$.
The {\it abelian integral} $\psi: ~\Gamma \to \R^g$ is given by
$$
P \mapsto \psi(P) 
= (\langle B_i, \overset{\curvearrowright}{P_0 P} \rangle)_{i=1,\ldots,g},
$$
where we choose a path $\overset{\curvearrowright}{P_0 P}$ from 
$P_0$ to $P$.
The map $\psi$ induces the map from a set of divisors $\mathrm{Div}(\Gamma)$
on $\Gamma$ to $\R^g;$
$$
  \sum_{i \in I} n_i P_i \mapsto \sum_{i \in I} n_i \, \psi(P_i),
$$
where $I$ is a finite set and $n_i \in \Z$.
\end{Definition}

\begin{Definition}\label{i:def:omega}
The period matrix $\Xi$ of $\Gamma$ is given by
\begin{align}\label{i:eq:tropOmega}
  \Xi = (\langle B_i, B_j\rangle)_{i,j=1,\ldots,g} \in M_g(\R).
\end{align}
The tropical Jacobian variety $J(\Gamma)$ of $\Gamma$ is the
$g$-dimensional real torus given by
\begin{align}\label{i:eq:tropJ}
  J(\Gamma) = \R^g / \Xi \Z^g.
\end{align}
\end{Definition}

\begin{Remark}
The matrix $\Xi$ is symmetric and positive definite by definition,
and $J(\Gamma)$ is a tropical analogue of Jacobian variety.
By removing all infinite edges of $\Gamma$,
we obtain the maximal compact subgraph $\Gamma^{\mathrm{cpt}}$  
of $\Gamma$.
Though the map $\psi$ depends on a choice of the path 
$\overset{\curvearrowright}{P_0 P}$,  
the induced map
$\Gamma^{\mathrm{cpt}} \to J(\Gamma)$ does not depend on the choice
and becomes injective.
When $g=1$,
$\psi$ induces the isomorphism from 
$\Gamma^{\mathrm{cpt}}$ to $J(\Gamma)$.
\end{Remark}

\begin{Example}
The tropical curve of genus $1$ depicted at
Figure \ref{i:fig:trop-examples} (ii)
has the period matrix $\Xi = 9$,
and the Jacobian variety $\R / 9 \Z$.
As for the tropical curve of genus $2$ depicted at Figure
\ref{i:fig:trop-example2},
the period matrix and the Jacobian variety are 
respectively obtained as  
$$
\Xi = \begin{pmatrix} 20 &-7 \\ -7 & 14 \end{pmatrix},
\qquad 
J(\Gamma) = \R^2 / \Xi \Z^2.
$$
\end{Example}

\subsubsection{Tropical Riemann theta function.}

Fix a positive integer $g$ and a symmetric and positive definite 
matrix $\Xi \in M_g(\R)$.
(Here the matrix $\Xi$ is not always 
a period matrix of some tropical curve.) 

\begin{Definition}\label{i:def:theta}
The {\it tropical Riemann theta function} $\Theta({\bf Z}; \Xi)$ 
of ${\bf Z}\in \R^g$ is defined by 
$$
\Theta({\bf Z};\Xi)
=
\min_{{\bf n} \in \Z^g} \left\{{\bf n} \cdot 
\bigl(\tfrac{1}{2}\, \Xi {\bf n} + {\bf Z}\bigr) \right\}.
$$   
We call the $g$-dimensional real torus given by
\begin{align}\label{i:eq:g-J}
J_\Xi = \R^g / \Xi \Z^g
\end{align} 
the {\it principally polarized tropical abelian variety}.
(If $\Xi$ is the period matrix of a tropical curve $\Gamma$,
then $J_\Xi$ is nothing but the tropical Jacobian variety $J(\Gamma)$.)
\end{Definition}

It is easy to see the following:
\begin{Lemma}
The function $\Theta({\bf Z}) = \Theta({\bf Z}; \Xi)$ 
satisfies the quasi-periodicity:
\begin{align}\label{i:theta-quasi}
\Theta({\bf Z} + \Xi {\bf m})
=
- {\bf m} \cdot \Bigl(\frac{1}{2} \Xi {\bf m} + {\bf Z}\Bigr) 
+ \Theta({\bf Z})
\qquad {\bf m} \in \Z^g.
\end{align}
\end{Lemma}

\begin{Remark}
Recall the Riemann's theta function:
\begin{align}
\theta({\bf z}; K) 
= 
\sum_{{\bf n} \in \Z^g} \exp \bigl(\pi {\rm i} \ {\bf n} \cdot 
(K {\bf n} +2 {\bf z})\bigr)
\qquad {\bf z} \in \mathbb{C}^g,
\end{align}
where $K \in M_g(\mathbb{C})$ is symmetric and ${\rm Im} K$ 
is positive definite.
The function $\theta({\bf z}; K)$ satisfies the periodicity and 
quasi-periodicity: 
\begin{align}
\begin{split}
&\theta({\bf z} + {\bf m}; K) = \theta({\bf z}; K),
\\
&\theta({\bf z}+ K {\bf m}; K) 
=  
\exp \bigl(-\pi {\rm i} \ {\bf m} \cdot (K {\bf m} +2 {\bf z})\bigr)\theta({\bf z}; K),
\end{split}
\end{align}
for ${\bf m} \in \Z^g$.
Remark that only the quasi-periodicity remains in the tropical case.
\end{Remark}

\subsection{General solution for tropical periodic Toda lattice}
\label{i:sec:pToda}

We briefly present the results on 
the general solution for the tropicalization of 
the $N$-periodic Toda lattice (trop-pToda) following \cite{IT08}.
This section offers not only an application of the tropical geometry,
but also a preparation for the next section
where we study the periodic BBS with the similar description 
to \S \ref{i:sec:TodaBBS}.
We emphasize that the independent variables 
of the trop-p Toda in this section are in $\R$,
whereas in the next section their integer parts
(i.e. ultradiscretization) will be studied.

The trop-pToda is given by 
the piecewise-linear evolution equation: 
\begin{align}\label{i:eq:UDpToda}
\begin{split}
&Q_j^{t+1} = \min(W_j^t, Q_j^t-X_j^t),
\qquad X_j^t = \min_{0 \leq k \leq N-1}
\bigl(\sum_{l=1}^k (W_{j-l}^t - Q_{j-l}^t)\bigr),
\\
&W_j^{t+1} = Q_{j+1}^t+W_j^t - Q_j^{t+1}
\end{split}
\end{align}
on the phase space
$\displaystyle{
\mathcal{T} 
= 
\{(Q_j,W_j)_{j \in \Z / N\Z} ~|~ \sum_{j=1}^N Q_j < \sum_{j=1}^N W_j \}
\subset \R^{2N}.}
$
(In \cite{IT08}, this system is called the ultradiscrete periodic Toda lattice,
where ``ultradiscrete'' means ``tropical " in our present terminology,
hence the naming ``trop p-Toda" here.)

\begin{Prop}{\rm \cite[Prop. 2.1]{KT02}}
Eq. \eqref{i:eq:UDpToda} is obtained 
as the tropicalization of the discrete Toda lattice 
\eqref{i:eq:Toda-q} and \eqref{i:eq:Toda-w} with
a periodic boundary condition $q_{j+N}^t = q_j^t$,
$w_{j+N}^t = w_j^t$ and the condition
$\prod_{j=1}^N w_j^t / q_j^t < 1$ so that the tropicalization of 
$(1- \prod_{j=1}^N w_j^t / q_j^t)$ is zero.
\end{Prop}
\proof
First we show that 
under $N$-periodic boundary condition,
\eqref{i:eq:Toda-q} and \eqref{i:eq:Toda-w} are expressed as
\begin{equation}\label{t:aug3b}
q^{t+1}_j = w^t_j + 
q^t_j\frac{1-\prod_{l=1}^N(w^t_{l}/q^t_{l})}
{\sum_{k=0}^{N-1}\prod_{l=1}^k(w^t_{j-l}/q^t_{j-l})}.
\end{equation}
With a common denominator, the RHS of (\ref{t:aug3b}) is rewritten as 
\begin{equation}\label{t:aug3c}
q^{t+1}_j = 
q^t_j \frac{\sum_{k=0}^{N-1}\prod_{l=1}^k(w^t_{j-l+1}/q^t_{j-l+1})}
{\sum_{k=0}^{N-1}\prod_{l=1}^k(w^t_{j-l}/q^t_{j-l})}.
\end{equation}
On the other hand, \eqref{t:eq:toda} 
are equivalent to \eqref{i:eq:Toda-q} and \eqref{i:eq:Toda-w} 
under the substitution of variables
$q^t_{j+1}=1/x_j, w^t_j = 1/y_j, q^{t+1}_j = 1/{\tilde x}_j, 
w^{t+1}_j = 1/{\tilde y}_j$.
Hence the birational $R$ given by
(\ref{t:eq:RPP}) is also equivalent to 
the $N(=n+1)$-periodic discrete Toda lattice equation.
From (\ref{t:eq:RPP}) with the above variable substitution
we obtain
\begin{align}
q^{t+1}_j &= 
q^t_{j+1} \frac{\sum_{k=1}^{N} \prod_{l=k}^N (1/q^t_{j+l+1})  \prod_{l=1}^k (1/w^t_{j+l})}
{\sum_{k=1}^{N} \prod_{l=k}^N (1/q^t_{j+l})  \prod_{l=1}^k (1/w^t_{j+l-1})} 
\nonumber \\
&= q^t_{j} \frac{\sum_{k=1}^{N} \prod_{l=k}^{N-1} (1/q^t_{j+l+1})  \prod_{l=0}^{k-1} (1/w^t_{j+l+1})}
{\sum_{k=1}^{N} \prod_{l=k}^{N-1} (1/q^t_{j+l})  \prod_{l=0}^{k-1} (1/w^t_{j+l})} \nonumber \\
&= q^t_{j} \frac{\sum_{k=1}^{N} \prod_{l=k}^{N-1} (w^t_{j+l+1}/q^t_{j+l+1})}
{\sum_{k=1}^{N} \prod_{l=k}^{N-1} (w^t_{j+l}/q^t_{j+l})}, \label{t:aug3d}
\end{align}
which becomes (\ref{t:aug3c})
by the replacements $k \rightarrow N-k, l \rightarrow N-l$.

Next we apply tropicalization to \eqref{t:aug3b}
and \eqref{i:eq:Toda-w}.
Since the numerator of the second term in (\ref{t:aug3b}) is a constant
with respect to $t$, 
the tropicalization of this numerator is constantly zero
under the given condition on $\prod_{j=1}^N w_j^t / q_j^t$.
Thus \eqref{t:aug3b} and \eqref{i:eq:Toda-w} yield \eqref{i:eq:UDpToda}.
\qed

The system \eqref{i:eq:UDpToda} has $N+1$ conserved quantities
$H_k ~(k=1,\ldots,N+1)$
given by the tropical polynomials on $\mathcal{T}$ as
\begin{align}\label{i:udpToda-H}
\begin{split}
&H_1 = \min_{1 \leq j \leq N}(Q_j,W_j),
\\  
&H_2 = \min\bigl(\min_{1\leq i<j \leq N} (Q_i+Q_j),
            \min_{1\leq i<j \leq N} (W_i+W_j),
            \min_{1\leq i,j\leq N, j \neq i,i-1} (Q_i+W_j)\bigr),
\\
&\ldots, \quad
H_N = \min \bigl(\sum_{j=1}^N Q_j, \sum_{j=1}^N W_j \bigr),
\quad 
H_{N+1} = \sum_{j=1}^N (Q_j + W_j).
\end{split}
\end{align}
Fix $C = (C_k)_{k=1,\cdots,N+1} \in \R^{N+1}$ and define the 
isolevel set $\mathcal{T}_C$ by
\begin{align}\label{i:TC}
\mathcal{T}_C 
= 
\{ \tau \in \mathcal{T} ~|~ H_k(\tau) = C_k ~(k=1,\cdots,N+1) \},
\end{align}
and the affine tropical curve $\Gamma_C$ given by the indifferentiable 
points of 
\begin{align}\label{i:trop-Toda}
F(X,Y)
= 
\min\bigl(2Y, Y+ \min(NX, (N-1)X+C_1, \ldots, X+C_{N-1}, C_N), C_{N+1}\bigr).
\end{align}
We call $\Gamma_C$ the spectral curve of the trop-pToda.
For the derivation of the conserved quantities \eqref{i:udpToda-H} 
and $F(X,Y)$ \eqref{i:trop-Toda}, see \cite[\S 3.1, \S 3.2]{IT08}.
We set $L$ and $\lambda_k$, $\eta_k$ for $k = 0,\ldots,N-1$ as 
\begin{align}\label{i:C-lambda}
\begin{split}
&L = C_{N+1} - 2 (N-1) C_1,
\\
&\lambda_0 = 0, \qquad \lambda_k = C_{k+1} - C_k
\qquad k=1,\ldots,N-1, 
\\
&\eta_0=L, \qquad \eta_k = L - 2 \sum_{j=1}^{N-1} \min(\lambda_k,\lambda_j) 
\qquad k=1,\ldots,N-1.
\end{split}
\end{align}
The curve $\Gamma_C$ is smooth if and only if 
$\lambda_1 < \lambda_2 < \cdots < \lambda_{N-1}$ and $\eta_k > 0$ for 
$k=1,\ldots,N-1$. The genus of the smooth $\Gamma_C$ is
$N-1$. See Figure 3 for $\Gamma_C$, where
we set $C_1 =0$ for simplicity.

Assume $\Gamma_C$ is smooth, and write $g$ for the genus, $g=N-1$. 
Fix the basis $B_1,\ldots,B_g$ of the fundamental group $\pi_1(\Gamma_C)$
as Figure 3.
In what follows we denote by $\Omega$ the period matrix \eqref{i:eq:tropOmega} 
of $\Gamma_C$. The matrix $\Omega = (\Omega_{ij})_{i,j = 1,\ldots,g}$ is 
explicitly written as 
\begin{align}\label{i:omega}
\Omega_{ij} = 
\begin{cases}
\eta_{i-1} + \eta_{i} + 2 (\lambda_{i} - \lambda_{i-1}) 
\qquad i=j 
\\
-\eta_i \qquad j=i+1
\\
-\eta_j \qquad i=j+1
\\
0 \qquad \text{otherwise}
\end{cases}
\end{align}
and we get the tropical Jacobian variety of $\Gamma_C$ as 
$J(\Gamma_C) = \R^g / \Omega \Z^g$.

\begin{figure}\label{i:fig:curveToda}
\unitlength=0.8mm
\begin{picture}(140,78)(-25,3)

\put(20,10){\vector(1,0){70}}
\put(0,10){\line(1,0){20}}
\put(91,9){\scriptsize $X$}

\thicklines

\put(15,74){\scriptsize $L$}
\put(20,6){\scriptsize $0$}
\put(28,10){\line(0,-1){1}}
\put(27,6){\scriptsize $\lambda_1$}
\put(36,10){\line(0,-1){1}}
\put(35,6){\scriptsize $\lambda_2$}
\put(45,6){$\cdots$}
\put(56,10){\line(0,-1){1}}
\put(55,6){\scriptsize $\lambda_{g-1}$}
\put(68,10){\line(0,-1){1}}
\put(67,6){\scriptsize $\lambda_{g}$}

\put(20,10){\line(-1,-2){4}}
\put(20,10){\line(4,5){8}}
\put(28,20){\line(1,1){8}}
\put(36,28){\line(2,1){10}}
\put(53,35){\line(5,1){15}}
\put(68,38){\line(1,0){15}}

\put(20,75){\line(-1,2){4}}
\put(20,75){\line(4,-5){8}}
\put(28,65){\line(1,-1){8}}
\put(36,57){\line(2,-1){10}}
\put(53,50){\line(5,-1){15}}
\put(68,47){\line(1,0){15}}

\put(20,10){\line(0,1){65}}
\put(28,20){\line(0,1){45}}
\put(36,28){\line(0,1){29}}
\put(56,35.5){\line(0,1){14}}
\put(68,38){\line(0,1){9}}
\put(46,42){$\cdots$}

\thinlines 

\put(24,42){\oval(4,45)}
\put(26,32){\vector(0,1){3}}
\put(22,29){\scriptsize $B_{\!1}$}
\put(32,42){\oval(4,27)}
\put(34,37){\vector(0,1){3}}
\put(30,34){\scriptsize $B_{\!2}$}
\put(62,42){\oval(8,8)}
\put(66,41){\vector(0,1){3}}
\put(61,41){\scriptsize $B_{\!g}$}

\end{picture}
\caption{Spectral curve for the trop-pToda}
\end{figure}
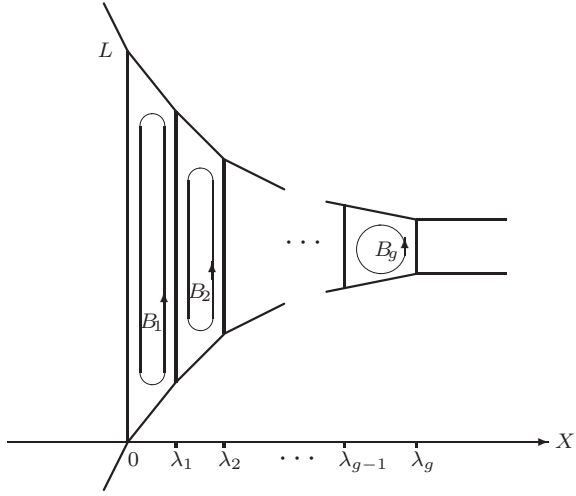

The general solution for the trop-pToda is obtained 
by the following theorem:

\begin{Theorem}\label{i:th:pToda} 
When $\Gamma_C$ is smooth, we have the following:
\\
(i) {\rm \cite[Th.~3.5]{IT09}} 
Fix ${\bf Z}_0 \in \R^g$ and define 
$T_n^t = \Theta({\bf Z}_0 + \boldsymbol{\lambda} t - L {\bf e}_1 n)$,
where 
$$
\boldsymbol{\lambda} = (\lambda_1, \lambda_2 -\lambda_1, \ldots,
\lambda_g-\lambda_{g-1}), ~
{\bf e}_1 = (1,0,\ldots,0) \in \R^g.
$$ 
The general solution for the trop-pToda is given by
\begin{align}\label{i:QWsol}
\begin{split}
  &Q_n^t = T_{n-1}^{t} + T_{n}^{t+1}
           - T_{n-1}^{t+1} - T_{n}^{t}+C_1,
  \\
  &W_n^t = T_{n-1}^{t+1} + T_{n+1}^t-
          T_{n}^{t} - T_{n}^{t+1} + L+C_1.
\end{split}
\end{align}
(ii) {\rm \cite[Th.~1.3]{IT09b}} This solution induces the isomorphism 
$J(\Gamma_C) \stackrel{\sim}{\rightarrow} \mathcal{T}_C$.
In particular, the time evolution of the trop-pToda is linearized
on $J(\Gamma_C)$, whose velocity is $\boldsymbol{\lambda}$.
\end{Theorem}

\begin{Example} 
The case of $N=2$.
In this simplest case we can explicitly construct the isomorphism
$\alpha$:
\begin{align*}
\begin{matrix}
\mathcal{T}_C & \stackrel{\alpha}{\to} & \Gamma_C^{\mathrm{cpt}}
& \stackrel{\psi}{\to} & J(\Gamma_C) \\
(Q_1,W_1,Q_2,W_2) & \mapsto & P=(\min(Q_2,W_1), Q_1+W_1) & \mapsto 
& \langle B_1,\overset{\curvearrowright}{P_0 P} \rangle
\end{matrix}.
\end{align*}
The solution \eqref{i:QWsol} induces the inverse map of
$\psi \circ \alpha$. 
Let us consider the case of $C=(0,3,8)$, where $\Gamma_C$ is depicted as 

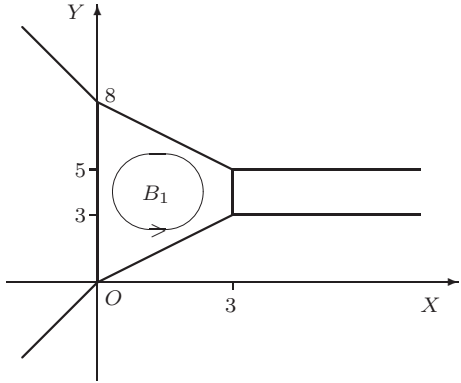
\unitlength=1.0mm
\begin{picture}(60,55)(-10,-12)

\put(3,5){\vector(1,0){60}}
\put(58,1){\scriptsize$X$}
\put(15,-8){\vector(0,1){50}}
\put(11,40){\scriptsize$Y$}
\put(16,2){\scriptsize$O$}

\put(16,29){\scriptsize$8$}
\put(15,20){\line(-1,0){1}}
\put(12,19){\scriptsize$5$}
\put(15,14){\line(-1,0){1}}
\put(12,13){\scriptsize$3$}
\put(33,5){\line(0,-1){1}}
\put(32,1){\scriptsize$3$}

\put(23,17){\oval(12,10)}
\put(22,11.2){\scriptsize$>$}
\put(21,16){\scriptsize${B_1}$}

\thicklines

\put(15,5){\line(2,1){18}}
\put(33,14){\line(1,0){25}}
\put(33,14){\line(0,1){6}}
\put(15,29){\line(2,-1){18}}
\put(33,20){\line(1,0){25}}
\put(15,29){\line(-1,1){10}}
\put(15,5){\line(-1,-1){10}}
\put(15,5){\line(0,1){24}}

\end{picture}

\noindent
The following is an example of linearization, 
where one sees $\boldsymbol{\lambda} = (3)$.
We set $P_0 = O$:
\end{Example}
\begin{align*}
\begin{matrix}
& & \mathcal{T}_C = \{(Q_1,W_1,Q_2,W_2)\} & \stackrel{\alpha}{\to} & 
\Gamma_C^{\mathrm{cpt}} & \stackrel{\psi}{\to} & J(\Gamma_C) \simeq \R/16 \Z 
\\[1mm]
\texttt{t=0}  & & (3,4,0,1) & & (0,7) & & 9
\\
\texttt{t=1} & & (3,1,0,4) & & (0,4) & & 12
\\
\texttt{t=2} & & (1,0,2,5) & & (0,1) & & 15
\\
\texttt{t=3} & & (0,2,3,3) & & (2,2) & & 2 \equiv 18
\\
\texttt{t=4} & & (0,5,3,0) & & (3,5) & & 5 \equiv 21
\end{matrix}
\end{align*}

For general $N>2$, the isomorphism 
$\mathcal{T}_C \stackrel{\sim}{\to} J(\Gamma_C)$
is regarded as a composition of the injective map
$\alpha: \mathcal{T}_C \to \mathrm{Div}_{\mathrm{eff}}^{g}(\Gamma_C)$
and the abelian integral $\psi$,
but $\alpha$ becomes too complicated.

\begin{Example}
The case of $N=3$ and $C=(0,2,6,19)$. 
We have $\Omega = \begin{pmatrix} 34 & -11 \\
                     -11 & 22 \end{pmatrix}$ and 
$\boldsymbol{\lambda}=(2,2)$.
Observe that the velocity in $J(\Gamma_C)$ is indeed $\boldsymbol{\lambda}$. 
\end{Example}
\begin{align*}
\begin{matrix}
    & \mathcal{T}_C = \{(Q_1,W_1,Q_2,W_2,Q_3,W_3) \} 
  & \simeq & J(\Gamma_C) \\[1mm] 
  \texttt{t=0} & (2, 1, 0, 9, 4,  3) & & (29, -3) \\
  \texttt{t=1} & (1, 0, 2, 11, 3, 2) & & (31, -1) \\
  \texttt{t=2} & (0, 2, 4, 10, 2, 1) & & (10, -10) \equiv (33,1) \\ 
  \texttt{t=3} & (1, 5, 4, 8, 1, 0) & & (12, -8) \\
  \texttt{t=4} & (2, 7, 4, 5, 0,  1) & & (14,-6) \\ 
  \texttt{t=5} & (2, 9, 4, 1, 0, 3) & & (16, -4) \\
\end{matrix}  
\end{align*} 

\begin{Remark}
Theorem \ref{i:th:pToda} corresponds to a tropical version
of \cite{DT76,KvM75} where the general solution for the periodic Toda lattice
is studied by using (complex) algebraic geometry. 
When $\Gamma_C$ is not smooth, neither the structure of $\mathcal{T}_C$ 
nor the solution has been clarified yet.
It requires a further study 
on a degeneration of the period matrix $\Omega$ and Jacobian variety
$J(\Gamma_C)$. 
\end{Remark}

\begin{Remark}
In \cite{IT07}, $\lambda_i$s and $\Omega$ were derived by 
directly ultradiscretizing  
abelian integrals on the spectral curve of the periodic discrete Toda lattice.
By combining this strategy and tropical curve theory,
Theorem \ref{i:th:pToda}(i) was proved in \cite{IT09}.
\end{Remark}

\subsection{Periodic BBS and tropical geometry}
\label{i:pBBS-Toda}

As a periodic version of \S \ref{i:sec:TodaBBS},
we have an embedding of the states of periodic BBS in those of 
the trop-pToda \cite{KT02,IT08}.
Differently from the case of the original BBS,
this embedding is not always consistent with
the time evolution of the trop-pToda.
We revisit the results in \S \ref{k:ss:nsol}
with this embedding and tropical geometry.

Let $\mathcal{P}_L(\mu)$ \eqref{k:ils} 
be the isolevel set of the $L$-periodic BBS 
with the configuration $\mu = (i_g^{m_{i_g}}, \ldots,i_1^{m_{i_1}})$.
(Recall we assume $i_1 < \cdots < i_g$.) 
Set $N=1+\sum_{k=1}^g m_{i_k}$ in \S \ref{i:sec:pToda},
and fix $C = (C_1,\ldots,C_{N+1})$ at \eqref{i:TC} as
\begin{align}\label{i:general-C}
\begin{split}
&C_1=0, ~~C_{N+1} = L, 
\\
&C_{m_{i_1}+\cdots+m_{i_{k-1}}+l+1} 
= 
\sum_{j=1}^{k-1}i_j m_{i_j} + i_k l\qquad k=1,\ldots,g, ~l=1,\ldots,m_{i_k}.
\end{split}
\end{align} 
Then the embedding $\eta : \mathcal{P}_L(\mu) \to \mathcal{T}_C$
is defined as follows:
among $L$ boxes, fix ``the leftmost box'' (it can be any box) 
and do the following procedure:
\begin{enumerate}
\item
if the leftmost box is occupied, then set 
$Q_1 =$(the number of the first consecutive balls from 
the left), otherwise set $Q_1=0$.
\item
Set $W_i=$(the number of $i$-th consecutive empty boxes from the left)
for $i=1,\cdots,N$.
If $Q_1 \neq 0$, set 
$Q_i=$(the number of the $i$-th consecutive balls from the left),
otherwise set 
$Q_i=$(the number of the ($i-1$)-th consecutive balls from the left)
for $i=2,\cdots,N$.
\end{enumerate}
Then we obtain
\begin{align*}
&\overbrace{1..11}^{W_1}\overbrace{2..22}^{Q_2}\overbrace{11...1}^{W_2}
........\overbrace{1...1}^{W_{N-1}}\overbrace{22..2}^{Q_{N}}
\overbrace{1...1}^{W_{N}}
& \text{when $Q_1 = 0$},
\\
& \overbrace{2..22}^{Q_1}\overbrace{11...1}^{W_1}
\overbrace{2...22}^{Q_2}\overbrace{1...1}^{W_2}
........\overbrace{1...1}^{W_{N-1}}\overbrace{22..2}^{Q_{N}}
& \text{when $Q_1 > 0$}.
\end{align*}
Note that we have $Q_1 = 0$ or $W_{N}=0$.

\begin{Example}\label{i:ex:pBBS-Toda}
The case of $L=9$, $\mu=(3,1)$ and $N=3$.
We have $C=(0,1,4,9)$. 
The evolution of the periodic BBS in $\mathcal{P}_L(\mu)$ is at the left,
and its embedding in 
$\mathcal{T}_C = \{(Q_1,W_1,Q_2,W_2,Q_3,W_3)\}$ is in the center.
The evolution of the trop-pToda in $\mathcal{T}_C$ is 
written at the right. 
\begin{verbatim} 
     t=0   122211211   (0,1,3,2,1,2)   (0,1,3,2,1,2) 
     t=1   111122122   (0,4,2,1,2,0)   (0,4,2,1,2,0) 
     t=2   222111211   (3,3,1,2,0,0)   (3,3,1,2,0,0) 
     t=3   111222121   (0,3,3,1,1,1)   (3,1,1,1,0,3) 
\end{verbatim} 
The time evolution does not agree with the embedding at $t=3$.
\end{Example}

Let $T_\infty$ and $T_{\text{Toda}}$ be the time evolution operators 
in $\mathcal{P}_L(\mu)$ and $\mathcal{T}_C$ respectively.
As one observes in the above example, in general
the following diagram is {\it not} commutative:
$$
\begin{matrix}
\mathcal{P}_L(\mu) & ~~\stackrel{\eta}{\to} & \mathcal{T}_C
\\
\downarrow_{T_\infty} & & ~~~~\downarrow_{T_{\rm Toda}}
\\
\mathcal{P}_L(\mu) & ~~\stackrel{\eta}{\to} & \mathcal{T}_C
\end{matrix}
$$
i.e. $\eta \circ T_\infty \neq T_{\rm Toda} \circ \eta$. 

\begin{Prop}{\rm \cite[Prop.~4.4]{IT08}}\label{i:Prop:T_Cs}
Let $s$ be a shift operator on $\mathcal{T}_C$ given by
$$
s : (Q_1,W_1,Q_2,W_2, \ldots, Q_{N},W_{N})
\mapsto (Q_2,W_2,\ldots, Q_{N},W_{N},Q_1,W_1).
$$
We write $\mathcal{T}_C / \sim_s$ for the quotient space 
of $\mathcal{T}_C$ with respect to 
the action of $s$.
Then the following diagram is commutative:
\begin{align}\label{i:eta-comm}
\begin{matrix}
\mathcal{P}_L(\mu) & ~~\stackrel{\eta^\ast}{\to} & \mathcal{T}_C / \sim_s
\\
\downarrow_{T_\infty} & & ~~~~\downarrow_{T_{\rm Toda}}
\\
\mathcal{P}_L(\mu) & ~~\stackrel{\eta^\ast}{\to} & \mathcal{T}_C / \sim_s
\end{matrix}
\end{align}
The induced map $\eta^\ast$ gives one-to-one correspondence 
between $\mathcal{P}_L(\mu)$ and 
$(\mathcal{T}_C \cap \Z^{2N}) / \sim_s$.
\end{Prop}

In Example \ref{i:ex:pBBS-Toda},
we have $s:(3,1,1,1,0,3) \mapsto (0,3,3,1,1,1)$
at $t=3$, which indicates the commutativity of the diagram
\eqref{i:eta-comm}.

Now we come to the final stage of this section.
Set $m_{i_k} = 1 ~(k=1,\ldots,g)$, 
thus \eqref{i:general-C} gives \eqref{i:C-lambda} 
with $\lambda_k=i_k$ and $\eta_k = p_{i_k}$ \eqref{k:ipj}.
Then the spectral curve $\Gamma_C$ is smooth,
and the isolevel set $\mathcal{T}_C$ is isomorphic with
$J(\Gamma_C)$ (Theorem \ref{i:th:pToda}).
On the other hand, 
we have the important result on the periodic BBS,
the one-to-one correspondence between 
$\mathcal{P}_L(\mu)$ and $\mathcal{J}_L(\mu)$ due to the map $\Phi$
(Theorems \ref{k:th:pmain}, \ref{k:th:udt}).
These results and Proposition \ref{i:Prop:T_Cs} are summarized as follows,
which gives a tropical geometrical 
explanation for $\mathcal{J}_L(\mu)$:

\begin{Prop}\label{i:prop:Omega-F}
(i){\rm \cite[Lemma 2.5]{IT08}} 
We keep the setting of 
$J(\Gamma_C)$ and $\mathcal{J}_L(\mu)$.
Let $J_F = \R^g / F \Z^g$ be the 
principally polarized tropical abelian variety 
where the matrix $F$ is defined by \eqref{k:jfred}.
Let $c$ be a shift operator on $J(\Gamma_C)$ given by
$$
c : [(Z_1,Z_2,\ldots,Z_g)] \mapsto [(Z_1+L,Z_2,\ldots,Z_g)].
$$
Then we have 
$
J_F \simeq J(\Gamma_C) / \sim_c,
$
where $J(\Gamma_C) / \sim_c$ 
is the quotient space of $J(\Gamma_C)$ with respect to $c$.
\\
(ii) {\rm \cite[Eq.~(4.5)]{IT08}}
We have the following commutative diagram:
$$
\begin{matrix}
\mathcal{P}_L(\mu) & ~~\stackrel{\eta^\ast}{\hookrightarrow} & 
\mathcal{T}_C / \sim_s & \twoheadleftarrow & \mathcal{T}_C
\\
_\Phi \updownarrow_{1:1} & & \downarrow_{\text{iso.}} & &
\downarrow_{\text{iso.}} 
\\
\mathcal{J}_L(\mu) & \subset & 
J_F \simeq J(\Gamma_C) / \sim_c & \twoheadleftarrow & J(\Gamma_C) 
\end{matrix}
$$
\end{Prop}

\begin{Example}
We illustrate Proposition \ref{i:prop:Omega-F} (i) by using 
Example \ref{i:ex:pBBS-Toda}.
The period matrices $\Omega$ \eqref{i:omega} and 
$F$ \eqref{k:jfred} of $J(\Gamma_C)$ and $J_F$ 
are respectively 
$$
\Omega 
= 
\begin{pmatrix} 16 & -5 \\ -5 & 10 \end{pmatrix},
\qquad 
F = \begin{pmatrix} 7 & 2 \\ 2 & 7 \end{pmatrix}.
$$
By changing the basis of $\pi_1(\Gamma_C)$ from $B_1, B_2$ to
$B_1, B_1+B_2$, the period matrix $\Omega$ is transformed into 
$$
\Omega' = 
\begin{pmatrix} 1 & 0 \\ 1 & 1 \end{pmatrix}
\Omega
\begin{pmatrix} 1 & 1 \\ 0 & 1 \end{pmatrix}
=
\begin{pmatrix} 16 & 11 \\ 11 & 16 \end{pmatrix}.
$$
(Of course we have $J_\Omega \simeq J_{\Omega'}$.)
The shift operator $c$ acts on $J_{\Omega'}$ as
$c : [(Z_1,Z_2)] \mapsto [(Z_1+9,Z_2+9)]$, and 
we have $\R^2 / F \Z^2 \simeq J_{\Omega'} / \sim_c$.
This indicates $J_F \simeq J(\Gamma_C) / \sim_c$.
\end{Example}

\ack

The authors thank Miyuki Takagi for valuable advice.
R.~I. is partially supported by Grant-in-Aid for Young Scientists (B)
(22740111).
A.~K. and T.~T. are partially supported by Grand-in-Aid for Scientific 
Research JSPS No. 195403931 and 22540241.


\section*{References}

\end{document}